\documentclass[11pt]{article}

\usepackage{hyperref}
\usepackage{amsmath}
\usepackage{amsfonts}
\usepackage{tikz}
\usepackage{authblk}

\graphicspath{{figpdf/}}

\newcommand{\ket}[1]{\left|  #1 \right>  }
\newcommand{\bra}[1]{\left<  #1 \right|  }

\newcommand{\SU}{\mathrm{SU}} 
\newcommand{\su}{\mathfrak{su}}
\newcommand{\suhat}{\widehat{\mathfrak{su}}}
\newcommand{\Hom}{\mathrm{Hom}}
\DeclareMathOperator{\Tr}{Tr}

\newcommand{\id}{\mathbf{1}}

\begin{document}

\title{Chiral $\SU(2)_k$ currents as local operators in vertex models \\ and spin chains}

\author[1]{R.~Bondesan\thanks{roberto.bondesan@uni-koeln.de}}
\author[2]{J.~Dubail}
\author[2]{A.~Faribault}
\author[3,4]{Y.~Ikhlef\thanks{ikhlef@lpthe.jussieu.fr}}
\affil[1]{Institute of Theoretical Physics, University of Cologne,
Z\"ulpicher Stra\ss{}e 77, D-50937 Cologne, Germany}
\affil[2]{IJL, CNRS \& Universit\'e de Lorraine, Boulevard des Aiguillettes
F-54506 Vand\oe{}uvre-l\`es-Nancy Cedex, France}
\affil[3]{Sorbonne Universit\'es, UPMC Univ Paris 06, UMR 7589, LPTHE, F-75005,
Paris, France}
\affil[4]{CNRS, UMR 7589, LPTHE, F-75005, Paris, France}

\date{}
\maketitle

\begin{abstract}
  The six-vertex model and its spin-$S$ descendants obtained from the
  fusion procedure are well-known lattice discretizations of the
  $\SU(2)_k$ WZW models, with $k=2S$. It is shown that, in these models,
  it is possible to exhibit a local observable on the lattice that
  behaves as the chiral current $J^a(z)$ in the continuum limit. The
  observable is built out of generators of the $\su(2)$ Lie algebra
  acting on a small (finite) number of lattice sites. The construction
  works also for the multi-critical quantum spin chains related to the
  vertex models, and is verified numerically for $S=1/2$ and $S=1$ using
  Bethe Ansatz and form factors techniques.
\end{abstract}

\tableofcontents


\section{Introduction: discretizing conformal blocks}


Two-dimensional conformal field theory (CFT) has proved to be an
extremely powerful tool in the study of many problems in theoretical
physics ranging from condensed matter to string theory. Its 
effectiveness is rooted in the infinite dimensional algebra of conformal
transformations, which is typically generated by two mutually commuting copies of the
Virasoro algebra, one holomorphic and the other anti-holomorphic. In a CFT
the operator product expansion (OPE) of two fields
decomposes generically into a direct sum of conformal families
indexed by primary fields \cite{BPZ}.  This fact leads to the notion
of conformal blocks which represent holomorphic (or chiral)
contributions to correlation functions
\begin{equation}
  \label{eq:Phi_correlator}
  \left< \Phi_{1} (z_1, \overline{z}_1) \Phi_{2} (z_2,
    \overline{z}_2) \dots \Phi_{n} (z_n, \overline{z}_n) \right>
\end{equation}
of primary fields $\Phi_i (z_i, \overline{z}_i)$. A conformal block is
specified by a choice of intermediate fusion channels, and can be
encoded in the following diagram:
\begin{align*}
  \begin{tikzpicture}[thick]
    \draw (-.25,0)--(-1-1.5-1-0.25,0) node[left] {$
      h_1$};
    \draw[dotted] (-.25,0)--(.25,0);
    \draw (.25,0)--(1+1.5+1.25,0)  node[right] {$h_n\, .$};
    \draw (-1-1.5,0)--(-1-1.5,1) node[above] {$h_2$};
    \draw (-1,0)--(-1,1) node[above] {$h_3$};
    \node at (-1-.75,0.25) {$h'_1$};
    \draw (1+1.5,0)--(1+1.5,1) node[above] {$h_{n-1}$};
    \draw (1,0)--(1,1) node[above] {$h_{n-2}$};
    \node at (1+.75,0.25) {$h'_{n-3}$};
  \end{tikzpicture}
\end{align*}
The conformal block
$\mathcal{F}(\{z_i\}|\{h_i\},\{h'_j\})$ is a function of the
holomorphic coordinates $z_1,\dots,z_n$, and depends on external chiral conformal
dimensions $h_1,\dots,h_{n}$, the intermediate dimensions
$h'_1,\dots,h'_{n-3}$, and the central charge of the theory.  The
correlation function \eqref{eq:Phi_correlator} is reconstructed by
gluing the holomorphic block with its anti-holomorphic counterpart,
$\overline{\mathcal{F}}
(\{\overline{z}_i\}|\{\overline{h}_i\},\{\overline{h}'_j\})$, and
summing over the intermediate channels weighted by the OPE
coefficients. (For standard textbooks on CFT, see e.g.,
\cite{ginsparg1990lecture,
  DMS97,henkel1999conformal,mussardo2010statistical}.)

For generic sets of primary operators $\Phi_i(z_i, \overline{z}_i)$,
the blocks have non-trivial monodromy. But, if the operators
$\Phi_i(z_i, \overline{z}_i)$ are mutually local, then
$\mathcal{F}(\{z_i\}|\{h_i\},\{h'_j\})$ (resp.~$\overline{\mathcal{F}}
(\{\overline{z}_i\}|\{\overline{h}_i\},\{\overline{h}'_j\})$)
is a meromorphic (resp.~anti-meromorphic) function of $z_i$, with poles located at the positions $z_j$, $j\neq
i$. When this is the case, it is a natural question to ask whether $\mathcal{F}(\{z_i\}|\{h_i\},\{h'_j\})$ can itself be
realized as a correlator of local observables, without its anti-holomorphic counterpart. One
can further wonder whether it is possible to construct {\it local
observables} in some {\it lattice model}, whose correlators would then
converge to this conformal block in the continuum limit. This is the
basic question that is motivating this paper. 

Perhaps the simplest situation where this question can be
asked is when the operators $\Phi_i(z_i, \overline{z}_i)$ are all
chiral currents $J^{a_i}(z_i)$ arising in a 
Wess-Zumino-Witten (WZW) model. These currents are primary operators
with respect to the Virasoro algebra, but not with respect to the
full chiral algebra, which they themselves generate---typically, a
Kac-Moody algebra. In this paper we restrict ourselves to 
correlators of the form
\begin{equation}
  \label{eq:block_generic}
  \left< J^{a_1}(z_1) J^{a_2}(z_2)  \dots J^{a_n}(z_n) \right>
\end{equation}
in $\SU(2)_k$ WZW models. We believe that the extension to other WZW models is relatively straightforward.
For simplicity, we assume that we are on a surface of genus zero, 
namely the Riemann sphere; on surfaces of higher genus, the correlators (\ref{eq:block_generic}) would depend on the boundary conditions around the different
cycles of the surface (see e.g.~\cite{bernard1988wess, bernard1988wess_g}). The question we wish to answer is the
following: {\it is it possible to find a two-dimensional lattice
  model, and a set of local observables in this lattice model, such
  that the continuum limit of their correlator is the conformal block
  of Eq.~\eqref{eq:block_generic}}?

The reason why this seems non-trivial to us is that 
lattice models at criticality are described by {\it non-chiral}
CFTs in the continuum limit, so the correlators of local
observables on the lattice typically become field theory 
correlators of the form \eqref{eq:Phi_correlator}, involving a sum of products of 
chiral and anti-chiral blocks. Separating the chiral from the anti-chiral part of local operators in lattice models appears to be difficult in general, and
typically leads to non-local operators attached to defect lines, usually called {\it parafermionic observables} \cite{kadanoff1971determination,fradkin1980disorder}. The latter do not appear in this paper though, since we are dealing with currents only. Notice, however, that parafermionic observables would appear if one wanted to construct lattice versions of the holomorphic $\SU(2)_k$ {\it primary fields} (primary with respect to the full chiral algebra). We hope to come back to this question in the near future. We also note that essentially the same program has been carried out independently by Mong {\it et al.} for a three-state Potts quantum spin chain \cite{mong2014parafermionic}.
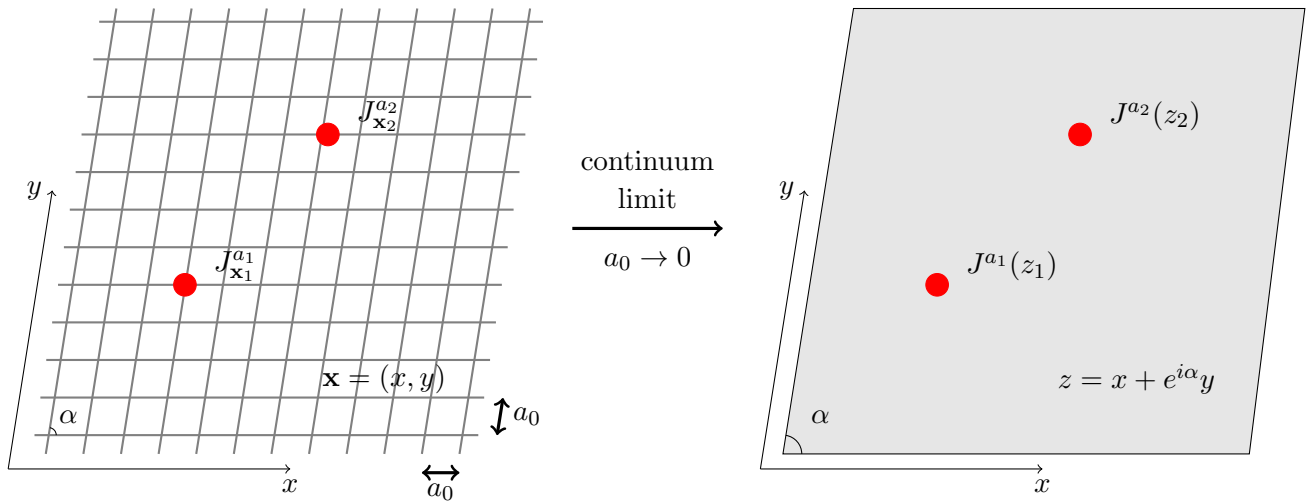
\begin{figure}[h]
  \begin{center}
    \begin{tikzpicture}
    \begin{scope}[scale=0.5]
      \foreach \x in {0,1,...,11}
      {
        \draw[thick,gray] (\x,0) -- (\x+0.1564*12,0.9877*12);
        \draw[thick,gray] (-0.3+0.1564*\x,\x+0.5) -- (11.5+0.1564*\x,\x+0.5);
      }
      \draw (0.0782,0.5) ++(0.2,0) arc (0:81:0.2);
      \draw (0.6,1) node{$\alpha$};
      
      \draw[very thick,<->] (10,-0.5) -- (11,-0.5);
      \draw (10.5,-1) node{$a_0$};
      \draw[very thick,<->] (12,0.5) -- (12+1*0.1564, 0.5+1*0.9877);
      \draw (12.8,1) node{$a_0$};
   
      \draw[->] (-1,-0.4) -- (6.5,-0.4) node[below]{$x$};
      \draw[->] (-1,-0.4) -- (-1+0.1564*7.5,-0.4+0.9877*7.5) node[left]{$y$};
      \filldraw[red] (3.7,4.5) circle (3mm) ++ (0.5,0.5) node[black,right]{$J^{a_1}_{{\bf x}_1}$};
      \filldraw[red] (7.5,8.5) circle (3mm) ++(0.5,0.5) node[black, right]{$J^{a_2}_{{\bf x}_2}$};
	\draw (9,2) node {${\bf x} = (x,y)$};
    \end{scope}
    
    \begin{scope}[xshift=10cm,scale=0.5]
 	\filldraw[fill=gray!20] (-0.4,0) -- (-0.4+0.1564*12,0.9877*12) -- (-0.4+12+0.1564*12,0.9877*12) -- (12,0) -- cycle;
	\draw[->] (-1,-0.4) -- (6.5,-0.4) node[below]{$x$};
	\draw[->] (-1,-0.4) -- (-1+0.1564*7.5,-0.4+0.9877*7.5) node[left]{$y$};
      \draw (0.1,0) arc (0:81:0.5);
      \draw (0.6,1) node{$\alpha$};
      \filldraw[red] (3.7,4.5) circle (3mm) ++ (0.5,0.5) node[black,right]{$J^{a_1}(z_1)$};
      \filldraw[red] (7.5,8.5) circle (3mm) ++(0.5,0.5) node[black, right]{$J^{a_2}(z_2)$};
	\draw (9,2) node {$z = x+e^{i\alpha} y$};
    \end{scope}

\draw[very thick,->] (7,3) -- (9,3);
\draw (8,3.9) node{continuum};
\draw (8,3.4) node{limit};
\draw (8,2.6) node{$a_0 \rightarrow 0$};
    \end{tikzpicture}
  \end{center}
  \caption{Correlators of lattice observables become CFT correlators as one sends the lattice spacing $a_0$ to zero. }
  \label{fig:lattice_continuum}
\end{figure}

In this paper, we consider the family of (integrable) spin-$k/2$ vertex models which descend
from the six-vertex model \cite{KRS81,Takhtajan1982,Babujian1982,Babujian1983}. 
These have long been known to
be multi-critical points of spin-$k/2$ models, whose continuum limit
is the $\SU(2)_k$ WZW model \cite{witten1984non, Affleck1985,
  affleck1986exact, affleck1987critical, di1988generalized,
  Affleck1989}. We construct a set of local lattice observables $J^a_{{\bf x}}$---where ${\bf x}$ is the position of a point on the lattice, and $a$ labels the three generators of the $\su(2)$ Lie algebra---which has the following property. As one sends the lattice spacing $a_0$ to zero, our lattice observables become the holomorphic currents generating $\suhat(2)_k$:
\begin{equation}
	\label{eq:goal}
	J^a_{{\bf x}} \; \underset{a_0 \rightarrow 0}{ = } \; J^a (z) \, + \, O(a_0^{\beta}) ,
\end{equation}
for some exponent $\beta >0$. As usual, this type of identity is meaningful only when the observables are inserted in correlators, namely
\begin{equation}
	\left<  J^{a_1}_{{\bf x}_1} \dots J^{a_n}_{{\bf x}_N} \right>_{\rm lattice} \; \, \underset{a_0 \rightarrow 0}{ = } \;  \, \left< J^{a_1} (z_1) \dots J^{a_n} (z_n) \right>_{{\rm CFT}} \, + \, O(a_0^{\beta}).
\end{equation}
Throughout the paper, the relation between the lattice position ${\bf x} = (x,y)$ and the complex coordinate $z$ is fixed as
\begin{equation}
	z \, = \, x + e^{i \alpha} y
\end{equation}
as illustrated in Fig. \ref{fig:lattice_continuum}. We will obtain our lattice observables (\ref{eq:goal}) from 
conserved currents in the vertex models, using a trick to isolate the holomorphic part.

The original motivation for the present paper comes from the analogy between two
classes of variational wave functions for quantum systems in two
dimensions: Tensor Network States (or Tensor Product States or Projected Entangled
Paired States) on the one hand, and the Moore-Read class of trial wave functions \cite{MooreRead} for chiral topological phases---e.g. quantum Hall systems---that are expressed by conformal blocks on the other hand. (See also the discussion in section V of \cite{dubail2012edge} about this analogy.)  
In this spirit, the case of $\suhat(2)_k$ ``lattice conformal blocks''
is related to the Read-Rezayi states of fractional quantum
Hall systems \cite{ReadRezayi}.  Applications of our work to Tensor
Network States and related topics will be discussed elsewhere. We note that lattice models related to
$\suhat(2)_k$ conformal blocks have been investigated recently in
\cite{greiter2011mapping,NCS11,Thomale2012}, using the original Moore-Read construction to produce wave functions that are
the ground states of long-range spin
systems of the Haldane-Shastry type. In contrast to the present paper, these references do not aim at the
discretization of the blocks themselves.
Finally, it is also worth mentioning that, apart from condensed matter applications,
renewed interest in conformal blocks has been triggered recently by
the AGT conjecture \cite{AGT}, which relates conformal blocks to
partition functions of $\mathcal{N} = 2$ four-dimensional
supersymmetric gauge theories.

The paper is organized as follows. In section 2 we review the necessary background material that is used later in the paper.
In section 3 we analyze the lattice spin operator $S^a_{{\bf x}}$ and
its $a_0$-expansion in terms of the local fields in the continuum. We
identify the first few coefficients in this expansion by symmetry
arguments. The knowledge of these coefficients allows us to construct
our lattice observable ($\ref{eq:goal}$) for all the descendants of
the six-vertex model, and the corresponding critical spin chains. This
is explained in section 4. Section 5 contains a few checks of our
predictions for the coefficients, which we obtain from numerical
evaluation of the form-factors for $k=1$ (spin-$1/2$) and $k=2$
(spin-$1$). We conclude in section 6. We also provide two
appendices. In the first one
we discuss logarithmic corrections and explain why they do not appear
at the leading order in correlations functions of the chiral current.  The second appendix contains
some details about the calculation of the form factors and other
technical aspects of the Bethe Ansatz solution for general $k \geq 1$.


\section{Background material}

\subsection{The $\suhat(2)_k$ current algebra}

Let us start by collecting the piece of information about the $\suhat(2)_k$
current algebra that will be needed in the rest of the paper.  We refer the
reader to \cite{knizhnik1984current,witten1984non,DMS97} for further details. $\suhat(2)_k$ is an affine
Lie algebra generated by the modes $J_n^a$ of the holomorphic currents
$J^a(z)$, defined by $J^a(z)=\sum_{n\in \mathbb{Z}}z^{-n-1}J_n^a$.
The index $a$ refers to a generator of the underlying Lie algebra
$\su(2)$.  The OPEs between the currents involve the structure
constants $f^{abc}$ and the Killing form $\kappa^{ab}$ of $\su(2)$, as
well as the {\it level} $k$:
\begin{equation}
  \label{eq:SU(2)k}
  J^a(z) J^b(w) \, = \, \frac{\frac{k}{2} \, \kappa^{ab}}{(z-w)^2} \, + \,
  \frac{i f^{ab}_{\phantom{a}\phantom{a}c}}{z-w}J^c(w) \, + \, {\rm
    regular \; terms}.
\end{equation}
It is sometimes convenient to fix a basis of $\su(2)$. We chose the one 
given by the Pauli matrices, $\frac{1}{2}\sigma^a$, $a=1,2,3$. Then the structure constants are the completely anti-symmetric tensor $f^{abc} = \epsilon^{abc}$, and the Killing form coincides with the Kronecker delta $\kappa^{ab} \, = \,\delta^{ab} $. The Lie algebra $\su(2)$ is embedded as the zero-modes subalgebra: $\left[ J_0^a , J^b_0 \right] \, = \, i f^{ab}_{\phantom{a}\phantom{a}c} J^c_0$.

The primary operators $\phi_j(w)$ for the affine Lie algebra (which
should not be confused with primaries for the Virasoro algebra) are
local fields with respect to the $\suhat(2)_k$ currents that satisfy
the following OPEs:
\begin{equation}
  \label{eq:primary}
    J^a(z) \phi_j(w) =  \frac{S^a \cdot \phi_j(w)}{z-w}   \, + \,
    {\rm regular \; terms}\, .
\end{equation}
In this formula we think of $\phi_j$ as vector-valued, with components $\phi_{j,j_3}$ with $j_3 = -j,-j+1, \dots, j-1,j$. $S^a$ is the 
spin-$j$ representation matrix.  The component of $\phi_j$ with $j_3=j$ is a highest weight vector for the affine Lie algebra and 
and the corresponding representation is generated by acting
on it with the lowering operators $J_0^1-iJ_0^2$ and
$J_{-n_1}^{a_1}\cdots J_{-n_k}^{a_k}$, $n_i> 0$.
(This procedure produces null vectors which have to be removed to obtain
an irreducible representation.)

In order for the theory to be unitary, $k$ must be a positive integer,
and the spin $j$ must be integer or half-integer, with the additional
restriction
\begin{equation}
  j \, \in \, \left\{ 0, \frac{1}{2}, 1 ,\dots , \frac{k}{2}  \right\}.
\end{equation}
Thus, at level $k$, there are exactly $k+1$ primary operators, and
$\phi_0$ is the identity field.  The Sugawara construction realizes
the stress-tensor of the theory as a bilinear in the currents,
\begin{equation}
  \label{eq:Sugawara}
  T(z) \, =\, \frac{1}{k+2}  : J^a (z) J_a(z) :  \,,
\end{equation}
and the following OPEs can be computed from
(\ref{eq:SU(2)k})-(\ref{eq:primary}) and (\ref{eq:Sugawara}):
\begin{subequations}
\begin{eqnarray}
  T(z) J^a(w) & = & \frac{1}{(z-w)^2} J^a(w) \, + \, \frac{1}{z-w} \partial J^a(w)  \, + \, {\rm regular \; terms} \\
  T(z) \phi_{j}(w) & = & \frac{h_j}{(z-w)^2} \phi_j(w) \, + \, \frac{1}{z-w} \partial \phi_j(w)  \, + \, {\rm regular \; terms} \\
  T(z) T(w) & = &  \frac{c/2}{(z-w)^4}  \, + \, \frac{2}{(z-w)^2} T(w) \, + 
  \, \frac{1}{z-w} \partial T(w)  \, + \, {\rm regular \; terms} .
\end{eqnarray}
\end{subequations}
with
\begin{equation}
  \label{eq:h_c}
  h_j \, =\, \frac{j(j+1)}{k+2} \qquad \quad c \, = \, \frac{3k}{k+2} .
\end{equation}
We see that $J^a(z)$ has conformal dimension $1$, as expected for a
current, that $\phi_j$ has conformal dimension $h_j$, and that $T(z)$
is the holomorphic stress-tensor of a conformal field theory with
central charge $c$ given in formula \eqref{eq:h_c}.  
In a similar way, $\overline{\suhat(2)}_k$ is generated
by the anti-holomorphic current $\bar{J}^a(\bar{z})$ satisfying
the anti-holomorphic counterpart of the above relations.


\subsection{The six-vertex model and its continuum limit}
\label{sec:6v}

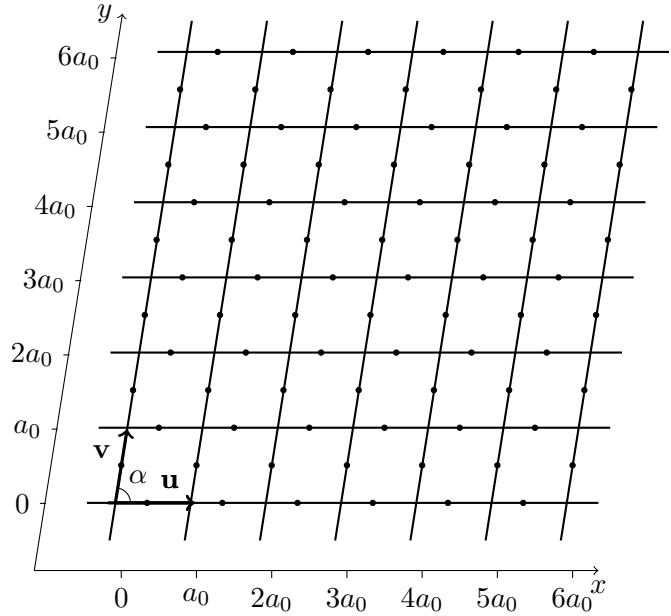
\begin{figure}[h]
  \begin{center}
    \begin{tikzpicture}
      \foreach \x in {0,1,...,6}
      {
        \draw[thick] (\x,0) -- (\x+0.1564*7,0.9877*7);
        \draw[thick] (-0.3+0.1564*\x,\x+0.5) -- (6.5+0.1564*\x,\x+0.5);
      }
      \draw[very thick,->] (-0.3+0.278,0.5) -- (1.135,0.5);
      \draw (0.8,1) node[below]{${\bf u}$};
      \draw[very thick,->] (0.0782,0.5) -- (1.135,0.5);
      \draw (0.8,1) node[below]{${\bf u}$};
      \draw[very thick,->] (0.0782,0.5) -- (0.235,1.4815);
      \draw (-0.1,1.4) node[below]{${\bf v}$};
      \draw (0.0782,0.5) ++(0.2,0) arc (0:81:0.2);
      \draw (0.38,0.85) node{$\alpha$};
      
      \draw[->] (-1,-0.4) -- (6.5,-0.4) node[below]{$x$};
      \draw[->] (-1,-0.4) -- (-1+0.1564*7.5,-0.4+0.9877*7.5) node[left]{$y$};
      \foreach \x in {1} \draw (\x+0.1564,-0.5) node[below]{$a_0$} -- (\x+0.1564,-0.4);
      \foreach \x in {2,...,6} \draw (\x+0.1564,-0.5) node[below]{$\x a_0$} -- (\x+0.1564,-0.4);
      \draw (0.1564,-0.5) node[below]{$0$} -- (0.1564,-0.4);
      \foreach \y in {1} \draw (-1+0.075+0.1564*\y,0.9877*\y+0.4937) node[left]{$a_0$} -- (-1+0.1564*\y+0.1564,0.9877*\y+0.4937);
      \foreach \y in {2,...,6} \draw (-1+0.075+0.1564*\y,0.9877*\y+0.4937) node[left]{$\y a_0$} -- (-1+0.1564*\y+0.1564,0.9877*\y+0.4937);
      \draw (-1+0.075,0.4937) node[left]{$0$} -- (-1+0.1564,0.4937);
      
      	\foreach \x in {1,2,...,6}
		{
		\foreach \y in {0,1,...,6}
			{
				\filldraw (\y+\x*0.1564,\x) circle (0.35mm);
				\filldraw (-0.5+\x+\y*0.1564,0.5+\y) circle (0.35mm);
			}
		}

    \end{tikzpicture}
  \end{center}
  \caption{The degrees of freedom of the six-vertex model are
    spin-$1/2$ living on
    the edges (black dots) of the lattice $\mathbb{Z} {\bf u} + \mathbb{Z} {\bf
      v}$, with $ \left|{\bf u}\right|= \left|{\bf v}\right| = a_0$.}
  \label{fig:lattice}
\end{figure}

Next, we review a few useful facts about the six-vertex model at the
$\SU(2)$ invariant point, which is well known to be a lattice
discretization of the diagonal $\suhat(2)_1 \otimes
\overline{\suhat(2)}_1$ CFT.  We consider the
six-vertex model defined on the lattice $\mathbb{Z} {\bf u} +
\mathbb{Z} {\bf v}$, see Figure \ref{fig:lattice}, where each edge
carries one spin-$1/2$ degree of freedom. The Boltzmann weights of the
different spin configurations are obtained from the $R$-matrix, which
is a tensor associated to every site {\bf x} of the lattice, involving
only the four spin $1/2$ representations living on the adjacent edges
at positions ${\bf x} \pm \frac{1}{2}{\bf u}, \; {\bf x} \pm
\frac{1}{2}{\bf v}$:
\begin{equation}
  \label{eq:R_x_tensor}
  R_{\bf x} \, \in \, 
  \left(\frac{1}{2}\right)_{{\bf x} + \frac{1}{2}{\bf v}} \otimes 
  \left(\frac{1}{2}\right)_{{\bf x} + \frac{1}{2}{\bf u}} \otimes 
  \left(\frac{1}{2}\right)^*_{{\bf x} - \frac{1}{2}{\bf u}} \otimes 
  \left(\frac{1}{2}\right)^*_{{\bf x} - \frac{1}{2}{\bf v}}\, .
\end{equation}
Here $\left(\frac{1}{2} \right)_e$ stands for the fundamental of
$\su(2)$ attached to the edge $e$ and $\left(\frac{1}{2} \right)^*_e$ for its
dual (which is isomorphic to the fundamental representation).
We represent graphically the $R$-matrix $(R_{\bf x})_{\sigma_1
  \sigma_2 \sigma_3 \sigma_4} $ as
\begin{equation}
\begin{tikzpicture}[scale=0.75]
	\label{eq:R6v0}
	 \begin{scope}
		\draw[very thick]  (-1,0) node[left]{$\sigma_3 \in \left(\frac{1}{2}\right)^*$} -- (1,0) node[right]{$\sigma_2 \in \left(\frac{1}{2}\right)$};
		\draw[very thick]  (0.1564,0.9877) node[above]{$\sigma_1 \in \left(\frac{1}{2}\right)$} -- (-0.1564,-0.9877) node[below]{$\sigma_4 \in \left(\frac{1}{2}\right)^*$};
		\draw (0.3,0.3) node{${\bf x}$};
	\end{scope}
	\draw (3.5,-0.5) node{.};
\end{tikzpicture}
\end{equation}
The total weight of a global configuration of spins is obtained by
contracting all the $R$-matrices, using the canonical pairing on each
edge. It is customary to use the canonical 
isomorphism $W\otimes V^*\simeq \Hom(V,W)$ to view the $R$-matrix as
the linear operator ({i.e.} as the complex matrix)
\begin{align}
	\label{eq:matrix_matrix}
  R_{\bf x} : \;  \left(\frac{1}{2}\right)_{{\bf x} - \frac{1}{2}{\bf u}} \otimes 
  \left(\frac{1}{2}\right)_{{\bf x} - \frac{1}{2}{\bf v}} 
  \longrightarrow \;
  \left(\frac{1}{2}\right)_{{\bf x} + \frac{1}{2}{\bf v}}\otimes 
  \left(\frac{1}{2}\right)_{{\bf x} + \frac{1}{2}{\bf u}} 
\end{align}
acting from the south-west to the north-east direction. Global $\SU(2)$ symmetry is imposed by requiring that the
$R$-matrix is an element of the spin-$0$ ({\it i.e.} invariant) subspace of the
tensor product (\ref{eq:R_x_tensor}). This subspace is
two-dimensional, so the R-matrix of Eq.~(\ref{eq:matrix_matrix})
can be written as a linear combination of the identity
$\mathbf{1}$ and the projector  $P_0$
onto the singlet contained in $(\frac{1}{2})\otimes(\frac{1}{2})$. 

Up to normalization,
this leaves us with one free parameter, which is the relative weight
of $\mathbf{1}$ and $P_0$. For the purposes of this paper, it is
convenient to parametrize the $R$-matrix directly by the geometric
angle $\alpha$, see Fig.~\ref{fig:lattice}. 
($\alpha$ is related to the spectral parameter as given in Appendix
\ref{sec:appendixA}.) Throughout the paper, the $R$-matrix will be the
same for every vertex {\bf x}, so we will often drop the subscript
{\bf x}. The explicit expression of $R$ together with its graphical
representation is:
\begin{equation}
  \label{eq:R6v1}
  \begin{tikzpicture}[scale=0.75]
  \draw (-1.25,0) node{$R(\alpha) \, = $};
  \draw[thick] (0,0) -- (2, 0);
  \draw[thick] (1-0.1564,-0.9877) -- (1+0.1564,0.9877);		
  \draw (-0.2+1,0) arc (-180:-99:0.2) ++(0,-0.1) node[left]{$\alpha$};
  \draw (5,0) node{$\displaystyle{\,=\,   \frac{\alpha}{\pi} \,\id \,+  
    \, 2\,\left(1- \frac{\alpha}{\pi}\right)\, P_0 \, .}$};
  \end{tikzpicture}
\end{equation}
This is related to the usual parametrization of the weights of the
six-vertex model,
$$
\begin{tikzpicture}[scale=0.75]
	\begin{scope}
		\draw[very thick,->]  (-1,0) -- (-0.45,0);
		\draw[very thick,->]  (0,0) -- (0.6,0);
		\draw[very thick]  (-1,0) -- (1,0);
		\draw[very thick,->]  (0,0) -- (0.55*0.1564,0.55*0.9877);
		\draw[very thick,->]  (-0.1564,-0.9877) -- (-0.45*0.1564,-0.45*0.9877);
		\draw[very thick]  (0.1564,0.9877) -- (-0.1564,-0.9877);
		\draw (0,-1.5) node{$a(\alpha)$};
		\draw (-0.2,0) arc (-180:-90:0.2) ++(0,-0.1) node[left]{$\alpha$};
	\end{scope}
	\begin{scope}[xshift=3cm]
		\draw[very thick,->]  (1,0) -- (0.45,0);
		\draw[very thick,->]  (0,0) -- (-0.6,0);
		\draw[very thick]  (1,0) -- (-1,0);
		\draw[very thick,->]  (0,0) -- (-0.55*0.1564,-0.55*0.9877);
		\draw[very thick,->]  (0.1564,0.9877) -- (0.45*0.1564,0.45*0.9877);
		\draw[very thick]  (0.1564,0.9877) -- (-0.1564,-0.9877);
		\draw (0,-1.5) node{$a(\alpha)$};
		\draw (-0.2,0) arc (-180:-90:0.2) ++(-0.06,-0.1) node[left]{$\alpha$};		
	\end{scope}
	\begin{scope}[xshift=6cm]
		\draw[very thick,->]  (1,0) -- (0.45,0);
		\draw[very thick,->]  (0,0) -- (-0.6,0);
		\draw[very thick]  (-1,0) -- (1,0);
		\draw[very thick,->]  (0,0) -- (0.1564*0.55,0.9877*0.55);
		\draw[very thick,->]  (-0.1564,-0.9877) -- (-0.45*0.1564,-0.45*0.9877);
		\draw[very thick]  (0.1564,0.9877) -- (-0.1564,-0.9877);
		\draw (0,-1.5) node{$b(\alpha)$};
		\draw (-0.2,0) arc (-180:-90:0.2) ++(-0.05,-0.1) node[left]{$\alpha$};
	\end{scope}
	\begin{scope}[xshift=9cm]
		\draw[very thick,->]  (-1,0) -- (-0.45,0);
		\draw[very thick,->]  (0,0) -- (0.6,0);
		\draw[very thick]  (-1,0) -- (1,0);
		\draw[very thick,->]  (0,0) -- (-0.55*0.1564,-0.55*0.9877);
		\draw[very thick,->]  (0.1564,0.9877) -- (0.45*0.1564,0.45*0.9877);
		\draw[very thick]  (0.1564,0.9877) -- (-0.1564,-0.9877);
		\draw (0,-1.5) node{$b(\alpha)$};
		\draw (-0.2,0) arc (-180:-90:0.2) ++(-0.05,-0.1) node[left]{$\alpha$};
	\end{scope}
	\begin{scope}[xshift=12cm]
		\draw[very thick,->]  (-1,0) -- (-0.45,0);
		\draw[very thick,->]  (1,0) -- (0.45,0);
		\draw[very thick]  (-1,0) -- (1,0);
		\draw[very thick,->]  (0,0) -- (0.55*0.1564,0.55*0.9877);
		\draw[very thick,->]  (0,0) -- (-0.55*0.1564,-0.55*0.9877);
		\draw[very thick]  (0.1564,0.9877) -- (-0.1564,-0.9877);
		\draw (0,-1.5) node{$c(\alpha)$};
		\draw (-0.2,0) arc (-180:-90:0.2) ++(-0.05,-0.1) node[left]{$\alpha$};
	\end{scope}
	\begin{scope}[xshift=15cm]
		\draw[very thick,->]  (0,0) -- (-0.6,0);
		\draw[very thick,->]  (0,0) -- (0.6,0);
		\draw[very thick]  (-1,0) -- (1,0);
		\draw[very thick,->]  (0.1564,0.9877) -- (0.45*0.1564,0.45*0.9877);
		\draw[very thick,->]  (-0.1564,-0.9877) -- (-0.45*0.1564,-0.45*0.9877);
		\draw[very thick]  (0.1564,0.9877) -- (-0.1564,-0.9877);
		\draw (0,-1.5) node{$c(\alpha)$};
		\draw (-0.2,0) arc (-180:-90:0.2) ++(-0.05,-0.1) node[left]{$\alpha$};
	\end{scope}
\end{tikzpicture}
$$
by $a(\alpha)= \frac{\alpha}{\pi}, b(\alpha) = \frac{\alpha}{\pi}-1,
c(\alpha)=1$. The $R$-matrix satisfies the Yang-Baxter equation
\begin{equation}
  \label{eq:YB}
  \raisebox{-1.5cm}{\begin{picture}(0,0)%
\includegraphics{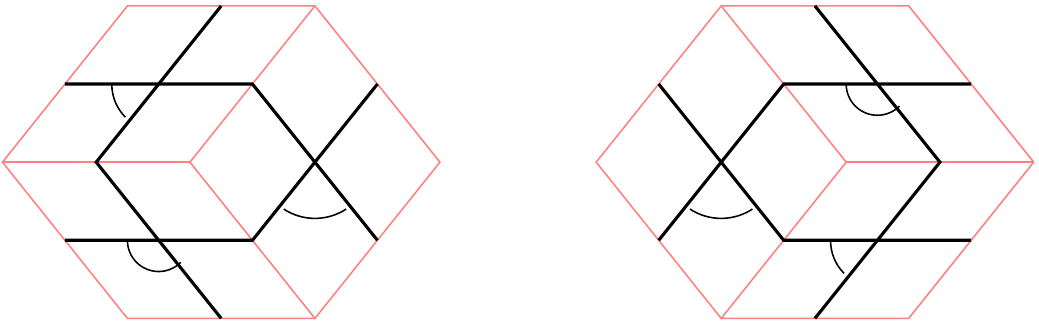}%
\end{picture}%
\setlength{\unitlength}{3947sp}%
\begingroup\makeatletter\ifx\SetFigFont\undefined%
\gdef\SetFigFont#1#2#3#4#5{%
  \reset@font\fontsize{#1}{#2pt}%
  \fontfamily{#3}\fontseries{#4}\fontshape{#5}%
  \selectfont}%
\fi\endgroup%
\begin{picture}(4974,1544)(439,-2633)
\put(2926,-1861){\makebox(0,0)[b]{\smash{{\SetFigFont{12}{14.4}{\familydefault}{\mddefault}{\updefault}{\color[rgb]{0,0,0}$=$}%
}}}}
\put(1126,-2461){\makebox(0,0)[rb]{\smash{{\SetFigFont{10}{12.0}{\familydefault}{\mddefault}{\updefault}{\color[rgb]{0,0,0}$\alpha$}%
}}}}
\put(1951,-2311){\makebox(0,0)[b]{\smash{{\SetFigFont{10}{12.0}{\familydefault}{\mddefault}{\updefault}{\color[rgb]{0,0,0}$\alpha-\beta$}%
}}}}
\put(901,-1711){\makebox(0,0)[rb]{\smash{{\SetFigFont{10}{12.0}{\familydefault}{\mddefault}{\updefault}{\color[rgb]{0,0,0}$\beta$}%
}}}}
\put(3901,-2311){\makebox(0,0)[b]{\smash{{\SetFigFont{10}{12.0}{\familydefault}{\mddefault}{\updefault}{\color[rgb]{0,0,0}$\alpha-\beta$}%
}}}}
\put(4351,-2461){\makebox(0,0)[rb]{\smash{{\SetFigFont{10}{12.0}{\familydefault}{\mddefault}{\updefault}{\color[rgb]{0,0,0}$\beta$}%
}}}}
\put(4576,-1711){\makebox(0,0)[rb]{\smash{{\SetFigFont{10}{12.0}{\familydefault}{\mddefault}{\updefault}{\color[rgb]{0,0,0}$\alpha$}%
}}}}
\end{picture}%
}
\end{equation}
and the inversion relation
\begin{equation}
	\label{eq:inversion}
	\begin{tikzpicture}
		\draw[very thick] (-1,-0.4) arc (150:30:2);
		\draw[very thick] (-1,0.4) arc (-150:-30:2);
		\draw (-0.83,-0.15) arc (-135:-225:0.2) ++(0,-0.1) node[left]{$\alpha$};	
		\draw (2,-0.15) arc (-135:-225:0.2) ++(0,-0.1) node[left]{$2\pi-\alpha$};
		\draw (4.1,0) node{$\displaystyle{= \, \frac{\alpha}{\pi}\left(\frac{2\pi-\alpha}{\pi}\right)}$};
		\draw[very thick] (5.5,-0.4) arc (150:30:2 and 0.3);
		\draw[very thick] (5.5,0.4) arc (-150:-30:2 and 0.3);
		\draw (9.2,-0.2) node{.};
	\end{tikzpicture}
\end{equation}
Now we replace the infinite lattice by a cylinder with $N$ sites in
the periodic direction. The transfer matrix acting on the space of
spins at fixed $y$-coordinate is:
\begin{equation}
	\label{eq:transfer_matrix_6v}
	\begin{tikzpicture}[scale = 0.8]
		\draw (-1.4,0) node{$T_L(\alpha) \, = $};
		\draw[thick] (0,0) -- (10, 0);
		\draw[thick] (1-0.1564,-0.9877) node[below]{$a_0$} -- (1+0.1564,0.9877);		
		\draw[thick] (2-0.1564,-0.9877) node[below]{$2a_0$} -- (2+0.1564,0.9877);		
		\draw[thick] (3-0.1564,-0.9877) -- (3+0.1564,0.9877);		
		\draw[thick] (4-0.1564,-0.9877) -- (4+0.1564,0.9877);		
		\draw[thick] (5-0.1564,-0.9877) -- (5+0.1564,0.9877);		
		\draw[thick] (6-0.1564,-0.9877) -- (6+0.1564,0.9877);		
		\draw[thick] (7-0.1564,-0.9877) -- (7+0.1564,0.9877);		
		\draw[thick] (8-0.1564,-0.9877)  node[below]{} -- (8+0.1564,0.9877);		
		\draw[thick] (9-0.1564,-0.9877)  node[below]{$a_0 N$} -- (9+0.1564,0.9877);		
		\draw[thick] (0,-0.2) -- ++(0.3,0.4);
		\draw[thick] (0.15,-0.2) -- ++(0.3,0.4);
		\draw[thick] (9.6,-0.2) -- ++(0.3,0.4);
		\draw[thick] (9.75,-0.2) -- ++(0.3,0.4);
		\draw (-0.2+1,0) arc (-180:-99:0.2) ++(0,-0.1) node[left]{$\alpha$};		
		\draw (-0.2+2,0) arc (-180:-99:0.2) ++(0,-0.1) node[left]{$\alpha$};		
		\draw (-0.2+3,0) arc (-180:-99:0.2) ++(0,-0.1) node[left]{$\alpha$};		
		\draw (-0.2+4,0) arc (-180:-99:0.2) ++(0,-0.1) node[left]{$\alpha$};		
		\draw (-0.2+5,0) arc (-180:-99:0.2) ++(0,-0.1) node[left]{$\alpha$};		
		\draw (-0.2+6,0) arc (-180:-99:0.2) ++(0,-0.1) node[left]{$\alpha$};		
		\draw (-0.2+7,0) arc (-180:-99:0.2) ++(0,-0.1) node[left]{$\alpha$};		
		\draw (-0.2+8,0) arc (-180:-99:0.2) ++(0,-0.1) node[left]{$\alpha$};		
		\draw (-0.2+9,0) arc (-180:-99:0.2) ++(0,-0.1) node[left]{$\alpha$};	
		\draw (10.5,0) node{$.$};	
	\end{tikzpicture}
\end{equation}
Our convention is that the transfer matrix acts from bottom to top. The length
of the system is $L = N a_0$, where $N$ is an integer. It is always a pleasure to observe that the Yang-Baxter equation
($\ref{eq:YB}$) and the inversion relation ($\ref{eq:inversion}$)
imply that, for any values of the spectral parameters $\alpha$ and
$\beta$,
\begin{equation}
  \left[ T_L(\alpha) , T_L(\beta) \right] \, = \, 0.
\end{equation}
The integrals of motion are defined as the logarithmic derivatives of $T_L$ at $\alpha=\pi$.
Since $R(\pi)=\id$, the first integral of motion is the operator $e^{-i a_0 P}$
translating one site to the right:
\begin{align}
	\label{eq:translation}
  T_L(\pi) = e^{-i a_0 P} \,.
\end{align}
The first derivative of the $R$-matrix is the {\it Hamiltonian density} $h \equiv R'(\pi) = \frac{1}{\pi} \mathbf{1} -\frac{2}{\pi} P_0$. 
For two spins-1/2, $\mathbf{S}_1$ and $\mathbf{S}_2$, we have $P_0=\frac{1}{4}\mathbf{1}  - \mathbf{S}_1 \cdot \mathbf{S}_2$.
Hence the next integral of motion is the spin-$\frac{1}{2}$ antiferromagnetic Heisenberg Hamiltonian
\begin{eqnarray}
 \nonumber H^{(1/2)}  & = & T^{\, '}_L(\pi) \cdot T_L^{-1}(\pi) \\
 \nonumber 	&=& \sum_{x/a_0 =1}^N  h_{x,x+a_0}  \\
  	&=& \frac{2}{\pi} \sum_{x/a_0 =1}^N \left( \mathbf{S}_x \cdot \mathbf{S}_{x+a_0} + \frac{1}{4} \mathbf{1}\right) \,,
\end{eqnarray}
with periodic boundary conditions $\mathbf{S}_{L+a_0}=\mathbf{S}_{a_0}$.
The spectrum of the Heisenberg Hamiltonian, and more generally the one
of the transfer matrix $T_L(\alpha)$ for any value of $\alpha$, can be obtained from the Bethe Ansatz (for a review, see \cite{Sierra_book}, or the information given in appendix  \ref{sec:appendixA}). It is known 
that, in the thermodynamic limit, the low-lying eigenvalues of $-\log
T_L(\alpha)$ match the ones of the CFT Hamiltonian:
\begin{equation}
  \label{eq:Hcft}
  \begin{aligned}
    &T_L(\alpha) \, \underset{L \gg a_0}{\simeq}
    \, \Sigma\,\times \, \exp\left( - a_0 \, \sin\alpha\,\left(E_\infty(\alpha) \,L \, +  H_{\rm CFT}\right) \, + \, i \,a_0  \,\cos \alpha  \,P_{\rm CFT}  \right) , \\
    &H_{\rm CFT}\, = \, \frac{2\pi}{L} \left( L_0 +\overline{L}_0 - \frac{c}{12} \right)
    \qquad \quad
    P_{\rm CFT}\, = \, \frac{2\pi}{L} \left( L_0 -\overline{L}_0 \right) .
  \end{aligned} 
\end{equation}
Here $E_\infty(\alpha)$ is the free energy per unit area in the thermodynamic
limit; it varies continuously with $\alpha$, and in particular, $\sin \alpha \times E_\infty(\alpha)$ vanishes when $\alpha \rightarrow 0$ or $\alpha \rightarrow \pi$. $L_0$ ($\overline{L}_0$) is the zero mode of the
holomorphic (anti-holomorphic) component of the Sugawara stress-tensor
(\ref{eq:Sugawara}) for $\suhat(2)_1$ ($\overline{\suhat(2)}_1$):
\begin{equation}
  \label{eq:Ln}
  T(z) \, = \, \sum_{n \in \mathbb{Z}} z^{-n-2} L_n   \qquad \quad
  \overline{T}(\overline{z}) \, = \, \sum_{n \in \mathbb{Z}}
  \overline{z}^{-n-2} \overline{L}_n .
\end{equation}
We have also introduced an operator $\Sigma$, which satisfies
$\Sigma^2 = 1$. Such an operator is needed because of the {\it
  staggering}: some of the low-energy states in the spectrum of the
Heisenberg Hamiltonian have a momentum close to $\pi$, rather than
to $0$. For instance, when $N \in 4 \mathbb{N}+2$, the ground
state itself has momentum $\pi$, and the corresponding eigenvalue of
$T_L$ is a negative real number. To match the lattice momentum $P$
with the CFT momentum operator $P_{\rm CFT}$, one needs to take this
sign into account; this is precisely what the operator $\Sigma$
does. Notice that if, instead of $T_L(\alpha)$, we were focusing on
the {\it double-row transfer matrix} $(T_L(\alpha))^2$, then an
identification of the form (\ref{eq:Hcft}) would still hold, and this
time no operator $\Sigma$ would be needed.

The identification of the spectrum of $T_L(\alpha)$ with the one of
$H_{\rm CFT}$ and $P_{\rm CFT}$ is valid for chains with an {\it even}
number of sites $N$. Then the spectrum is the one of the operator in
(\ref{eq:Hcft}) acting on
\begin{equation}
  \mathcal{H}_{\rm CFT} \, = \,	\left[ \phi_0\right] \otimes
    \left[\,\overline{\phi}_0\right] 
    \, \oplus \, \left[\phi_{\frac{1}{2}}\right] \otimes 
      \left[\,\overline{\phi}_{\frac{1}{2}} \right] .
\end{equation}
Here $\left[\phi_j\right]$ denotes the $\suhat(2)_k$ irreducible
highest weight representation associated to $\phi_j$.  (Recall that
there are only two primary fields at level $k=1$, $\phi_{0}$ and
$\phi_{\frac{1}{2}}$.)  In other words, the continuum limit of the
six-vertex model is the diagonal $\suhat(2)_1 \otimes
\overline{\suhat(2)}_1$ CFT. We will use this well-known result as our
starting point in the construction of the lattice holomorphic
$\suhat(2)_1$ current in section \ref{sec:chiral_current}.


\subsection{Spin-$k/2$ descendants of the six-vertex model and $\suhat(2)_k$}

In this paper, we are interested in constructing a lattice version of
the holomorphic $\suhat(2)_k$ current for general $k \geq 1$; thus,
we need lattice discretizations of the $\suhat(2)_k \otimes
\overline{\suhat(2)}_k$ CFT. As reviewed in the previous section, a
discretization at level $k=1$ is given to us by the six-vertex
model. For $k>1$, it turns out that the lattice discretizations
existing in the literature are also related to the six-vertex model:
they are spin-$k/2$ {\it descendants} of the six-vertex model obtained
from the {\it fusion procedure} of Kulish-Reshetikhin-Sklyanin \cite{KRS81}. One
starts from the $\SU(2)$-invariant $R$-matrix
(\ref{eq:R_x_tensor})-(\ref{eq:R6v1}), and constructs a new
$\SU(2)$-invariant $R$-matrix for higher spin representations as
follows. The new $R$-matrix is a tensor
\begin{equation}
	\label{eq:R_x_tensor_k}
	R^{(k/2)}_{\bf x} \, \in \, 
  \left(\frac{k}{2}\right)_{{\bf x} + \frac{1}{2}{\bf v}} \otimes 
  \left(\frac{k}{2}\right)_{{\bf x} + \frac{1}{2}{\bf u}} \otimes 
  \left(\frac{k}{2}\right)^*_{{\bf x} - \frac{1}{2}{\bf u}} \otimes 
  \left(\frac{k}{2}\right)^*_{{\bf x} - \frac{1}{2}{\bf v}}\, ,
\end{equation}
which is best represented pictorially as
\begin{equation}
	\label{eq:fused_R}
	\begin{tikzpicture}
	\begin{scope}[xshift=0cm]
		\draw[line width=2.5pt]  (-1,0) -- (1,0);
		\draw[line width=2.5pt]  (-0.1564,-0.9877) -- (0.1564,0.9877);
		\draw (0,-0.2) arc (-90:-180:0.2) ++(-0.1,-0.3) node{$\alpha$};
	\end{scope}
	\draw (2,0) node{$=$};
	\begin{scope}[xshift=5.5cm]
		\draw[thick]  (-2.5-0.1564,-1) -- (2.5-0.1564,-1);
		\draw[thick]  (-2.5-0.5*0.1564,-0.5) -- (2.5-0.5*0.1564,-0.5);
		\draw[thick]  (-2.5,0) -- (2.5,0);
		\draw[thick]  (-2.5+0.5*0.1564,0.5) -- (2.5+0.5*0.1564,0.5);
		\draw[thick]  (-2.5+0.1564,1) -- (2.5+0.1564,1);
		
		\draw[thick]  (-1-2.5*0.1564,-2.5*0.9877) -- (-1+2.5*0.1564,2.5*0.9877);
		\draw[thick]  (-0.5-2.5*0.1564,-2.5*0.9877) -- (-0.5+2.5*0.1564,2.5*0.9877);
		\draw[thick]  (0-2.5*0.1564,-2.5*0.9877) -- (0+2.5*0.1564,2.5*0.9877);
		\draw[thick]  (0.5-2.5*0.1564,-2.5*0.9877) -- (0.5+2.5*0.1564,2.5*0.9877);
		\draw[thick]  (1-2.5*0.1564,-2.5*0.9877) -- (1+2.5*0.1564,2.5*0.9877);
		
		\draw (-1.1564-0.15,-1*0.9877) arc (-180:-99:0.15) ++(-0.2,-0.12) node{\footnotesize $\alpha_1$};
		\draw (0.5-1.1564-0.15,-1*0.9877) arc (-180:-99:0.15) ++(-0.2,-0.12) node{\footnotesize $\alpha_2$};
		\draw (1-1.1564-0.15,-1*0.9877) arc (-180:-99:0.15) ++(-0.2,-0.12) node{\footnotesize $\alpha_3$};
		\draw (2-1.1564-0.15,-1*0.9877) arc (-180:-99:0.15) ++(-0.2,-0.12) node{\footnotesize $\alpha_k$};

		\draw (-1-0.5*0.1564-0.15,-0.5*0.9877) arc (-180:-99:0.15) ++(-0.2,-0.12) node{\footnotesize $\alpha_2$};
		\draw (-0.5-0.5*0.1564-0.15,-0.5*0.9877) arc (-180:-99:0.15) ++(-0.2,-0.12) node{\footnotesize $\alpha_3$};
		\draw (-1-0.15,0) arc (-180:-99:0.15) ++(-0.2,-0.12) node{\footnotesize $\alpha_3$};

		\draw (-1+0.1564-0.15,0.9877) arc (-180:-99:0.15) ++(-0.2,-0.12) node{\footnotesize $\alpha_k$};
		\draw (1+0.1564-0.15,0.9877) arc (-180:-99:0.15) ++(-0.25,-0.12) node{\footnotesize $\alpha_{2k-1}$};
		
		\filldraw[fill=white] (-2,0) ellipse (0.15 and 1.2);
		\filldraw[fill=white] (2,0) ellipse (0.15 and 1.2);
		\filldraw[fill=white] (0+2.1*0.1564,2.1) ellipse (1.2 and 0.15);
		\filldraw[fill=white] (0-2.1*0.1564,-2.1) ellipse (1.2 and 0.15);
	\end{scope}
	\end{tikzpicture}
\end{equation}
where the big ellipses stand for the projector onto the
spin-$k/2$ representation of $\SU(2)$, namely the full symmetrizer in
$\bigotimes_{i=1}^k \mathbb{C}^2$. 
Here $\alpha$ is again the geometric angle. The parameters $\alpha_n$ all depend on $\alpha$ in a specific way. 
The choice of these parameters is a crucial step in the fusion procedure. It turns out that the correct choice is
\begin{equation}
	\alpha_n \, = \, \alpha + (k-n) \pi\, .
\end{equation}
It ensures that the following two identities hold:
\begin{subequations}
	\label{eq:2relations}
\begin{eqnarray}
	\begin{tikzpicture}
	\begin{scope}[xshift=0cm]
		\draw[thick]  (-1-1*0.1564,-1*0.9877) -- (-1+1*0.1564,1*0.9877);		
		\draw[thick]  (-0.5-1*0.1564,-1*0.9877) -- (-0.5+1*0.1564,1*0.9877);		
		\draw[thick]  (-1*0.1564,-1*0.9877) -- (1*0.1564,1*0.9877);		
		\draw[thick]  (0.5-1*0.1564,-1*0.9877) -- (0.5+1*0.1564,1*0.9877);		
		\draw[thick]  (1-1*0.1564,-1*0.9877) -- (1+1*0.1564,1*0.9877);		
		\draw[thick]  (-1.3,0) -- (1.3,0);

		\draw (-1.02,-0.17) arc (-99:-180:0.17) ++(-0.08,-0.3) node{\footnotesize $\alpha_1$};
		\draw (0.5-1.02,-0.17) arc (-99:-180:0.17) ++(-0.08,-0.3) node{\footnotesize $\alpha_2$};
		\draw (2-1.02,-0.17) arc (-99:-180:0.17) ++(-0.08,-0.3) node{\footnotesize $\alpha_k$};

		\filldraw[fill=white] (0.1,0.7) ellipse (1.2 and 0.15);
		\filldraw[fill=white] (-0.1,-0.7) ellipse (1.2 and 0.15);
	\end{scope}
	\draw (2,0) node{$=$};
	\begin{scope}[xshift=4cm]		
		\draw[thick]  (-1-1*0.1564,-1*0.9877) -- (-1+1*0.1564,1*0.9877);		
		\draw[thick]  (-0.5-1*0.1564,-1*0.9877) -- (-0.5+1*0.1564,1*0.9877);		
		\draw[thick]  (-1*0.1564,-1*0.9877) -- (1*0.1564,1*0.9877);		
		\draw[thick]  (0.5-1*0.1564,-1*0.9877) -- (0.5+1*0.1564,1*0.9877);		
		\draw[thick]  (1-1*0.1564,-1*0.9877) -- (1+1*0.1564,1*0.9877);		
		\draw[thick]  (-1.3,0) -- (1.3,0);

		\draw (-1.02,-0.17) arc (-99:-180:0.17) ++(-0.08,-0.3) node{\footnotesize $\alpha_1$};
		\draw (0.5-1.02,-0.17) arc (-99:-180:0.17) ++(-0.08,-0.3) node{\footnotesize $\alpha_2$};
		\draw (2-1.02,-0.17) arc (-99:-180:0.17) ++(-0.08,-0.3) node{\footnotesize $\alpha_k$};

		\filldraw[fill=white] (-0.1,-0.7) ellipse (1.2 and 0.15);
	\end{scope}
	\end{tikzpicture} \\
	\begin{tikzpicture}
	 \begin{scope}[xshift=0cm]
		\draw[thick]  (-1-1*0.1564,-1*0.9877) -- (1-1*0.1564,-1*0.9877);		
		\draw[thick]  (-1-0.5*0.1564,-0.5*0.9877) -- (1-0.5*0.1564,-0.5*0.9877);
		\draw[thick]  (-1-0*0.1564,-0*0.9877) -- (1-0*0.1564,-0*0.9877);
		\draw[thick]  (-1+0.5*0.1564,0.5*0.9877) -- (1+0.5*0.1564,0.5*0.9877);
		\draw[thick]  (-1+1*0.1564,1*0.9877) -- (1+1*0.1564,1*0.9877);
		
		\draw[thick]  (0+-1.3*0.1564,-1.3*0.9877) -- (1.3*0.1564,1.3*0.9877);
		
		\draw (-.1564-0.15,-1*0.9877) arc (-180:-99:0.15) ++(-0.2,-0.12) node{\footnotesize $\alpha_1$};
		\draw (-0.5*.1564-0.15,-0.5*0.9877) arc (-180:-99:0.15) ++(-0.2,-0.12) node{\footnotesize $\alpha_2$};
		\draw (.1564-0.15,1*0.9877) arc (-180:-99:0.15) ++(-0.2,-0.12) node{\footnotesize $\alpha_k$};
		
		\filldraw[fill=white] (-0.6,0) ellipse (0.15 and 1.2);
		\filldraw[fill=white] (0.6,0) ellipse (0.15 and 1.2);
	\end{scope}
	\draw (2,0) node{$=$};	
	 \begin{scope}[xshift=4cm]
		\draw[thick]  (-1-1*0.1564,-1*0.9877) -- (1-1*0.1564,-1*0.9877);		
		\draw[thick]  (-1-0.5*0.1564,-0.5*0.9877) -- (1-0.5*0.1564,-0.5*0.9877);
		\draw[thick]  (-1-0*0.1564,-0*0.9877) -- (1-0*0.1564,-0*0.9877);
		\draw[thick]  (-1+0.5*0.1564,0.5*0.9877) -- (1+0.5*0.1564,0.5*0.9877);
		\draw[thick]  (-1+1*0.1564,1*0.9877) -- (1+1*0.1564,1*0.9877);
		
		\draw[thick]  (0+-1.3*0.1564,-1.3*0.9877) -- (1.3*0.1564,1.3*0.9877);
		
		\draw (-.1564-0.15,-1*0.9877) arc (-180:-99:0.15) ++(-0.2,-0.12) node{\footnotesize $\alpha_1$};
		\draw (-0.5*.1564-0.15,-0.5*0.9877) arc (-180:-99:0.15) ++(-0.2,-0.12) node{\footnotesize $\alpha_2$};
		\draw (.1564-0.15,1*0.9877) arc (-180:-99:0.15) ++(-0.2,-0.12) node{\footnotesize $\alpha_k$};
		
		\filldraw[fill=white] (-0.6,0) ellipse (0.15 and 1.2);
	\end{scope}
	\draw (5.3,-1.2) node{.};
	\end{tikzpicture}	
\end{eqnarray}
\end{subequations}
These two identities may be proved inductively.
It is yet another pleasant exercise to check that the above two
relations, together with \eqref{eq:YB}, imply that $R^{(k/2)}$ is
itself a solution to the Yang-Baxter equation for higher spin:
\begin{equation}
	\label{eq:YB_high_spin}
        \raisebox{-1.5cm}{\begin{picture}(0,0)%
\includegraphics{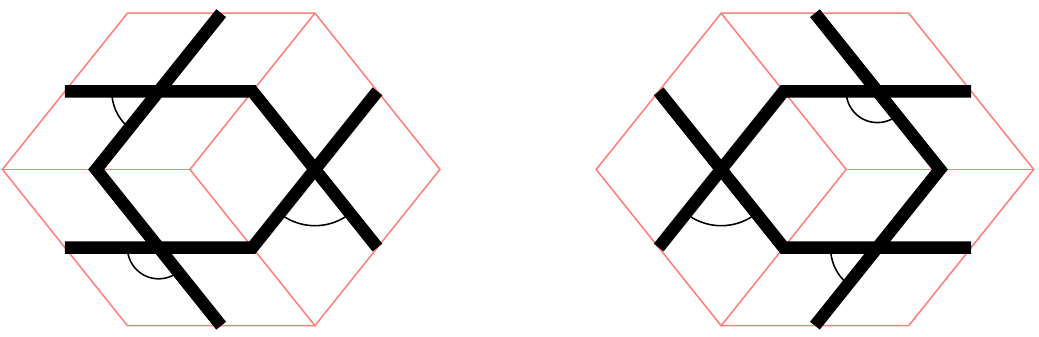}%
\end{picture}%
\setlength{\unitlength}{3947sp}%
\begingroup\makeatletter\ifx\SetFigFont\undefined%
\gdef\SetFigFont#1#2#3#4#5{%
  \reset@font\fontsize{#1}{#2pt}%
  \fontfamily{#3}\fontseries{#4}\fontshape{#5}%
  \selectfont}%
\fi\endgroup%
\begin{picture}(4974,1608)(439,-2665)
\put(2926,-1861){\makebox(0,0)[b]{\smash{{\SetFigFont{12}{14.4}{\familydefault}{\mddefault}{\updefault}{\color[rgb]{0,0,0}$=$}%
}}}}
\put(1126,-2461){\makebox(0,0)[rb]{\smash{{\SetFigFont{10}{12.0}{\familydefault}{\mddefault}{\updefault}{\color[rgb]{0,0,0}$\alpha$}%
}}}}
\put(1951,-2311){\makebox(0,0)[b]{\smash{{\SetFigFont{10}{12.0}{\familydefault}{\mddefault}{\updefault}{\color[rgb]{0,0,0}$\alpha-\beta$}%
}}}}
\put(901,-1711){\makebox(0,0)[rb]{\smash{{\SetFigFont{10}{12.0}{\familydefault}{\mddefault}{\updefault}{\color[rgb]{0,0,0}$\beta$}%
}}}}
\put(3901,-2311){\makebox(0,0)[b]{\smash{{\SetFigFont{10}{12.0}{\familydefault}{\mddefault}{\updefault}{\color[rgb]{0,0,0}$\alpha-\beta$}%
}}}}
\put(4351,-2461){\makebox(0,0)[rb]{\smash{{\SetFigFont{10}{12.0}{\familydefault}{\mddefault}{\updefault}{\color[rgb]{0,0,0}$\beta$}%
}}}}
\put(4576,-1711){\makebox(0,0)[rb]{\smash{{\SetFigFont{10}{12.0}{\familydefault}{\mddefault}{\updefault}{\color[rgb]{0,0,0}$\alpha$}%
}}}}
\end{picture}%
}\, .
\end{equation}
Similarly, one can check that the relation (\ref{eq:2relations}) and the inversion relation (\ref{eq:inversion}) for the original $R$-matrix, imply that $R^{(k/2)}$ satisfies the same relation, up to a global coefficient:
\begin{equation}
	\label{eq:inversion_k}
	\begin{tikzpicture}
          \draw[line width=2.5pt] (-1,-0.4) arc (150:30:2);
          \draw[line width=2.5pt] (-1,0.4) arc (-150:-30:2);
          \draw (-0.83,-0.15) arc (-135:-225:0.2) ++(0,-0.1) node[left]{$\alpha$};	
          \draw (2,-0.15) arc (-135:-225:0.2) ++(0,-0.1) node[left]{$2\pi-\alpha$};
          \draw (6,0) node{$\displaystyle =\,
            \prod_{p=-k+1}^{k-1}
            \left[\left(\frac{\alpha}{\pi}+p\right)\left(\frac{2\pi-\alpha}{\pi}+p\right)\right]^{k-|p|}$};
          \draw[line width=2.5pt] (9.5,-0.4) arc (150:30:2 and 0.3) ++(0.2,-0.1) node{.};
          \draw[line width=2.5pt] (9.5,0.4) arc (-150:-30:2 and 0.3);
	\end{tikzpicture}
\end{equation}
As in the spin-$1/2$ case, the $R$-matrix is used to construct the
one-parameter family of commuting transfer matrices
$T^{(k/2)}_L(\alpha)$ on $N = L/a_0$ sites, with periodic boundary conditions;
the spectrum of the transfer matrix can again be
obtained from the Bethe Ansatz (see \cite{Sierra_book} or appendix \ref{sec:appendixA}).  The
low-lying spectrum of $-\log T^{(k/2)}_L(\alpha)$ has been identified
with the spectrum of the CFT Hamiltonian (\ref{eq:Hcft}), where $L_0$
and $\overline{L}_0$ are the zero modes of the Sugawara stress-tensor
in the $\suhat(2)_k \otimes \overline{\suhat(2)}_k$ theory (for more details about
the general $k$ case, see
e.g.~\cite{Affleck1989}). Like in the spin-$1/2$ case, this
identification holds when the number of sites $N$ is $\it even$, and
when the Hilbert space of the CFT is the diagonal module
\begin{equation}
  \label{eq:diag_module}
  \mathcal{H}_{\rm CFT} \, = \, \bigoplus_{j= 0}^{k/2} \left[
    \phi_{j}\right] \otimes \left[\,\overline{\phi}_{j} \right] .
\end{equation}
Again, the known identification of the diagonal $\suhat(2)_k \otimes
\overline{\suhat(2)}_k$ theory as the continuum limit of the
spin-$\frac{k}{2}$ descendant of the six-vertex model is the starting
point of this paper.

Let us conclude this section with the explicit formulas for the $R$-matrix
and for the critical Hamiltonian, which are often useful in calculations.
A tedious but straightforward computation leads to
\begin{equation}
  R^{\left(k/2 \right)}(\alpha) \, = \, 
  \frac{1}{k!}  \sum_{j=0}^k  
  \left[  \, \prod_{p=0}^{j-1} \left(p+\frac{\alpha}{\pi}\right) 
    \, \prod_{q=j+2}^{k+1} \left(q-\frac{\alpha}{\pi}\right) 
  \right] P_j  \, ,
\end{equation}
where $P_j$ is the projector on spin $j$. $P_j$ can be expressed as a polynomial in the Heisenberg coupling $S^a\otimes S_a$ in
$\left(\frac{k}{2}\right)\otimes\left(\frac{k}{2}\right)$ as
\begin{align}
  P_j = \prod_{\substack{\ell=0\\\ell\neq j}}^k \frac{S^a\otimes S_a-
    x_\ell}{x_j-x_\ell}\, ,
\end{align}
with $x_\ell = \frac{1}{2}\ell(\ell+1)-\frac{k}{2}(\frac{k}{2}+1)$. When $\alpha \rightarrow \pi$, we have
\begin{equation}
	R^{(k/2)}(\alpha) \, \underset{\alpha \rightarrow \pi}{=} \, 1 \, + \, (\alpha - \pi) \frac{2}{\pi} \left[ \sum_{j=1}^k c_j P_j \; - \, \frac{1}{2} c_k \right] \, + \, O((\alpha - \pi)^2),
\end{equation}
with $c_j  =\sum_{p=1}^j
  \frac{1}{p}$ the harmonic number. As in the $k=1$ case, it is convenient to first define the {\it Hamiltonian density} $h \equiv R'(\pi)$, and then observe that
\begin{eqnarray}
  \label{eq:H}
\nonumber  H^{(k/2)} 
   &\equiv  & 
  \left( \frac{\partial}{\partial \alpha} \log T_N \right)\Bigg|_{\alpha = \pi} \\
\nonumber  & = & \sum_{x/a_0 = 1}^N h_{x,x+a_0} \\
  &=&
  \frac{2}{\pi}
  \sum_{x/a_0=1}^N
  \left(
  \sum_{j=1}^k \, c_{j}\, (P_j)_{x,x+a_0}   \, - \,
  \frac{1}{2}c_k 
  \right) \,.
\end{eqnarray}


\section{Expansion of the lattice spin operator}
\label{sec:lattice_spin_op}

In the thermodynamic limit all observables in the vertex model can be
expressed as polynomials in the spin operators $S^a_{{\bf x}}$, which
act on the spin-$k/2$ representations of $\SU(2)$ that live on the
edges of the lattice.  $S^a_{{\bf x}}$ itself can be expanded in terms
of the primary fields and their descendants
\begin{equation}
  \label{eq:S_J}
  S^a_{{\bf x}} \quad \underset{a_0 \rightarrow 0}{=}  
  \quad \sum_{j \in \{0, \frac{1}{2}, \dots \frac{k}{2}\}} C_{{\bf x}}^{(j)} \, a_0^{2 h_{j}} \,  P_1 \left[  \phi_{j}(z) \otimes \overline{\phi}_{j}(\bar{z}) \right]^a \;
  + \; {\rm descendants},
\end{equation}
where, again, ${\bf x } =(x,y)$ and $z=x + e^{i \alpha} y$. As
remarked in the introduction, this type of identity makes sense when
it is inserted inside correlators. Note that the currents $J^a(z)$ and 
$\overline{J}^a(\overline{z})$ contribute to the descendants.
The notation $P_1 \left[ . \right]^a$ requires some explanation. By
definition, the left-hand side transforms in the adjoint (spin-$1$)
representation under the action of $\mathfrak{su}(2)$. The terms that
appear in the right-hand side must therefore transform in the adjoint
representation as well. Now note that the field $\phi_{j}(z) \otimes
\overline{\phi}_{j}(\bar{z}) $ is a matrix-valued primary field, with
$(2j+1) \times (2j+1)$ entries, which transforms in the representation
$(j) \otimes (j)$ under the action of the diagonal $\mathfrak{su} (2)$
subalgebra generated by the zero modes $J^a_0 + \overline{J}^a_0$.
Then $P_1 \left[ . \right]$ stands for the projector onto the unique spin-$1$
irreducible representation occurring in $(j) \otimes (j)
= \bigoplus_{p=0}^{2j}(p)$ if $j>0$. For $j=0$, $P_1[.]$ just vanishes.
Now, if $j>0$, $P_1[\phi_j \otimes \overline{\phi}_j]$ has three components transforming in the adjoint
representation, so one may identify them as the three components $a=1,2,3$ in a unique way (up to a global
factor that is irrelevant for our purposes).
 
\subsection{The coefficients $C^{(j)}_{{\bf x}}$}
 
The coefficients $C^{(j)}_{{\bf x}}$ are non-universal; yet, they are
crucial when one tries to match the correlation functions computed on
the lattice with the ones in the continuum limit. In general,
computing these factors is a difficult task. It has nevertheless been
carried out up to some extent in the literature, at least for $k=1$
\cite{Affleck_JPA98,Lukyanov97,LukyanovTerras}. To our knowledge, no
explicit form is known for arbitrary $k$ and arbitrary $j$. 
Fortunately though, the explicit values of these coefficients won't be needed
for the purposes of this paper. The only thing we need is the fact
that roughly half of the coefficients are {\it staggered}, and that
the remaining half vanishes, as we now argue.

It is known (see for instance \cite{Affleck1985,affleck1987critical,
  affleck1988critical}) that one site lattice translations by 
either of the two vectors ${\bf u}$ or ${\bf v}$
 corresponds to changing the sign of the matrix-valued WZW
field $g(z,\bar{z})=\phi_{\frac{1}{2}}(z)\otimes\overline{\phi}_
{\frac{1}{2}}(\overline{z})$:
\begin{equation}
	\label{eq:g_m_g}
	g(z,\bar{z}) \; \mapsto \; -g(z,\bar{z}).
\end{equation}
(As before we identify $z=x+e^{i\alpha}y$,
$\overline{z}=x+e^{-i\alpha}y$ for $(x,y)$ a point of the lattice.)
In the $k=1$ case, this fact in particular prevents the {\it relevant}
$\SU(2)$-symmetric perturbation $\Tr \left[ g(z,\bar{z}) \right]$ to
appear in the effective action that describes the spin-$\frac{1}{2}$
Heisenberg chain.  This perturbation would drive the system away from
criticality but since it breaks translation-invariance, which is a
symmetry of the Heisenberg Hamiltonian, its appearance in the
effective theory is prohibited.  When instead translation symmetry is
explicitly broken, the spin-$\frac{1}{2}$ chain typically dimerizes.
 
Now, since the matrix-valued field $\phi_j (z)\otimes
\overline{\phi}_j(\bar{z})$ can be obtained by fusing
$g(z, \bar{z}) = \phi_{\frac{1}{2}}(z) \otimes \overline{\phi}_{\frac{1}{2}}
(\bar{z})$ with itself $2j$ times, we see that a translation by ${\bf u}$ or by
$\bf v$ on the lattice must act as
\begin{equation}
	\phi_{j}(z) \otimes \overline{\phi}_{j} (\bar{z}) \; \mapsto \; (-1)^{2j} \phi_{j}(z) \otimes \overline{\phi}_{j} (\bar{z}).
\end{equation}
This means that all the coefficients $C^{(j)}$ with half-integer $j$ are {\it staggered}:
\begin{equation}
	\label{eq:Cxj_halfint}
	({\rm half-integer } \,j)  \qquad \quad C_{\bf x}^{(j)} \, = \, \left\{  \begin{array}{lrl}  C^{(j)}_{\rm horiz.} \, (-1)^{\frac{x+y}{a_0}-\frac{1}{2}}   &&  ({\rm horizontal \, edges}) \\
				 C^{(j)}_{\rm vert.} \, (-1)^{\frac{x+y}{a_0}-\frac{1}{2}}    &&  ({\rm vertical \, edges}) , \end{array} \right.
\end{equation}
while the coefficients for integer $j$ are not. Actually, the latter simply vanish:
\begin{equation}
	\label{eq:Cxj_int}
	({\rm integer } \,j)  \qquad \qquad \quad C_{\bf x}^{(j)} \, = 0 \, .
\end{equation}
This may be justified as follows. First, we observe that since for
integer $j$ the coefficient $C_{\bf x}^{(j)}$ is not staggered, it
must be independent of ${\bf x}$. Then, without loss of generality,
one may focus on this coefficient at a specific point ${\bf x} =
(x,y)$ with $y=0$, such that ${\bf x}$ is a fixed point of the spatial
inversion $z = x +  e^{i \alpha} y \longmapsto \bar{z}  = x + e^{-i
  \alpha} y$. Under this transformation, $S_{\bf x}^a$ is mapped to
itself, so any non-zero term appearing in the right-hand side of
(\ref{eq:S_J}) must be invariant. On the other hand, spatial inversion
is a symmetry of the continuous euclidean field theory, which
exchanges the chiral and anti-chiral sectors $\phi_j$ and
$\overline{\phi}_j$  in $P_1\left[ \phi_j \otimes \overline{\phi}_j
\right]$. Since the spin-$(1)$ representation appearing in the
decomposition  $\displaystyle (j) \otimes (j)$ is always {\it anti-symmetric} for integer spin $j$, then  $P_1\left[ \phi_j \otimes \overline{\phi}_j \right]$ must be odd under spatial inversion. Therefore it cannot contribute to $S^a_{\bf x}$ in the continuum limit.

The structure (\ref{eq:Cxj_halfint})-(\ref{eq:Cxj_int}) of the expansion of $S_{\bf x}^a$ is related to properties of the Bethe states and of the associated form factors. These are discussed for $k=1$ and $k=2$ in Refs. \cite{Hagemans_Caux, Caux_spin_1}, which also make crucial use of spatial inversion. More details about this point are given in appendix \ref{sec:appendixA}.

\subsection{Contribution of the current to $S_{\bf x}^a$}

Next, we analyze the second part of the right-hand side of
(\ref{eq:S_J}), namely the contribution of the descendants. The latter are generated by the action of the chiral and
anti-chiral currents on the primary fields. They always
have a scaling dimension that is the one of the primary operator they
descend from, plus some positive integer number. The currents $J^a(z)$
and $\overline{J}^a(\bar{z})$ themselves are descendants of the
identity, and have scaling dimension one. All other descendants appearing
in the r.h.s. of (\ref{eq:S_J}) have strictly larger scaling
dimensions; the smallest possible one being $\beta \equiv 2
h_{\frac{1}{2}} +1 \, >\,1$, for the first descendants of the primary
field $\phi_{\frac{1}{2}}(z) \otimes
\overline{\phi}_{\frac{1}{2}}(\bar{z})$. Thus,
\begin{equation}
	\label{eq:S_J_desc}
	 S^a_{{\bf x}} \quad \underset{a_0 \rightarrow 0}{=}  \quad \sum_{j \; {\rm half-int.}} C^{(j)}_{{\bf x}} \, a_0^{2 h_{j}} \,  P_1 \left[  \phi_{j}(z) \otimes \overline{\phi}_{j}(\bar{z}) \right]^a \; + \;  a_0 \left[ C^J_{{\bf x}} \, J^a(z) \,  + \,  C^{\bar{J}}_{{\bf x}} \, \overline{J}^a(\bar{z}) \right] \; + \; O(a_0^\beta) .
\end{equation}
The next step is to fix the coefficients $C^J_{{\bf x}}$ and
$C^{\bar{J}}_{{\bf x}}$, using the $\SU(2)$ symmetry of the vertex
model. By construction, the $R$-matrix $R^{(\frac{k}{2})}_{\bf x}$ is
an $\SU(2)$-invariant tensor, see
Eq.~\eqref{eq:R_x_tensor_k}, and as such it is annihilated
by the total spin operator
\begin{align}
\label{eq:Stot}
  S_{{\bf x}+\frac{1}{2}{\bf u}}^a+S_{{\bf x}+\frac{1}{2}{\bf v}}^a
  -
  (S^a_{{\bf x}-\frac{1}{2}{\bf u}})^t-
  (S^a_{{\bf x}-\frac{1}{2}{\bf v}})^t
  \, ,
\end{align}
for any generator $S^a$ of $\mathfrak{su}(2)$. One may interpret this
as the fact that the following {\it discrete contour integral} around
a vertex at position ${\bf x}$ vanishes:
\begin{equation}
\label{eq:little_contour}
\begin{tikzpicture}[scale=0.75]
  \begin{scope}
    \draw[line width = 2.2pt]  (-1,0) -- (1,0);
    \draw[line width = 2.2pt]  (0.1564,0.9877) -- (-1*0.1564,-1*0.9877);
    \draw[very thick,red] (-0.6-0.6*0.1564,-0.6*0.9877) -- (-0.6+0.6*0.1564,0.6*0.9877) -- (0.6+0.6*0.1564,0.6*0.9877) -- (0.6-0.6*0.1564,-0.6*0.9877) -- cycle;
    \draw (-0.2,0) arc (-180:-90:0.2) ++(0,-0.1) node[left]{$\alpha$};
    \filldraw (0.5,-0.1) rectangle ++(0.2,0.2); 
  \end{scope}
  \node at (1.5,0) {$+$};
\end{tikzpicture} 
\begin{tikzpicture}[xshift=1cm,scale=0.75]
	\begin{scope}
		\draw[line width = 2.2pt]  (-1,0) -- (1,0);
		\draw[line width = 2.2pt]  (0.1564,0.9877) -- (-1*0.1564,-1*0.9877);
		\draw[very thick,red] (-0.6-0.6*0.1564,-0.6*0.9877) -- (-0.6+0.6*0.1564,0.6*0.9877) -- (0.6+0.6*0.1564,0.6*0.9877) -- (0.6-0.6*0.1564,-0.6*0.9877) -- cycle;
		\draw (-0.2,0) arc (-180:-90:0.2) ++(0,-0.1) node[left]{$\alpha$};
		\filldraw (-0.12+0.7*0.1564,-0.1+0.6*0.9877) rectangle ++(0.2,0.2); 
	\end{scope}
  \node at (1.5,0) {$+$};
\end{tikzpicture} 
\begin{tikzpicture}[xshift=3cm,scale=0.75]
	\begin{scope}
		\draw[line width = 2.2pt]  (-1,0) -- (1,0);
		\draw[line width = 2.2pt]  (0.1564,0.9877) -- (-1*0.1564,-1*0.9877);
		\draw[very thick,red] (-0.6-0.6*0.1564,-0.6*0.9877) -- (-0.6+0.6*0.1564,0.6*0.9877) -- (0.6+0.6*0.1564,0.6*0.9877) -- (0.6-0.6*0.1564,-0.6*0.9877) -- cycle;
		\draw (-0.2,0) arc (-180:-90:0.2) ++(0,-0.1) node[left]{$\alpha$};
		\filldraw (-0.5,-0.1) rectangle (-0.7,0.1); 
	\end{scope}
  \node at (1.5,0) {$+$};
\end{tikzpicture} 
\begin{tikzpicture}[xshift=5cm,scale=0.75]
  \begin{scope}
    \draw[line width = 2.2pt]  (-1,0) -- (1,0);
    \draw[line width = 2.2pt]  (0.1564,0.9877) -- (-0.1564,-0.9877);
    \draw[very thick,red] (-0.6-0.6*0.1564,-0.6*0.9877) -- (-0.6+0.6*0.1564,0.6*0.9877) -- (0.6+0.6*0.1564,0.6*0.9877) -- (0.6-0.6*0.1564,-0.6*0.9877) -- cycle;
    \draw (-0.2,0) arc (-180:-90:0.2) ++(0,-0.1) node[left]{$\alpha$};
    \filldraw (-0.075-0.7*0.1564,-0.7*0.9877) rectangle ++(0.2,0.2); 
  \end{scope}
  \node at (1.75,0) {$=0$.};
\end{tikzpicture}
\end{equation}
In Eq.~\eqref{eq:Stot}, $-(S^a)^t$ corresponds to the dual action on
$(\frac{k}{2})^*$. If instead we regarded the $R$-matrix
as the linear operator of Eq.~\eqref{eq:matrix_matrix} (generalized
for arbitrary $k$), then $\SU(2)$ invariance would read as
the vanishing of
\begin{align}
  \label{eq:Stot_2}
  S_{{\bf x}+\frac{1}{2}{\bf u}}^a+S_{{\bf x}+\frac{1}{2}{\bf v}}^a
  -
  S_{{\bf x}-\frac{1}{2}{\bf u}}^a-S_{{\bf x}-\frac{1}{2}{\bf v}}^a
  \, ,
\end{align}
when inserted in a correlator.  Note that the presence of the
transpose in Eq.~\eqref{eq:Stot} is due to the fact that vectors in
the dual representation are regarded as column vectors on which the
matrix acts on the left.  However, when the $R$-matrix is regarded as an 
operator on the Hilbert space of the vertex model, as in
Eq.~\eqref{eq:Stot_2}, the action by the commutator has to be used.  
In particular this second point of view has to be used when we consider
Eq.~\eqref{eq:S_J_desc} in a correlator.

\begin{figure}[h]
\centering
	\begin{tabular}{lcl}
\begin{tikzpicture}[scale=1]
	\foreach \x in {-3,-2,...,3} 	\draw[thick] (\x-3.5*0.1564,-3.5*0.9877) -- (\x+3.5*0.1564,3.5*0.9877);
	\foreach \y in {-3,-2,...,3} 	\draw[thick] (-3.5+\y*0.1564,\y*0.9877) -- (3.5+\y*0.1564,\y*0.9877);
	\draw[very thick,red] (2.5+1.5*0.1564,1.5*0.9877) -- (2.5-0.5*0.1564,-0.5*0.9877) -- (0.5-0.5*0.1564,-0.5*0.9877) -- (0.5-1.5*0.1564,-1.5*0.9877) -- (-1.5-1.5*0.1564,-1.5*0.9877) -- (-1.5+1.5*0.1564,1.5*0.9877) -- cycle node[above]{$\Gamma$};
	\foreach \x in {-2.5,-1.5,...,2.5}
		{
		\foreach \y in {-3,-2,...,3}
			{
				\filldraw (\x+\y*0.1564,\y*0.9877) circle (0.4mm);
				\filldraw (\y+\x*0.1564,\x*0.9877) circle (0.4mm);
			}
		}
	\draw (-3.2-3*0.1564,-3*0.9877) arc (-180:-99:0.2) ++(-0.3,0.05) node[below]{$\alpha$};
\end{tikzpicture}
	&& 
\begin{tikzpicture}[scale=0.5]
	\foreach \x in {-6,-5,...,7} 	\draw[thin] (\x-6.5*0.1564,-6.5*0.9877) -- (\x+7.5*0.1564,7.5*0.9877);
	\foreach \y in {-6,-5,...,7} 	\draw[thin] (-6.5+\y*0.1564,\y*0.9877) -- (7.5+\y*0.1564,\y*0.9877);
	\draw[very thick,blue] (2.5+1.5*0.1564,1.5*0.9877) -- (2.5-2.5*0.1564,-2.5*0.9877) -- (0.5-2.5*0.1564,-2.5*0.9877) -- (0.5-1.5*0.1564,-1.5*0.9877) -- (-1.5-1.5*0.1564,-1.5*0.9877) -- (-1.5+1.5*0.1564,1.5*0.9877) --cycle node[above]{$\Gamma'$};
	\draw[very thick,red] (4.5+3.5*0.1564,3.5*0.9877) -- (4.5-3.5*0.1564,-3.5*0.9877) -- (2.5-3.5*0.1564,-3.5*0.9877) -- (2.5-4.5*0.1564,-4.5*0.9877) -- (-2.5-4.5*0.1564,-4.5*0.9877) -- (-2.5+3.5*0.1564,3.5*0.9877) -- (0.5+3.5*0.1564,3.5*0.9877) -- (0.5+5.5*0.1564,5.5*0.9877) -- (4.5+5.5*0.1564,5.5*0.9877) -- cycle node[above]{$\Gamma$};
	\foreach \x in {-5.5,-4.5,...,6.5}
		{
		\foreach \y in {-6,-5,...,7}
			{
				\filldraw (\x+\y*0.1564,\y*0.9877) circle (0.2mm);
				\filldraw (\y+\x*0.1564,\x*0.9877) circle (0.2mm);
			}
		}
	\filldraw (1.5+0*0.1564,0*0.9877) circle (0.8mm) node[above]{$\mathcal{O}_{{\bf x}_1}$};
	\filldraw (-1-0.5*0.1564,-0.5*0.9877) circle (0.8mm) node[right]{$\mathcal{O}_{{\bf x}_2}$};
	\filldraw (-5-2.5*0.1564,-2.5*0.9877) circle (0.8mm) node[right]{$\mathcal{O}_{{\bf x}_3}$};
	\filldraw (-2.5+5*0.1564,5*0.9877) circle (0.8mm) node[below]{$\mathcal{O}_{{\bf x}_4}$};
\end{tikzpicture} \\
	(a) && (b)
	\end{tabular}
	\caption{(a) The discrete contour $\Gamma$ is a closed curve on the edges of the dual lattice. (b) A configuration with two contours $\Gamma$ and $\Gamma'$, and some local operators $\mathcal{O}_{{\bf x}_1}, \mathcal{O}_{{\bf x}_2}$, \dots None of these operators lies in the annular region between $\Gamma$ and $\Gamma'$.}
	\label{fig:contour_Gamma}
\end{figure}
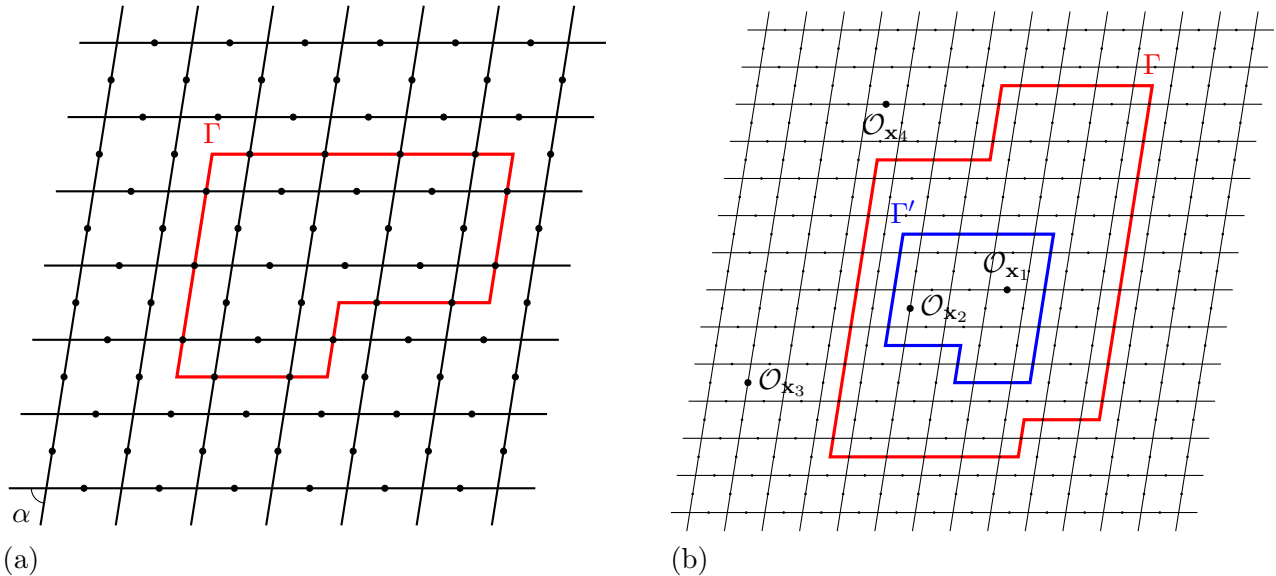

The observation \eqref{eq:little_contour} extends to larger contours $\Gamma$ on the dual lattice, as illustrated in Fig.~\ref{fig:contour_Gamma}.(a). Let us introduce the following notation for this discrete contour sum:
\begin{equation}
	\mathcal{S}^a_\Gamma \, \equiv \, \sum_{{\bf x} \in E(\Gamma)} S_{{\bf x}}^a\, ,
\end{equation}
where $E(\Gamma)$ is the set of dual edges visited by the contour and
$S_{{\bf x}}^a$ is understood as acting either on
$(\frac{k}{2})$ or as its dual representation on $(\frac{k}{2})^*$ at
${\bf x}$. As long as $\Gamma$ does not enclose any operator
(see Fig. \ref{fig:contour_Gamma}.(a)), the discrete contour integral
$\mathcal{S}^a_\Gamma$ vanishes, as a direct consequence of global
$\SU(2)$ symmetry. 
More generally, $\SU(2)$ invariance allows us to
deform the contour $\Gamma$ without changing the correlators in which
$\mathcal{S}^a_\Gamma$ is inserted. This implies that, for two
contours $\Gamma$ and $\Gamma'$ without operators between them (as
illustrated in Fig. \ref{fig:contour_Gamma}.(b)), one can deform
$\Gamma'$ to $\Gamma$ and obtain the relation
\begin{equation}
  \label{eq:Sgamma_comm}
  \left[ \mathcal{S}^a_\Gamma , \mathcal{S}^b_{\Gamma'} \right]
  \, = \, i f^{ab}_{\phantom{a}\phantom{a} c}\, \mathcal{S}^c_{\Gamma} \, ,
\end{equation}
which, again, holds when it is inserted inside a correlator.

It is known that some version of this actually survives when the Lie
algebra $\mathfrak{su}(2)$ is replaced by a quantum group. This was
discussed in full generality by Bernard and Felder \cite{Bernard1991};
more recently, the existence of such conserved (non-local) currents on
the lattice was used to enlighten the topic of ``lattice
holomorphicity'' \cite{Ikhlef2013}. We note that, in our case,
everything is much simpler than in the quantum group case, and that
the very existence of vanishing discrete contour integrals boils down
to the $\SU(2)$ symmetry of the model. These exist even when the model
is not integrable---one could choose any other $\SU(2)$-invariant
$R$-matrix that would not satisfy the Yang-Baxter relation---and
independently of criticality of the model in the continuum limit.

However, if one {\it knows} already that the model is critical---as is
the case here, thanks to integrability results---, then the existence
of the vanishing discrete contour integrals does have some
consequences. One of the consequences is that those discrete contour
integral must have a continuous counterpart in the CFT describing the
continuum limit. The relation between the lattice observables and the
fields in the continuum is constrained.  Indeed then
Eq.~\eqref{eq:Sgamma_comm} strongly suggests that, in the continuum
limit $a_0 \rightarrow 0$, we should identify the {\it discrete
  contour integral} $\mathcal{S}_\Gamma$ with the {\it contour
  integral}
\begin{equation}
  \label{eq:contour_int}
  \frac{1}{2\pi i} \oint_\Gamma {\rm d}z J^a(z) \, - \, \frac{1}{2\pi i}
  \oint_{\Gamma} {\rm d}\bar{z} \bar{J}^a(\bar{z}) \, = \, J_0^a +
  \overline{J}_0^a \, ,
\end{equation}
which generates the $\su(2)$ subalgebra in $\suhat(2)_k\otimes
\overline{\suhat(2)}_k$. The discretized line element ${\rm d}z$
around a vertex is $\{a_0,a_0e^{i\alpha},-a_0,-a_0e^{i\alpha}\}$, where the
contour is read counterclockwise starting from the edge below the
vertex. Then the identification of $\mathcal{S}^a_\Gamma$ with (\ref{eq:contour_int}) is possible {\it if} the contribution
of the currents to the spin operator on the lattice is
\begin{equation}
  \label{eq:expansion_S_phase}
	S^a_{\bf x} \, = \, \dots \, + \, \frac{a_0}{2\pi i}  \left[ e^{i \varphi_{\bf x}} J^a(z)  \, - \, e^{-i \varphi_{\bf x}} \bar{J}^a(\bar{z})  \right] \, + \, \dots\, ,
\end{equation}
where the factor $e^{i \varphi_{\bf x}}$ is a phase that depends on the orientation of the edge ${\bf x}$:
\begin{equation}
	e^{i \varphi_{\bf x}} \, = \, \left\{ \begin{array}{lcl} -1 &&
          ({\bf x} \text{ vertical edge}) \\ e^{i \alpha} && ({\bf
            x}\text{ horizontal edge}) .\end{array} \right.
\end{equation}
Indeed, if the currents $J^a$ and $\overline{J}^a$ appear in the expansion of the lattice spin
operator in the form (\ref{eq:expansion_S_phase}), then they contribute to $\mathcal{S}^a_\Gamma$ as a Riemann sum, giving precisely the contour
integral (\ref{eq:contour_int}) in the limit $a_0 \rightarrow 0$. The contribution of all the other terms appearing in the expansion (\ref{eq:S_J_desc}) vanishes in this limit. 
This is because, as we have seen previously, the non-zero terms are
either staggered (primary operators with half-integer spin $j$), leading to an alternating sum, or subleading (descendants that are not the currents themselves). Either way,
their contribution to the discrete contour integral goes to zero when $a_0 \rightarrow 0$.

In summary, the argument can be formulated as follows: the discrete contour integral should become the contour integral (\ref{eq:contour_int}) in the continuum limit, and this allows us to fix the coefficients $C^J_{\bf x}$ and $C^{\overline{J}}_{\bf x}$. We end up with the following expression of the lattice spin operator:
\begin{equation}
	\label{eq:S_J_final}
	 S^a_{{\bf x}} \quad \underset{a_0 \rightarrow 0}{=}  \quad \sum_{j \; {\rm half-int.}} C^{(j)}_{{\bf x}} \, a_0^{2 h_{j}} \,  P_1 \left[  \phi_{j}(z) \otimes \overline{\phi}_{j}(\bar{z}) \right]^a \; + \; \frac{a_0}{2\pi i}  \left[ e^{i \varphi_{\bf x}} J^a(z)  \, - \, e^{-i \varphi_{\bf x}} \bar{J}^a(\bar{z})  \right] \; + \; O(a_0^\beta) .
\end{equation}
We note that the amplitude of the coefficients $C^J_{\bf x}$ and
$C^{\overline{J}}_{\bf x}$ was known previously in the literature: it appeared already in \cite{Affleck1989}, and perhaps in earlier references; there, the argument to fix this amplitude is similar to the one we just used. However, the position-dependent complex phase of these coefficients is, to our knowledge, a new result. It is the key point of this paper, which allows us to cook up a lattice observable that behaves as the chiral current in the continuum limit.

\section{Chiral currents as local lattice operators}

\label{sec:chiral_current}

After the preparatory steps of the previous section, we are finally
ready to construct a lattice observable $J^a_{\bf x}$ which has the
behavior of Eq.~(\ref{eq:goal}). We first do it for the vertex model with an
arbitrary geometric angle $\alpha$, and then compute an expression for
the chiral current in the spin chain, by considering the anisotropic
limit $\alpha \rightarrow \pi$.

\subsection{Chiral current in the vertex model}

We have reached the expression (\ref{eq:S_J_final}) for the expansion
of the lattice spin operator in terms of the CFT operators. Clearly,
we can take linear combinations of this expression at different sites
${\bf x}$ in order to cook up new expressions where the leading
contribution in the limit $a_0 \rightarrow 0$ is nothing but the
chiral current $J^a(z)$ itself. One simple possibility is

\begin{equation}
\label{eq:Jlat}
J^{a}_{\bf x} (\alpha) \, = \,
\frac{\pi}{2 a_0 \sin \alpha} \left[
S^a_{{\bf x}+\frac{1}{2}{\bf u}} \, +
\, S^a_{{\bf x}-\frac{1}{2}{\bf u}} \, +
\, e^{-i \alpha}\, S^a_{{\bf x}+\frac{1}{2}{\bf v}} \, +
\,e^{-i \alpha} \, S^a_{{\bf x}-\frac{1}{2}{\bf v}} \right]\, .
\end{equation}
This expression is chosen such that all the primary fields in
(\ref{eq:S_J_final}) cancel because of the staggering---more
precisely, our expression picks up their first derivative, which has
scaling dimension $\beta = 2h_j + 1 > 1$, so it is less relevant than
the current---, and the same happens for the anti-chiral current
$\overline{J}^a(\bar{z})$, thanks to the complex phases in
(\ref{eq:S_J_final}). We have normalized our expression such that
\begin{equation}
J^{a}_{{\bf x}}(\alpha) \; \underset{a_0\rightarrow 0}{=} \;
J^a(z)\, + \, O(a_0^\beta).
\end{equation}
(Once again, this identification makes sense when inserted in
correlation functions.) More complicated linear combinations,
involving more lattice sites, could be used as well; the only
constraint is that the terms that could be more relevant (in the RG
sense) than, or as relevant as, $J^a(z)$, cancel. Similarly, the
anti-chiral current on the lattice can be identified as
\begin{equation}
\label{eq:Jlat_b}
\overline{J}^{a}_{\bf x} (\alpha) \, = \,
\frac{\pi}{2 a_0 \sin \alpha} \left[
S^a_{{\bf x}+\frac{1}{2}{\bf u}} \, +
\, S^a_{{\bf x}-\frac{1}{2}{\bf u}} \, +
\, e^{i \alpha}\, S^a_{{\bf x}+\frac{1}{2}{\bf v}} \, +
\,e^{i \alpha} \, S^a_{{\bf x}-\frac{1}{2}{\bf v}} \right]\, .
\end{equation}
Equations (\ref{eq:Jlat}) and (\ref{eq:Jlat_b}) are the main result of this paper. They answer the question whether
or not it is possible to realize the CFT chiral current in the lattice model (see the introduction) in the affirmative
way, and tell us explicitly how one can do it.


\subsection{Chiral current in the spin chain from the anisotropic limit}
\label{sec:anisotropic}

We just exhibited a lattice observable $J^{a}_{{\bf x}}(\alpha)$ which
becomes the holomorphic current $J^a(z)$ in the continuum limit. It is
then natural to ask whether one can construct a similar operator
$J^a_{x}$ acting directly on the Hilbert space $(\frac{k}{2})^{\otimes
  N}$ of the spin chain. To answer this question, we go back to the
transfer matrix formulation and to the description of the $R$-matrix
as an operator $R_{x,x'}$---we drop the superscript $(\frac{k}{2})$ in
this section---acting on spaces $(\frac{k}{2})_{x}\otimes
(\frac{k}{2})_{x'}$.

Before we take the anisotropic limit, it is convenient to do a small
manipulation, and replace the above expression of $J^a_{\bf
  x}(\alpha)$ by $i$ times this expression. The reason is the
following.  The conformal mapping from the plane (with complex
coordinate $z$) to the cylinder (complex coordinate $w = x+iy$, with
$x$ defined modulo $L$) is $z \, = \, \exp \left(-i \,2\pi \frac{
    w}{L}\right)$; the Jacobian of this transformation leads to
$J(x,y) \,=\, J(w) \, = \, \frac{2\pi}{L} \frac{z}{i} J(z) \, = \,
\frac{1}{i} \frac{2\pi}{L} \sum_n z^{-n} J_n$.  Because of the factor
$\frac{1}{i}$ coming from the Jacobian, one
sees that the operator $J(x)$ (in Schr\"odinger picture) is $\it
anti-Hermitian$. Since we find it more natural to work with a {\it
  Hermitian} operator $J(x)$, we simply make the replacement $J(x)
\rightarrow i J(x)$ when we work on the vertical cylinder. Taking this
additional factor $i$ into account, our expression for the lattice
chiral current (\ref{eq:Jlat}) at position $x$, becomes, in the transfer matrix formalism:
\begin{equation}
	\label{eq:Jstep1}
  \begin{tikzpicture}[scale = 0.8]
    \node at (-2.25,0)
    {$\displaystyle{J_x^a(\alpha) = \frac{i\,\pi}{2a_0 \sin\alpha} \times}$};
    \node at (10.3,-0.2)
    {$\displaystyle{\,,}$};
    \draw[line width=2pt] (0,0) -- (10, 0);
    \draw[line width=2pt] (1-0.1564,-0.9877) node[below]{$a_0$} -- (1+0.1564,0.9877);
    \draw[line width=2pt] (2-0.1564,-0.9877) node[below]{$2a_0$} -- (2+0.1564,0.9877);
    \draw[line width=2pt] (3-0.1564,-0.9877) -- (3+0.1564,0.9877);		
    \draw[line width=2pt] (4-0.1564,-0.9877) node[below]{$x$} -- (4+0.1564,0.9877);
    \draw[line width=2pt] (5-0.1564,-0.9877) -- (5+0.1564,0.9877);		
    \draw[line width=2pt] (6-0.1564,-0.9877) -- (6+0.1564,0.9877);		
    \draw[line width=2pt] (7-0.1564,-0.9877) -- (7+0.1564,0.9877);		
    \draw[line width=2pt] (8-0.1564,-0.9877)  -- (8+0.1564,0.9877);
    \draw[line width=2pt] (9-0.1564,-0.9877)  node[below]{$a_0 N$} -- (9+0.1564,0.9877);
    \draw[very thick,red]
    (3-0.1564+0.5,-0.9877)--(3+0.1564+0.5,0.9877)
    --(3+0.1564+0.5+1,0.9877)--(3-0.1564+1+0.5,-0.9877)
    --(3-0.1564+0.5,-0.9877);
    \filldraw (3.43,-0.1) rectangle ++(0.2,0.2);
    \filldraw (3.43+1,-0.1) rectangle ++(0.2,0.2);
    \filldraw (4-0.1564-0.1,-0.9877-0.1) rectangle ++(0.2,0.2);
    \filldraw (4+0.1564-0.1,0.9877-0.1) rectangle ++(0.2,0.2);
    \draw[thick] (0,-0.2) -- ++(0.3,0.4);
    \draw[thick] (0.15,-0.2) -- ++(0.3,0.4);
    \draw[thick] (9.6,-0.2) -- ++(0.3,0.4);
    \draw[thick] (9.75,-0.2) -- ++(0.3,0.4);
    \draw (-0.2+1,0) arc (-180:-99:0.2) ++(0,-0.1) node[left]{$\alpha$};
    \draw (-0.2+2,0) arc (-180:-99:0.2) ++(0,-0.1) node[left]{$\alpha$}; 
    \draw (-0.2+3,0) arc (-180:-99:0.2) ++(0,-0.1) node[left]{$\alpha$};
    \draw (-0.2+4,0) arc (-180:-99:0.2) ++(0,-0.1) node[left]{$\alpha$};
    \draw (-0.2+5,0) arc (-180:-99:0.2) ++(0,-0.1) node[left]{$\alpha$};
    \draw (-0.2+6,0) arc (-180:-99:0.2) ++(0,-0.1) node[left]{$\alpha$}; 
    \draw (-0.2+7,0) arc (-180:-99:0.2) ++(0,-0.1) node[left]{$\alpha$};
    \draw (-0.2+8,0) arc (-180:-99:0.2) ++(0,-0.1) node[left]{$\alpha$};
    \draw (-0.2+9,0) arc (-180:-99:0.2) ++(0,-0.1) node[left]{$\alpha$};
  \end{tikzpicture}
\end{equation}
where
\begin{align}
  \begin{tikzpicture}[scale = 0.8]
    \node at (10.25,0) {$\displaystyle{
        = (e^{-i\alpha} S^a_x + S^a_{x'}) R_{x,x'}(\alpha)
        + R_{x,x'}(\alpha) (S^a_x + e^{-i\alpha} S^a_{x'})\, .
      }$};
    \draw[line width=2pt] (4.43,0)--(3.43,0) node[left] {$x$};
    \draw[line width=2pt] (4-0.1564,-0.9877) node[below]{$x'$} -- (4+0.1564,0.9877);
    \draw[very thick,red]
    (3-0.1564+0.5,-0.9877)--(3+0.1564+0.5,0.9877)
    --(3+0.1564+0.5+1,0.9877)--(3-0.1564+1+0.5,-0.9877)
    --(3-0.1564+0.5,-0.9877);
    \filldraw (3.43,-0.1) rectangle ++(0.2,0.2);
    \filldraw (3.43+1,-0.1) rectangle ++(0.2,0.2);
    \filldraw (4-0.1564-0.1,-0.9877-0.1) rectangle ++(0.2,0.2);
    \filldraw (4+0.1564-0.1,0.9877-0.1) rectangle ++(0.2,0.2);
    \draw (-0.2+4,0) arc (-180:-99:0.2) ++(0,-0.1) node[left]{$\alpha$};
  \end{tikzpicture}
\end{align}
The operator $J_x^a(\alpha)$ is a mixture of the transfer matrix of the vertex model with the lattice current operator.
As such, it is usually {\it not} a local operator acting on the
spin chain. This is no different, of course, from the standard observation that the transfer matrix
itself is not a local operator acting on the spin chain. 
At this point, we simply hope that the expression
becomes a genuinely local operator acting on a finite number of sites in the spin chain when one takes the anisotropic limit $\alpha \rightarrow \pi$. However, this clearly cannot hold, since the transfer 
matrix itself does not become the identity in that limit; instead, it becomes the translation operator (\ref{eq:translation}), which is non-local. Therefore, if we took the limit $\alpha \rightarrow \pi$ directly in (\ref{eq:Jstep1}), we would get a result which is a mixture
of our lattice current operator with the translation operator, and this has to be non-local.
This is not quite what we are looking for. So, before we take the anisotropic limit, we multiply $J_x^a(\alpha)$ by the inverse of
the transfer matrix $T_N(\alpha)$, and this naturally compensates all the unwanted part of the operator (\ref{eq:Jstep1}). 
Thus, we define our local current in the spin chain as the anisotropic limit $\alpha \to \pi$ of the following combination,
\begin{equation}
J_x^a \, \equiv \, \lim_{\alpha \to \pi} J_x^a(\alpha) \cdot T^{-1}_N(\alpha).
\end{equation}
The explicit evaluation of the limit shows that this operator is indeed local, as expected:
\begin{equation} \label{eq:J_chain}
J_x^a \, =\, \frac{\pi}{2a_0} \left(
S^a_x+S^a_{x+a_0} \,+\,i \,[h_{x,x+a_0}\, ,\, S^a_{x+a_0}-S^a_{x}\, ]
\,
\right) \, ,
\end{equation}
where $h_{x,x'} = (dR_{x,x'}/d\alpha)(\pi)$, as in Sec.~\ref{sec:6v}. The fact that the Hamiltonian density appears here is not a surprise: indeed, it is exactly the dynamic data that is needed in order to separate the left- from the right-moving excitations, and therefore, allows one to isolate the {\it chiral} part of the current. Note also, that, because of the conventions used in this paper, the velocity of the excitations in the effective low-energy theory is always $v=1$, but if this was not the case, then the velocity would appear in (\ref{eq:J_chain}), namely $h_{x,x+a_0}$ would be replaced by $\frac{1}{v} h_{x,x+a_0}$.

The explicit form of the chiral current in the spin chain is the second main result of this paper. It is the spin chain counterpart of the more general expression (\ref{eq:Jlat}). The latter is more general in the sense that it is valid for an arbitrary angle $\alpha$ in the vertex model, while the expression (\ref{eq:J_chain}) involves only the limit $\alpha \rightarrow \pi$. It is possible to get to the expression for the spin chain in a more direct way, without relying on the result for the vertex model; we will explain this alternative derivation shortly. But, before we do so, let us give the explicit form of the current operator.

Plugging the explicit form of the Hamiltonian density $h_{x,x+a_0}$ leads to the following expressions. In the spin-1/2 case, $h_{x,x+a_0} =
\frac{2}{\pi}(\mathbf{S}_x \cdot \mathbf{S}_{x+a_0} + \frac{1}{4}
\id)$, so one finds
\begin{equation}
(k=1)
\qquad \qquad
J^a_x \, = \,\frac{\pi}{2a_0} \left(
S^a_x+S^a_{x+a_0} \,+ \,\frac{4}{\pi} (\mathbf{S}_{x} \times \mathbf{S}_{x+a_0})^a
\right) \,,
\end{equation}
where $(\mathbf{S}_{x} \times \mathbf{S}_{x+a_0})^a \, \equiv \, f^a_{\phantom{a}bc} S^b_x S^c_{x+a_0}$.
For spin-$1$, one has $h_{x,x+a_0} = \frac{1}{2\pi} ( \mathbf{S}_x \cdot \mathbf{S}_{x+a_0}  - (\mathbf{S}_x \cdot \mathbf{S}_{x+a_0} )^2 + 3\, \mathbf{1} )$, which gives
\begin{eqnarray}
(k=2)
\qquad \qquad
J^a_x & = &\frac{\pi}{2a_0} \left(
S^a_x+S^a_{x+a_0} \,+ \,\frac{1}{\pi}  (\mathbf{S}_x \times \mathbf{S}_{x+a_0})^a \,   \right. \\
\nonumber && \left. \quad -\,	\frac{1}{\pi} \left[   (\mathbf{S}_x \cdot \mathbf{S}_{x+a_0}) (\mathbf{S}_{x} \times \mathbf{S}_{x+a_0})^a  +  (\mathbf{S}_{x} \times \mathbf{S}_{x+a_0})^a (\mathbf{S}_x \cdot \mathbf{S}_{x+a_0}) \right] \right) \,.
\end{eqnarray}
For spin-$3/2$, $h_{x,x+a_0} = \frac{1}{\pi} ( -\frac{1}{8} \mathbf{S}_x \cdot \mathbf{S}_{x+a_0}  + \frac{1}{27} (\mathbf{S}_x \cdot \mathbf{S}_{x+a_0} )^2 + \frac{2}{27} (\mathbf{S}_x \cdot \mathbf{S}_{x+a_0} )^3  + \frac{13}{12} \mathbf{1} )$. One gets
\begin{eqnarray}
(k=3)
\qquad \qquad
J^a_x & = &\frac{\pi}{2a_0} \left(
S^a_x+S^a_{x+a_0} \,- \,\frac{1}{4\pi}  (\mathbf{S}_x \times \mathbf{S}_{x+a_0})^a \,   \right. \\
\nonumber &&  +\,	\frac{2}{27\pi} \left[   (\mathbf{S}_x \cdot \mathbf{S}_{x+a_0}) (\mathbf{S}_{x} \times \mathbf{S}_{x+a_0})^a  +  (\mathbf{S}_{x} \times \mathbf{S}_{x+a_0})^a (\mathbf{S}_x \cdot \mathbf{S}_{x+a_0}) \right]  \\
\nonumber &&  +\, \frac{4}{27\pi} \left[   (\mathbf{S}_x \cdot \mathbf{S}_{x+a_0})^2 (\mathbf{S}_{x} \times \mathbf{S}_{x+a_0})^a  +   (\mathbf{S}_x \cdot \mathbf{S}_{x+a_0}) (\mathbf{S}_{x} \times \mathbf{S}_{x+a_0})^a (\mathbf{S}_x \cdot \mathbf{S}_{x+a_0})  \right. \\
\nonumber	&& \qquad \qquad \qquad \qquad \left. \phantom{\frac{1}{1}}  \left.+   (\mathbf{S}_{x} \times \mathbf{S}_{x+a_0})^a (\mathbf{S}_x \cdot \mathbf{S}_{x+a_0})^2\right] \right) \,.
\end{eqnarray}

\subsection{Alternative derivation}

There is alternative way of getting to the formula (\ref{eq:J_chain}). We adapt 
an idea suggested to us by Hubert Saleur (private communication), which
builds upon earlier work of Koo and Saleur about the lattice realization of the
stress-tensor \cite{SaleurKoo}. It is much more straightforward than going first through
the derivation of the main result (\ref{eq:Jlat}), and then extracting the spin chain
expression through the limiting procedure exposed in section \ref{sec:anisotropic}.
On the other hand, as we will see, the different steps
in this alternative derivation are perhaps not under as good a control as in the route we took to
arrive at the main result (\ref{eq:Jlat}).
But we view this alternative route as complementary,
and as a further evidence that our scheme for taking the anisotropic limit is meaningful. Besides, we
find this way of looking at formula (\ref{eq:J_chain}) physically illuminating.

One proceeds as follows. First, we know that the critical Hamiltonian for
the spin chain is
a finite-size version of the CFT Hamiltonian. Namely,
\begin{eqnarray}
\label{eq:ident_H}
\frac{2\pi a_0}{L} \left( L_0 + \overline{L}_0 - \frac{c}{12} \right)
& \simeq & H^{(k/2)} \,-\, E_{\infty} L \, = \, \sum_x h_{x,x+a_0} \,
- \, E_{\infty} L ,
\end{eqnarray}
where $h_{x,x+a_0}$ is the Hamiltonian density. Similarly, it is
natural to make an identification of the form
\begin{eqnarray}
\label{eq:Hubert1_wrong}
J_n^a \, + \, \overline{J}^a_{-n} & \simeq & \sum_{x} e^{-i \frac{2\pi
    n}{L }x } \,S^a_x\, .
\end{eqnarray}
Notice that there is some freedom in this identification, however. For instance, the discrepancy between the following
formula 
\begin{eqnarray}
\label{eq:Hubert1}
J_n^a \, + \, \overline{J}^a_{-n} & \simeq & \sum_{x} e^{-i \frac{2\pi
    n}{L }(x+a_0/2) } \,\frac{S^a_x + S^a_{x+a_0}}{2}\, ,
\end{eqnarray}
and (\ref{eq:Hubert1_wrong}) vanishes as $a_0/L$ when $L/a_0\rightarrow \infty$.
Different identifications may lead to different results in the end, so one needs to be careful here. We find that it is best to use (\ref{eq:Hubert1}). This may be justified in a way that is similar to the discussion of section \ref{sec:lattice_spin_op}: the spin operator has an expansion in terms of the CFT fields, in particular in terms of the primary fields $\phi_{j} \otimes \overline{\phi}_{j}$ and the currents.
In order to get rid of the unwanted primary operators, one can make use of the staggering, and average the spin operator over two neighboring sites. So we believe that (\ref{eq:Hubert1}) is the safer starting point here.

From the two expressions (\ref{eq:ident_H}) and (\ref{eq:Hubert1}), one constructs a third one, that is a
lattice version of the combination of modes $J_n - \overline{J}_{-n}$:
\begin{eqnarray}
\label{eq:Hubert2}
\nonumber -n\left(J_n^a \, - \, \overline{J}^a_{-n} \right) & = &
\left[ L_0 + \overline{L}_0 , J^a_n + \overline{J}^a_{-n} \right] \\
\nonumber & \simeq & \frac{L}{2\pi a_0} \sum_{x,x'} e^{-i \frac{2\pi n
  }{L} (x' + a_0/2)}  \left[ h_{x,x+a_0}, \,(S_{x'}^a + S_{x'+a_0}^a)/2\, \right] \\
& \simeq & -i \, \frac{n}{2} \sum_{x} e^{-i \frac{2\pi n}{L} (x+a_0/2)} \left[
  h_{x,x+a_0}, S_{x+a_0}^a - S_x^a\right] .
\end{eqnarray}
In the last line we have used $e^{-i \frac{2\pi n }{L}a_0} \simeq 1
-i \frac{2\pi n a_0}{L}$, which is valid if one fixes $n$ and then
take $L \gg a_0$, and then $ \left[ h_{x,x+a_0}, S_{x}^a + S_{x+a_0}^a
\right] =0$, which is simply stating that $h_{x,x+a_0}$ is
$\SU(2)$-invariant. Combining (\ref{eq:Hubert1}) and
(\ref{eq:Hubert2}), one gets the following approximation for $J^a_n$:
\begin{equation}
J^a_n \, \simeq \, \frac{a_0}{2\pi} \sum_x e^{-i \frac{2\pi n}{L} (x+a_0/2)}\;
\frac{\pi}{2a_0}\left( S_x^a + S_{x+a_0}^a \, + \, i \, \left[
    h_{x,x+a_0}, S_{x+a_0}^a-S_x^a \right] \right)\, ,
\end{equation}
which is nothing but the Fourier mode of $J^a_x$ obtained from the anisotropic limit. So
one recovers (\ref{eq:J_chain}) as claimed. Notice that it is important to use (\ref{eq:Hubert1}) instead of (\ref{eq:ident_H}) when we recombine the expressions for $J_n + J_{-n}$ and $J_n - J_{-n}$; otherwise we would actually get a different result. Again, this is a crucial point, in light of the discussion in section \ref{sec:lattice_spin_op}, since what we need is to suppress the contribution of the primary operators $\phi_{j} \otimes \overline{\phi}_{j}$ with $j$ half-integer. But, apart from this point, which was already encountered in the construction of the chiral current in the vertex model, what is interesting here is the way the chiral component of the current is isolated. In the vertex model, this was done by fine-tuning the phases of the spin operator on neighboring sites, depending on the geometric angle $\alpha$; in that sense, the chiral and anti-chiral components were separated thanks to the geometry of the two-dimensional model. Here, the left- and right-movers are separated by their dynamic properties, which one probes with the commutator of the Hamiltonian with the spin operator. We find that, even though one should certainly expect it, it is quite a non-trivial check that the two viewpoints match so perfectly in the anisotropic limit. 

Let us finally emphasize that, in
the construction of the lattice operator $J^a_x$ for the spin chain, we did not use
much, apart from the fact that we have a critical Hamiltonian, which we normalized such that the velocity of the
excitations is one, and the identification of the modes of the Lie
algebra generators $S^a_{x}$ with the Kac-Moody modes in the
continuum. The latter is the crucial step, and one needs to be careful in this
identification, as the most obvious guess for the lattice analogue of the Kac-Moody
modes may not be the correct expression, and might lead to wrong results.
This step is then a bit subtle, and involves a discussion of the relation between
the lattice observables and the field theory operators similar to the one in section \ref{sec:lattice_spin_op}.
But, apart from these subtleties, we only used the fact that the commutator of $L_0$ with $J^a_n$ is proportional to $n$,
which is a completely general feature of the $L_0$-operator, 
and in particular, we didn't make use of the existence of a Sugawara construction here.
This means that such an approach might be generalizable to many different
critical spin chains including those with a Lie (super-)group symmetry and
non-integrable ones.

\section{Numerical checks}

\begin{figure}[th]
  \begin{center}
    \begin{tabular}{lcl}
      \includegraphics[width=0.47\textwidth]{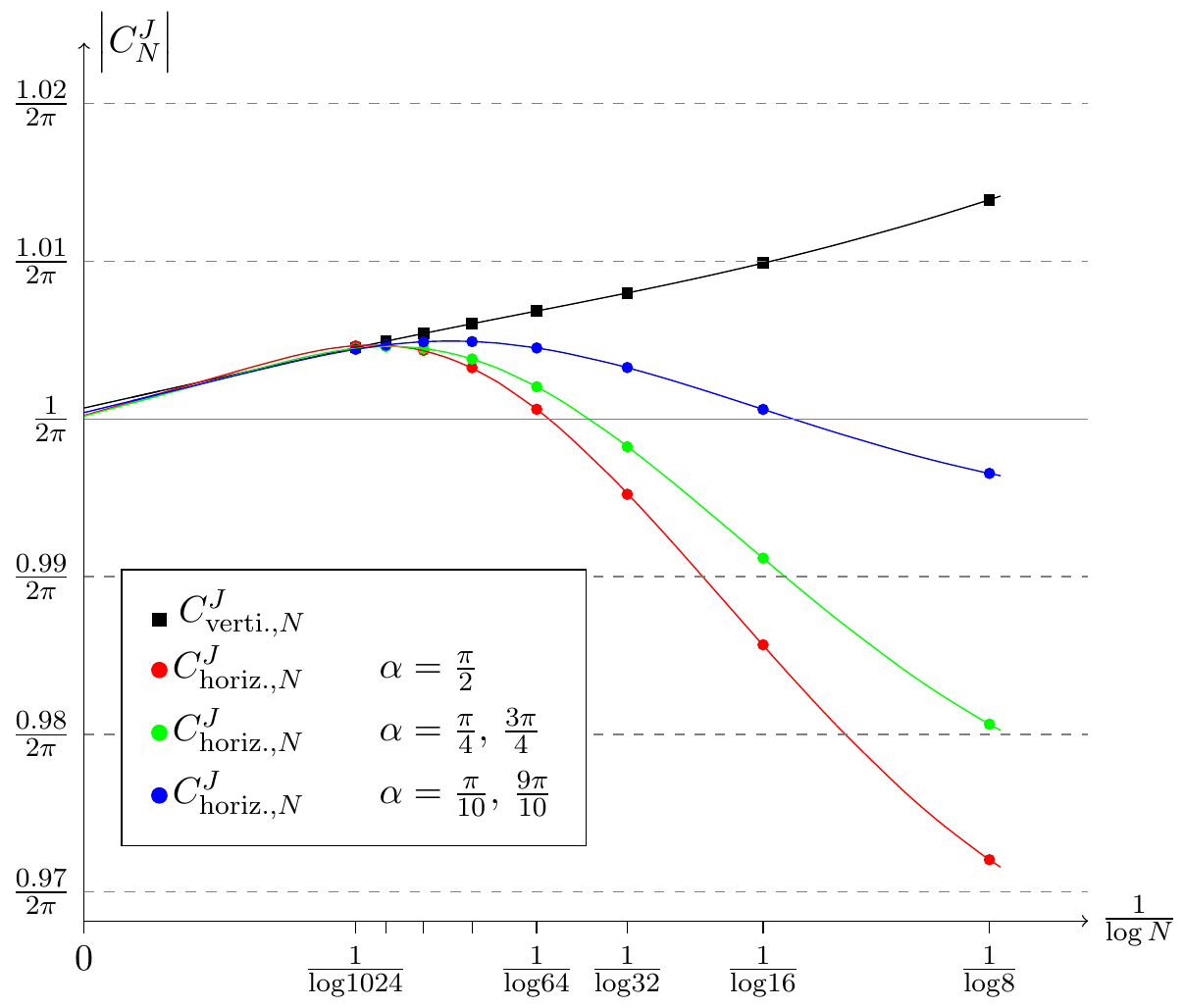}  && \includegraphics[width=0.46\textwidth]{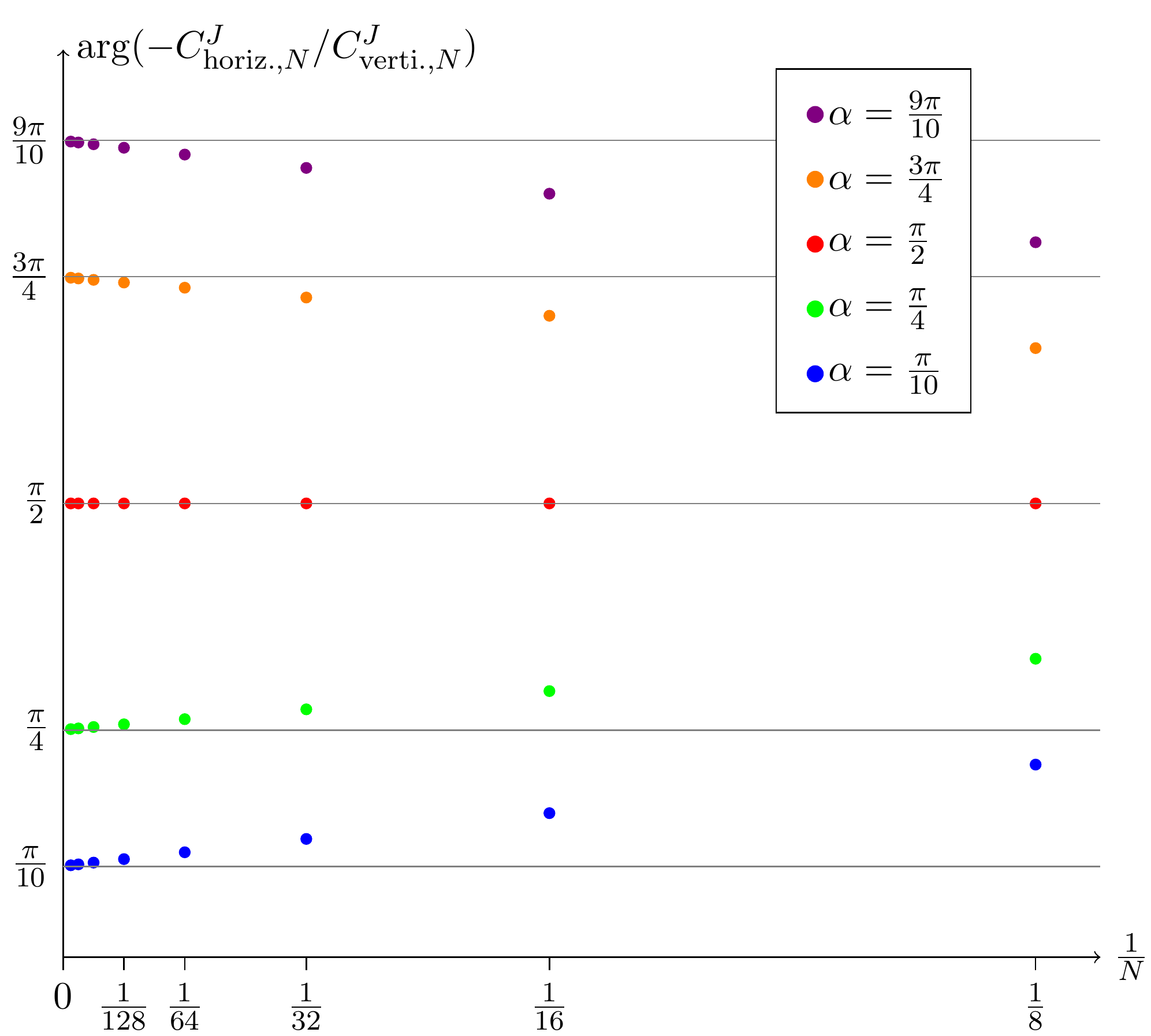} \\
      (a) && (b)
    \end{tabular}
    \caption{Measure of the coefficient $C^J$ for $k=1$ ({\it i.e.}~in the six-vertex model). It supports our formula (\ref{eq:conj_CJ}). We compute $C^J_{{\rm verti.},N}$ and $C^J_{{\rm horiz.},N}$ for $N=8,16,32,64,256,512,1024$. Notice that, by construction, $C^J_{{\rm horiz.},N}$ depends on the geometric angle $\alpha$, while $C^J_{{\rm verti.},N} $ does not. \newline
    \small  (a) Numerical values of $\left| C^J_{{\rm verti.},N} \right|$, $\left| C^J_{{\rm horiz.},N} \right|$, fitted with a function $f(N) \, = \, \alpha_1 \, + \, \frac{\alpha_2}{\log N} \, + \, \frac{\alpha_3}{N} \, + \, \frac{\alpha_3}{N \, \log N}$. The extrapolation is in good agreement with $\left| C^J_{{\rm horiz.}} \right| = \left| C^J_{{\rm verti.}} \right| = \frac{1}{2\pi}$ in the thermodynamic limit ($N \rightarrow \infty$).  \newline
      (b) Measure of the relative phase on horizontal/vertical edges, in agreement with $C^J_{\rm horiz.} / C^J_{{\rm verti.}}\, = \, -e^{i\alpha} $ in the thermodynamic limit.
    }
    \label{fig:results_CJ_k1}
  \end{center}
\end{figure}

\begin{figure}[th]
  \begin{center}
    \begin{tabular}{lcl}
      \includegraphics[width=0.47\textwidth]{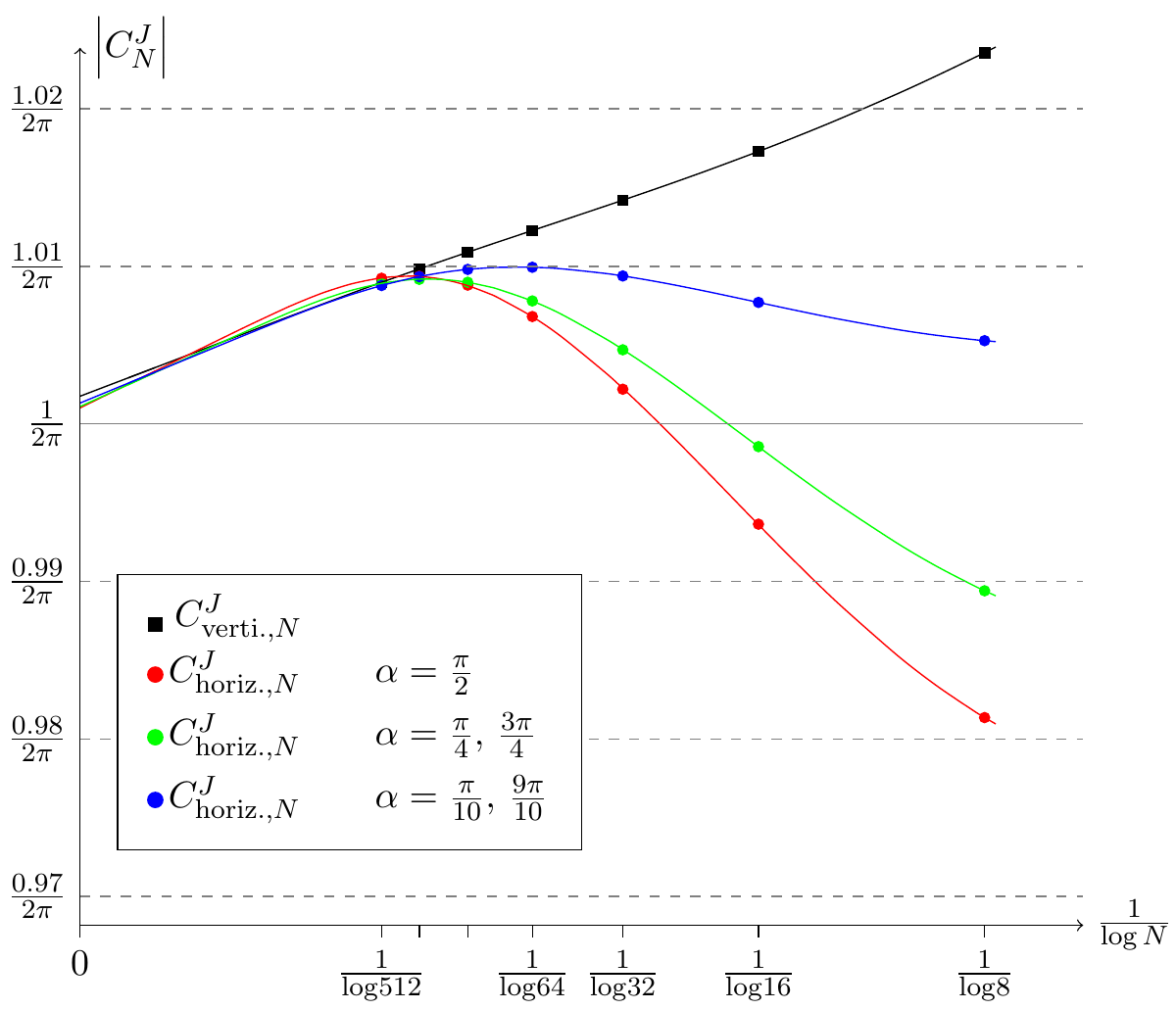}  && \includegraphics[width=0.46\textwidth]{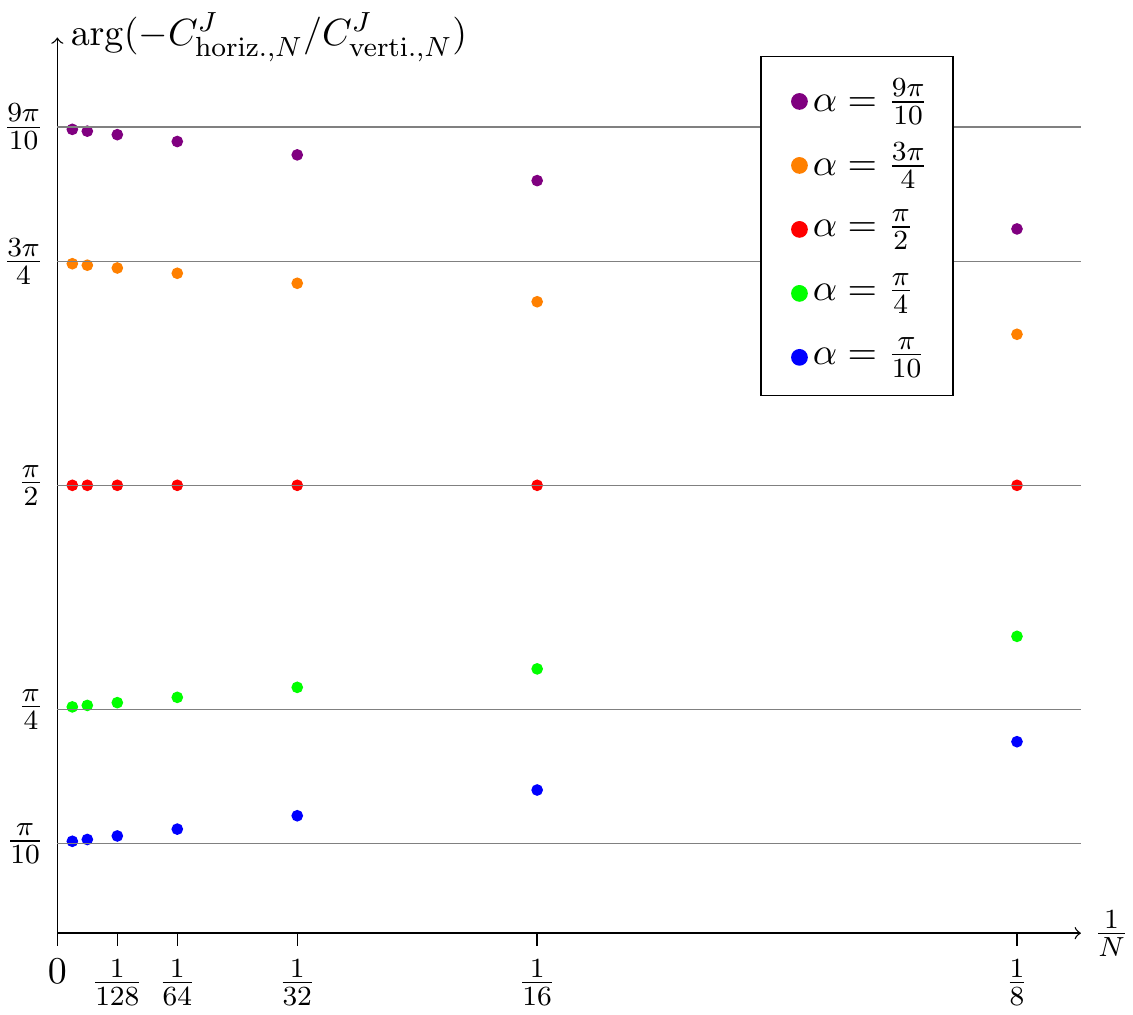} \\
      (a) && (b)
    \end{tabular}
    \caption{Same as Fig. \ref{fig:results_CJ_k1}, this time for $k=2$. Again, the results support our formula (\ref{eq:conj_CJ}). We compute $C^J_{{\rm verti.},N}$ and $C^J_{{\rm horiz.},N}$ for $N=8,16,32,64,256,512$.}
    \label{fig:results_CJ_k2}
  \end{center}
\end{figure}

The coefficients appearing in the expansion of the lattice spin operator~\eqref{eq:S_J} can in principle be related to form factors. This is what we discuss in this section. The relation with form factors gives us a way of checking numerically that the coefficients $C^J_{{\bf x}}$ are indeed given by
\begin{equation}
  \label{eq:conj_CJ}
  C^{J}_{\bf x} \, = \, \left\{Â  \begin{array}{cll}
      -\frac{a_0}{2\pi i}  && ({\bf x} \; {\rm vertical \; edge})  \\ \\
      \frac{a_0}{2\pi i} \, e^{i\alpha} &\quad&   ({\bf x} \; {\rm horizontal \; edge}) ,
    \end{array}   \right.
\end{equation}
as we argued in Eq. (\ref{eq:S_J_final}). The numerical results are plotted in Fig.~\ref{fig:results_CJ_k1} for $k=1$ and in Fig.~\ref{fig:results_CJ_k2} for $k=2$. We find that they are in perfect agreement with \eqref{eq:conj_CJ}.

The relation between form factors and the coefficients $C^J_{{\bf x}}$ goes as follows. Consider once again the infinitely long cylinder of circumference $L=a_0 N$ generated by the transfer matrix $T^{(k/2)}_L(\alpha)$. Let $\ket{0}$ be the eigenstate of $T_L^{(k/2)}(\alpha)$ with the largest eigenvalue (in absolute value). This state is also the {\it ground state} of the Hamiltonian $H^{(k/2)}$. Let $\ket{s}$ be some other eigenstate of $T^{(k/2)}_L$ or, equivalently, of $H^{(k/2)}$. One can always chose $\ket{s}$ to be a state with the following fixed quantum numbers:
\begin{itemize}
\item momentum $P$,
\item total $\SU(2)$ spin $(\sum_x \mathbf{S}_x)^2 = S(S+1)$,
\item magnetization $\sum_x S^z_x = S^z$.
\end{itemize}
In addition to these three quantum numbers, the state $\ket{s}$ has some energy $E_s$, which is the eigenvalue of $H^{(k/2)}$. These four quantities
uniquely identify an eigenstate
of the transfer matrix. The ground state itself is identified as follows: it has energy $E_0$, total spin and magnetization zero, and momentum $P_0 =0$ or $P_0 = \pi/a_0$, depending on the parity of $k N/2$ (recall that $L = a_0 N$ and that we assume that $N$ is even throughout the paper).

\noindent Next, we focus on the unique eigenstate characterized by:
\begin{itemize}
    \item $P = P_0 + \frac{2\pi}{L}$
    \item total spin $S  =  1$
    \item magnetization $S^z = 0$
    \item it is the lowest-energy state with the above three quantum numbers.
\end{itemize}
We call this state $\ket{J^3}$. This notation comes from the fact that
one wants to identify the spectrum of $H^{(k/2)}$ with the one of
$E_\infty L+H_{\rm CFT} =E_\infty L+  \frac{2\pi}{L}(L_0+\bar{L}_0 -
\frac{c}{12} )$ in the continuum limit, and the lowest energy state
with these quantum numbers in the CFT is $J^3_{-1}\ket{0}$. In particular, this means that the energy of the state $\ket{J^3}$ must behave as
\begin{equation}
    E_{J^3}(L) -  E_{0}(L) \, \underset{L \rightarrow \infty}{=} \, \frac{2\pi}{L} \, + \, o(L^{-1}),
\end{equation}
if $E_0(L)$ is the energy of the ground state. Note also that, with our convention for the OPEs of the currents, the norm of the CFT state $J^3_{-1}\ket{0}$ is $\left< J^3_{1} J^3_{-1} \right> \, = \, k/2$
so we fix the normalization of our lattice state $\left| J^3 \right>$ to be
\begin{equation}
    \label{eq:norm_Jz}
    \left< J^3 \left| J^3 \right>\right. \,  = \, \frac{k}{2}
\end{equation}
as well. Of course, alternatively, one could just identify the eigenstate $\ket{J^3}$ by specifying its Bethe roots configuration; we give more details about this in appendix \ref{sec:appendixA}.

At last, we arrive at the connection with the coefficients $C^J_{{\bf x}}$. Let us focus first on a point ${\bf x} = (x,y)$ on a vertical edge, so $(x,y) \in a_0\mathbb{Z}_N \times  a_0 (\mathbb{Z} + \frac{1}{2})$. We are interested in the form factor $\left< J^3 \right| S^a_{x}  \left| 0\right>$, which is now almost well-defined for any finite size $L$. Namely, it is defined up to a phase, since the phases of the ground state $\left| 0\right>$ and of $\left| J^3\right>$ are arbitrary. But we will come back to this question later. For now, we simply observe that, since $S^a_{{\bf x}}$ admits an expansion of the form
\begin{equation}
    S^a_{{\bf x}} \, = \, \dots  \, + \, a_0 \left[ C^J_{\bf x} \, J^a(z) +  C^{\bar{J}}_{\bf x} \, \bar{J}^a(\bar{z}) \right]\, + \, \dots,
\end{equation}
and since the mode expansion of $J^a(z)$ on the cylinder is
\begin{equation}
    J^a(z) \, = \, \frac{2\pi}{L} \sum_n e^{i n \frac{2\pi z}{L}} J_n^a \,,
\end{equation}
we expect the following behavior of the form factor (up to an undetermined phase):
\begin{eqnarray}
    \left< J^3 \right| S^a_{{\bf x}}  \left| 0\right>  & \underset{L \rightarrow \infty}{=} & \frac{2\pi \, a_0\, C^J_{\bf x}}{L} \sum_n e^{i n \frac{2\pi z}{L}}\, \left< 0 \right|  J^3_1 \, J^a_{n} \left| 0\right>  \: + \, o(L^{-1}) \\
\nonumber    &=& \frac{2\pi \, a_0\, C^J_{\bf x}}{L}  e^{-i \frac{2\pi z}{L}}\, \left< 0 \right|  J^3_1 \, J^a_{-1} \left| 0\right>  \: + \, o(L^{-1}).
\end{eqnarray}
Thus, the following quantity
\begin{equation}
    \label{eq:CJ_verti}
    C^J_{{\rm verti.},N} \, \equiv \,\frac{2}{k} \, \frac{L}{2\pi \, a_0}\, e^{i\frac{2\pi x}{L}}  \, \bra{J^3}  S^3_x  \ket{0}  \qquad \quad ({\rm vertical \; edge})
\end{equation}
is independent of $x$, and should converge (in amplitude) to the coefficient $C^J_{{\bf x}}$,
\begin{equation}
    \left|  C^J_{{\rm verti.},N} \right| \, \underset{N \rightarrow \infty}{\longrightarrow} \, \left|  C^J_{{\bf x}} \right| .
\end{equation}
The overall factor $2/k$ in (\ref{eq:CJ_verti}) comes from the normalization (\ref{eq:norm_Jz}). This quantity can be computed in finite size, and then extrapolated to the thermodynamic limit, which gives an estimate of the amplitude of the coefficient $C^J_{{\bf x}}$ on vertical edges.

Second, consider the case of a point ${\bf x} = (x,y)$ on a horizontal edge, so $(x,y) \in a_0(\mathbb{Z}_N+ \frac{1}{2}) \times  a_0 \mathbb{Z}$.
The coefficient $C^J_{x,y}$ may still be related to a finite-size quantity involving the lattice ground state $\ket{0}$ and our state $\ket{J^3}$, but this quantity is not, strictly speaking, a form factor. Instead, it is the matrix element of the following operator (we do not add a superscript $(k/2)$ here, but this operator of course depends on $k$):
\begin{equation}
  \begin{tikzpicture}[scale = 0.8]
    \draw (-1.4,0) node{$T^{(S^a)}_{L,x}(\alpha) \, = \;$};
    \draw[line width=2pt] (0,0) -- (10, 0);
    \draw[line width=2pt] (1-0.1564,-0.9877) node[below]{$a_0$} -- (1+0.1564,0.9877);        
    \draw[line width=2pt] (2-0.1564,-0.9877) node[below]{$2a_0$} -- (2+0.1564,0.9877);        
    \draw[line width=2pt] (3-0.1564,-0.9877) -- (3+0.1564,0.9877);        
    \draw[line width=2pt] (4-0.1564,-0.9877) -- (4+0.1564,0.9877);        
    \draw[line width=2pt] (5-0.1564,-0.9877) -- (5+0.1564,0.9877);        
    \draw[line width=2pt] (6-0.1564,-0.9877) -- (6+0.1564,0.9877);        
    \draw[line width=2pt] (7-0.1564,-0.9877) -- (7+0.1564,0.9877);        
    \draw[line width=2pt] (8-0.1564,-0.9877)  -- (8+0.1564,0.9877);
    \draw[line width=2pt] (9-0.1564,-0.9877)  node[below]{$a_0 N$} -- (9+0.1564,0.9877);
    \filldraw (3.43,-0.1) rectangle ++(0.2,0.2) node[above]{$S^a$} ++(-0.1,-0.3) node[below]{$x$};
    \draw[thick] (0,-0.2) -- ++(0.3,0.4);
    \draw[thick] (0.15,-0.2) -- ++(0.3,0.4);
    \draw[thick] (9.6,-0.2) -- ++(0.3,0.4);
    \draw[thick] (9.75,-0.2) -- ++(0.3,0.4);
    \draw (-0.2+1,0) arc (-180:-99:0.2) ++(0,-0.1) node[left]{$\alpha$};
    \draw (-0.2+2,0) arc (-180:-99:0.2) ++(0,-0.1) node[left]{$\alpha$};
    \draw (-0.2+9,0) arc (-180:-99:0.2) ++(0,-0.1) node[left]{$\alpha$};
    \draw (10.5,0) node{$,$};
  \end{tikzpicture}
\end{equation}
which we use to form the ratio
\begin{equation}
  \label{eq:CJ_horiz}
  C^J_{{\rm horiz.},N} \, \equiv \,  \frac{2}{k} \, \frac{L}{2\pi\, a_0}\, e^{i\frac{2\pi x}{L}}  \,
  \frac{ \bra{J^3}  T^{(S^3)}_{L,x}(\alpha)  \ket{0} }{ \bra{0}  T^{(k/2)}_L(\alpha)  \ket{0} }
  \qquad \quad ({\rm horizontal \; edge}).
\end{equation}
Again, the global factor $2/k$ comes from the normalization (\ref{eq:norm_Jz}), and again, this quantity is defined only up to a phase, coming from the undetermined phases of the ground state $\ket{0}$ and of $\ket{J^3}$. The quantity $C^J_{{\rm horiz.},N} $ does not depend on $x$, and its amplitude $\left| C^J_{{\rm horiz.},N} \right|$ allows us to estimate the amplitude of the coefficient $C^{J}_{{\bf x}}$ on horizontal edges.

Finally, note that, although $C^J_{{\rm verti.},N} $ and $C^J_{{\rm horiz.},N} $ are defined up to a phase, their {\it relative phase is well-defined}, and is a quantity which can be measured,
\begin{equation}
    \label{eq:CJ_phase}
    \frac{C^J_{{\rm horiz.},N}}{ C^J_{{\rm verti.},N} }  \, = \, \frac{ e^{i\frac{\pi a_0}{L}}  \bra{J^3}  T^{(S^3)}_{L,x+\frac{a_0}{2}}(\alpha)  \ket{0} }{ \bra{0}  T_L^{(k/2)}(\alpha)  \ket{0} \, \bra{J^3}  S^3_x  \ket{0} } ,\qquad \quad x \in a_0 \mathbb{Z}_N .
\end{equation}
We thus have access to the {\it relative phase} between horizontal and vertical edges, if we compute this phase for finite $N$ and then extrapolate the results to $N \rightarrow \infty$.

\vspace{0.3cm}

Numerically, we compute the matrix elements $ \bra{J^3} S^3_x
\ket{0}$, $ \bra{J^3} T^{(S^3)}_{L,x}(\alpha) \ket{0} $ using form
factor techniques and the integrable structure of the six-vertex
model. We make use of Slavnov's determinant formula \cite{Slavnov1989}
applied to the matrix elements of spin-1/2 \cite{KMT} and spin-1
\cite{Olalla_Maillet} chains. More details about these points are
given in appendix \ref{sec:appendixA}, where for convenience the
discussion is carried out for $J^+,S^+$ instead of $J^3,S^3$. (Due to
$\SU(2)$ invariance this change affects our formulas only in the
normalization.)  These techniques allow us to go to large system
sizes, as opposed to naive exact diagonalization. Large system sizes
are really needed here: the finite-size corrections decay very slowly
because of subleading logarithmic corrections, see \cite{Affleck1989}
and appendix \ref{sec:appendix_log}.  In Fig. \ref{fig:results_CJ_k1}
and \ref{fig:results_CJ_k2}, we plot our results for the finite-size
observables (\ref{eq:CJ_verti}), (\ref{eq:CJ_horiz}), which converge
towards the coefficient $C^J$ on horizontal and vertical edges.

\section{Conclusion}

We considered a class of lattice models that are known to be
discretizations of the $\SU(2)_k$ WZW model---the descendants of the
six-vertex model---and identified local observables on the lattice
that behave as the components of the chiral current $J^a(z)$ in the
continuum limit. These observables are constructed using a combination
of lattice spin operators on neighboring edges. We started by a
careful analysis of the expansion of the lattice spin operator
$S_{{\bf x}}^a$ in terms of the fields in the continuum limit, and
were able to put some constraints on the operators that can appear. We
found that primary operators $\phi_j$ with half-integer spin $j$ all
come with staggered coefficients---a fact that has long been known
when $j = \frac{1}{2}$, but which holds in full generality---while the
primary operators with integer spin $j$ are absent from the
expansion. Most importantly, we argued that the chiral and anti-chiral
components of the current appear in the expansion, with coefficients
that can be determined by combining $\SU(2)$-symmetry and some
intuition about how quantities that are automatically conserved on the
lattice become the zero modes of the currents in the continuum
limit. We provided numerical checks that support the identification of
these coefficients. The analysis of the expansion of $S^a_{{\bf x}}$
was finally used to produce a new observable involving the spin
operator on a few neighboring edges, and such that this observable
itself admits an expansion in which the most relevant operator is the
chiral current. This new observable therefore is one local lattice
operator that achieves the goal we were aiming for. In particular,
multi-point correlators of this observable do become the correlators
of the chiral currents~\eqref{eq:block_generic}, as we
wanted. Clearly, many other observables with this property could be
constructed, using more lattice sites. The one we exhibited here is
minimal in the sense it involves only four neighboring lattice
points. This observable was constructed for an arbitrary geometric
angle $\alpha$. Looking at the anisotropic limit, we were able to
obtain an expression for the chiral current in the spin-$\frac{1}{2}$
Heisenberg spin chain, and more generally, in the multi-critical
spin-$\frac{k}{2}$ spin chain, in terms of a commutator of the local
spin operator with the Hamiltonian density.

The present work can be extended to other lattice models possessing a
continuous symmetry, in a way that seems relatively
straightforward. One could, for instance, consider vertex models
discretizing $\SU(N)_k$ WZW models, and construct lattice versions of
the chiral currents of such theories.  Another interesting direction
would be to study the chiral current in supersymmetric spin
systems such as those investigated in \cite{Read2001}. 
Finally, as we already mentioned in the introduction, it would
be interesting to have lattice versions of other (non-local) chiral
observables, such as the chiral part of the primary fields,
$\phi_j(z)$. These must be non-local, and should typically be
associated with defects or local dislocations of the lattice. We hope
to come back to this question soon.

\paragraph{Acknowledgments:} We are grateful to T. Quella for several
enlightening discussions about key aspects of this paper, and to
H. Saleur for sharing his unpublished results and for pointing out
relevant references. We also thank R. Vlijm and J.-S. Caux for help
with the Bethe Ansatz solution, and P. Fendley for correspondence. JD
would like to thank B. Bradlyn and N. Read for collaboration on
related topics, which motivated this work.
RB and JD are grateful for the warm hospitality of the Erwin-Schr\"odinger
International Institute for
Mathematical Physics (Vienna) during the last stages of this work.

\appendix

\section{Appendix: current-current perturbation and logarithmic corrections}
\label{sec:appendix_log}
In the main text, we overlooked the role of logarithmic corrections
which appear in all of the spin-$\frac{k}{2}$ descendants of the
six-vertex. We were able to do so because such logarithmic corrections
do not affect the correlations of the observable we constructed, at
least not at the leading order. They appear only in subleading
corrections. This is in strong contrast to the case of the spin-spin
correlation, for example, which is well-known to pick a logarithmic
prefactor at the leading order:
\begin{equation}
	\label{eq:spin_spin_log}
	\left<  S^a_{{\bf x}}  S^b_{{\bf x}'}  \right> \, \underset{|{\bf x}-{\bf x}'| \rightarrow \infty}{\propto} \,  \epsilon_{{\bf x},{\bf x}'} \frac{(\log | z-z' |)^{\frac{1}{2}}}{ | z-z' |} \, \delta^{ab} \, ,
\end{equation}
where $\epsilon_{{\bf x},{\bf x}'} = \pm 1$ depends on the positions and takes care of the staggering. We have again used the notation
${\bf x} = (x,y)$, $z=x+ e^{i \alpha} y$. The purpose of this appendix is to explain why this happens for the spin operator, but not for the
lattice chiral observable we have constructed. We follow the beautiful treatment of Affleck, Gepner, Schulz and Ziman \cite{Affleck1989}.

The critical vertex models we are interested in can be described by the $\SU(2)_k$ WZW model, with perturbations. Since we know that these
models are at the critical point, no relevant perturbation is allowed. Irrelevant perturbations are certainly allowed, but also marginal ones. The right and left currents lead to a marginal perturbation
\begin{equation}
	S_{{\rm WZW}} \; \rightarrow \;  S_{{\rm WZW}} \, + \, g_0
        \int \frac{d^2z}{2\pi} 
        {\bf J}(z) \cdot \overline{{\bf J}}(\bar{z})\, .
\end{equation}
One could also wonder whether terms like ${\bf J}\cdot {\bf J}$ or
$\overline{{\bf J}}\cdot\overline{{\bf J}}$ could appear (we suppress
the $z,\bar{z}$ dependence when clear from the context), but these are
just the chiral and the anti-chiral components of the stress-tensor,
and these would change the metric, namely the geometric angle $\alpha$
in the main text. We have already fixed our conventions such that
$\alpha$ is properly taken into account, so these two perturbations
are actually not present here.  Now let us come back to the ${\bf
  J}\cdot \overline{{\bf J}}$ term, which has much less trivial
effects on the effective theory describing the critical point. The
$\beta$-function for the renormalized coupling constant $g$ can be
computed as follows (for examples of such calculations, see
e.g. \cite{LudwigWiese}). We introduce an UV (IR) cutoff $a_0$ ($L$),
and look at how $g_0$ is modified by
higher order terms in the expansion of the exponential
of the perturbation:
\begin{align}
	\left<  \dots e^{g_0 \int \frac{d^2 z}{2 \pi}  {\bf J}\cdot
            \overline{{\bf J}}}\right>_{\text{WZW}} 
        &= \left<  \dots \left( 1 +
            g_0 \int \frac{d^2z}{2 \pi} 
{\bf J}\cdot \overline{{\bf J}} + \frac{g_0^2}{2}  \int \frac{d^2z}{2
  \pi} 
{\bf J}\cdot \overline{{\bf J}}  \int \frac{d^2z'}{2 \pi}  {\bf J}\cdot \overline{{\bf J}}  + O(g_0^3) \right) \right>_{\text{WZW}}   \\
\nonumber	&=  \left<  \dots \left( 1 + \left( g_0  -  g_0^2 \log(L/a_0) \right)  \int \frac{d^2z}{2 \pi} {\bf J}\cdot \overline{{\bf J}} \right) + O(g_0^3)   \right>_{\text{WZW}}\, .
\end{align}
Here the dots stand for arbitrary operator insertions and the
expectation values are computed in the {\it unperturbed} theory.  From
the first to the second line, we have used the OPE $J^a(z)
\bar{J}^a(\bar{z}) J^b(z') \bar{J}^b(\bar{z}') \simeq (i
\epsilon^{abc})^2 /|z-z'|^2 J^c(z) \bar{J}^c (\bar{z}) = -2/|z-z'|^2
J^c(z) \bar{J}^c (\bar{z})$, and we have evaluated the integral $\int
\frac{d^2 z'}{2\pi |z-z'|^2}$ over $a_0<|z-z'|<L$, which gives $
\log(L/a_0)$. Thus, at the leading order, the coupling is renormalized
from $g_0$ to $g = g_0 - g_0^2 \log(L/a_0)+ O(g_0^3)$, which gives
\begin{equation}
	\beta(g) \, \equiv \,  \frac{\partial g}{\partial \log a_0} \, = \, g^2 \, + \, O(g^3).
\end{equation}
The perturbation ${\bf J}\cdot \overline{{\bf J}}$ is thus {\it
  marginally irrelevant} for $g_0>0$ and {\it marginally relevant} for
$g_0 <0$. For the vertex models we consider, it is known that we are
in the marginally irrelevant case, so for small $g_0$ we find
\begin{equation}
	g \, \underset{a_0 \rightarrow 0}{\propto} \, (-\log a_0)^{-1} .
\end{equation}
Variations of the correlation functions with the UV cutoff $a_0$ can be estimated thanks to the Callan-Symanzik equation. For example, for the two-point function of a (Virasoro) primary field $\phi(z,\bar{z})$ with scaling dimension $\Delta$, we have
\begin{equation}
	\label{eq:CallanSymanzik}
	\left( \frac{\partial }{ \partial \log a_0} +  \beta(g) \frac{\partial }{ \partial g} - 2 \Delta(g)  \right) \left< \phi(z_1,\bar{z}_1)  \phi(z_2,\bar{z}_2) \right> \, = \, 0.
\end{equation}
Importantly, the anomalous scaling dimension $\Delta (g)$ is different from the scaling dimension $\Delta$ in the pure CFT:
\begin{equation}
	\label{eq:anomalous_Delta}
	\Delta(g) \, = \, \Delta \, - \, g\sum_{n \in \mathbb{Z}} 
\left<{\bf J}_{n} \cdot \overline{{\bf J}}_n\right>_\phi \, + \, O(g^2).
\end{equation}
(Above and below $\left<{\bf J}_{n} \cdot \overline{{\bf
      J}}_n\right>_\phi$ is the expectation value of
this operator in the state $|\phi\rangle$ with scaling dimension
$\Delta$.) This is because the Hamiltonian is affected by the
current-current perturbation:
\begin{equation}
	H_{\rm CFT} \, \rightarrow \, H_{\rm CFT} \, - \, g \int \frac{dx}{2\pi} {\bf J}(e^{ix})\cdot \overline{{\bf J}}(e^{-ix})\, .
\end{equation}
For instance, the primary field $\phi (z , \bar{z}) = P_{j'}\left[ \phi_{j} \otimes \overline{\phi}_{j} \right]$ has the anomalous scaling dimension
\begin{equation}
	\Delta(g) \, = \, \Delta \, - \, \frac{j'(j'+1) - 2 j (j+1)}{2}\, g \, + \, O(g^2).
\end{equation}
Plugging (\ref{eq:anomalous_Delta}) into (\ref{eq:CallanSymanzik}), dropping terms of order $O(g^2)$, and integrating the differential equation, one finds that the two-point function becomes
\begin{equation}
	\left< \phi(z_1,\bar{z}_1)  \phi(z_2,\bar{z}_2) \right> \,
        \propto \, \left( \log \frac{|z_1-z_2|}{a_0} \right)^{2 
\left<{\bf J}_0 \cdot \overline{{\bf J}}_0\right>_\phi} \, \left( \frac{|z_1-z_2|}{a_0} \right)^{-2 \Delta}.
\end{equation}
In particular, since the leading contribution to the lattice spin
operator $S^a_{{\bf x}}$ is the primary operator $P_{1}\left[
  \phi_{\frac{1}{2}} \otimes \overline{\phi}_{\frac{1}{2}} \right]^a$,
we have 
$2\left<{\bf J}_0 \cdot \overline{{\bf J}}_0\right>_\phi \, = \, \frac{1}{2}$, in agreement with (\ref{eq:spin_spin_log}).

Finally, we see the reason why there are no multiplicative logarithmic
corrections to the chiral current-current correlation $\left< J^a(z_1)
  J^b(z_2) \right>$: it is because $\sum \left<{\bf J}_n \cdot
  \overline{{\bf J}}_n\right>_J= 0$, so there is no
correction to the scaling dimension at order $O(g)$. Note that the
structure factors are also affected by these logarithmic
corrections. Since they are related to one-point functions, one can
evaluate their variation with the system size $L/a_0$ from the
Callan-Symanzik equation for the one-point function. For instance, the
following structure factor must scale as
\begin{equation} 
	\left. \left<  \left. P_{1}\left[ \phi_{j} \otimes \overline{\phi}_{j} \right]^a \right| S^b_x \right| 0 \right> \, \propto \, (-1)^x \frac{\left[ \log (L/a_0) \right]^{1 - j(j+1)}}{(L/a_0)^{\Delta_j}} \delta^{ab}
\end{equation}
for half-integer $j$. We used the notation $\Delta_j = h_j + \bar{h}_j = 2\frac{j(j+1)}{k+2}$. For integer $j$, the structure factor vanishes, as discussed in the main text (see also the discussion in \cite{Caux_spin_1}). For the structure factor computed in the main text (see Fig.\ref{fig:results_CJ_k1} and Fig.\ref{fig:results_CJ_k2}), we do not find a
logarithmic correction, again because
$\sum \left<{\bf J}_n \cdot
  \overline{{\bf J}}_n\right>_J= 0$.

\section{Appendix: Bethe Ansatz and form factors}
\label{sec:appendixA}

\subsection{Bethe Ansatz of the spin-$k/2$ Heisenberg chain}

In this appendix, we prefer to parametrize the weights of the six-vertex model with $u$ rather than with the geometric angle $\alpha$:
\begin{equation}
  a(u) \, =\,1-u \qquad \quad b(u) \, = \, -u \qquad \quad c(u) \, =\,1.
\end{equation}
The spectral parameter $u$ is related to the geometric angle by $u = 1-\frac{\alpha}{\pi}$.
We use the standard notations for the elements of the monodromy matrix:
\begin{equation}
  \begin{aligned}
    & A(u) = \raisebox{-0.8cm}{\includegraphics{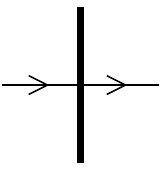}} 
    \qquad\qquad B(u) = \raisebox{-0.8cm}{\includegraphics{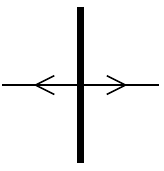}}  \\
    \\
    & C(u) = \raisebox{-0.8cm}{\includegraphics{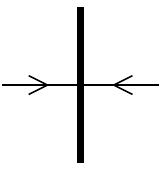}} 
    \qquad\qquad D(u) = \raisebox{-0.8cm}{\includegraphics{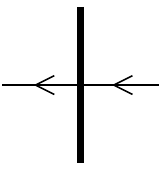}} \quad \,,
  \end{aligned}
\end{equation}
where a thin horizontal line carries a spin-$1/2$ representation with spectral parameter $u$, and a thick vertical line carries an arbitrary representation, typically a tensor product of a spin-$k/2$ representations.
These operators satisfy the ``RTT'' relations, which directly follow from the Yang-Baxter equation.
For instance:
\begin{subequations}
  \label{eq:YangBaxter_algebra}
  \begin{eqnarray}
    B(u) B(v) & = & B(v) B(u) \,, \\
    A(u) B(v) & = & \frac{a(v-u)}{b(v-u)} B(v) A(u) \, - \, \frac{c(v-u)}{b(v-u)} B(u) A(v) \,, \\
    D(u) B(v) & = & \frac{a(u-v)}{b(u-v)} B(v) D(u) \, - \, \frac{c(u-v)}{b(u-v)} B(u) D(v) \,,
  \end{eqnarray}
\end{subequations}
and so on. For a set of complex numbers $\{\lambda_1, \dots, \lambda_M\}$, we define the
corresponding Bethe state as
\begin{equation} \label{eq:state}
  \ket{\{\lambda_q\}} = B(u_1) \dots B(u_M) \ket{\Uparrow} \,,
\end{equation}
where
\begin{equation}
  u_q =-\frac{k}{2} +1 - i\frac{\lambda_q}{2} \,.
\end{equation}
A Bethe state~\eqref{eq:state} is an eigenstate of the transfer matrix
$T^{(k/2)}_N$, and therefore of the Hamiltonian $H^{(k/2)}$
of Eq.~\eqref{eq:H}, only if the Bethe equations are satisfied:
\begin{equation}
  \left[ \frac{\lambda_j \, + \, i k}{\lambda_j \, - \, i k} \right]^N \, = \, \prod_{q\neq j} \frac{\lambda_j - \lambda_q + 2i}{\lambda_j - \lambda_q - 2i}\, .
\end{equation}
One can check that the momentum and energy---{\it i.e.} the eigenvalue of the Hamiltonian $H^{(k/2)}$---of a Bethe state $\ket{\{\lambda_q\}}$ is:
\begin{align}
  P(\{ \lambda_q \}) &=
  \frac{1}{a_0}\sum_{q=1}^M \left[ \pi - 2\,{\rm arctan}\frac{\lambda_q}{k} \right] \qquad {\rm mod} \; \frac{2 \pi}{a_0} \,, \\
  E(\{ \lambda_q \}) &= \sum_{q=1}^M \frac{-4k}{k^2+\lambda_q^2} \, + \, c_k \,.
\end{align}

\subsection{Root configurations corresponding to some low-energy states of interest}

\begin{itemize}
\item For any even $N$ the ground state is a singlet and corresponds
  to a configuration of $N/2$ $k$-strings, with Bethe-Takahashi numbers
  \cite{Takahashi1972,Caux_spin_1} $\left\{ - \frac{N-2}{4}, -
    \frac{N-2}{4} +1 , \dots , \frac{N-2}{4} -1, \frac{N-2}{4}
  \right\}$. The ground state has momentum $P_0 = 0$ if $k N/2$ is
  even, and momentum $P_0 = \pi/a_0$ if $k N/2$ is odd.

\item The Bethe state that would correspond to the CFT state $\ket{ P_1 \left[ \phi_{\frac{1}{2}} \otimes \overline{\phi}_{\frac{1}{2}} \right] }$ in the continuum limit is the one with $N/2-1$ $k$-strings, and one $(k-1)$-string, with Bethe-Takahashi numbers $\{ - \frac{N}{4} + 1, - \frac{N}{4} +2 , \dots , \frac{N}{4} -2, \frac{N}{4} -1 \}$ and $\{ 0 \}$. This Bethe state is, as usual, a highest weight state. Acting on it with the $\su(2)$ generators, one generates a triplet of degenerate excited states. One can check that the momentum of these three eigenstates is $P= \left(P_0 + \frac{\pi}{a_0}\right) \; {\rm mod}\, \frac{2\pi}{a_0}$.

\item The above facts generalize to $\phi_j$ for $j > \frac{1}{2}$ as follows. We claim that the Bethe state which corresponds to the (highest weight) CFT state $\ket{ P_1 \left[ \phi_j \otimes \overline{\phi}_j \right] }$ is the one that can be obtained from the ground-state by taking out two $k$-strings, and replacing them by one $(k-2j)$-string, and one $(k+2j-1)$-string. The Bethe-Takahashi numbers are $\{ - \frac{N-2}{2} +1 ,- \frac{N-2}{2} +2 , \dots, \frac{N-2}{2} -2 ,\frac{N-2}{2} -1   \}$, $\{0 \}$ and $\{ 0\}$ for the newly created strings. This gives us a triplet of degenerate states, as it should. Also, notice that such a Bethe state exists only if $j \leq \frac{k}{2}$, which is consistent with the fact that the primary fields for the (chiral) Kac-Moody algebra must have $\SU(2)$-spin $j \leq \frac{k}{2}$. The momentum $P$ of these states can be checked to be
  \begin{equation}
    \label{eq:singular_BS}
    a_0 \times (P-P_0) \equiv \begin{cases}
      \pi \quad \mod 2\pi  & \text{if $j$ is half-integer,} \\
      0   \quad \mod 2\pi  & \text{if $j$ is integer.}
    \end{cases}
  \end{equation}
For integer $j$, this follows from a subtle fact about the Bethe equations, which arises when one looks for solutions with a $p$-string, $p$ odd and $2p \geq k$. In that case, it turns out that there are solutions to the Bethe equations with a pair of the Bethe roots that are {\it exactly} equal to $\pm i k$. This phenomenon is known in the literature: the corresponding Bethe states are dubbed 'singular' for instance in Refs. \cite{Hagemans_Caux, Caux_spin_1}. These Bethe states are still perfectly valid eigenstates of the Hamiltonian, however, some particular limiting procedure must be used when one calculates the corresponding energies or momenta. The fact that the Bethe states for integer spin $j$ correspond precisely to these 'singular' states is crucial; this is what leads to (\ref{eq:singular_BS}), in agreement with what we claimed in the main parts of the paper. We note that essentially the same discussion appeared already in \cite{Affleck1989}.

\item Finally, the Bethe state which we identify with the CFT state $\ket{J^+}$ is the one with one $(k-1)$-string, and $N/2-1$ $k$-strings, and Bethe-Takahashi numbers $\{ 0 \}$ and $\{ -\frac{N}{4},-\frac{N}{4}+1, \dots , \frac{N}{4}-3,\frac{N}{4}-2 \}$. This state has momentum $P \,= \, P_0 + \frac{2\pi}{L} \; {\rm mod}\, \frac{2\pi}{a_0} $, as required.
\end{itemize}

\newpage

\begin{figure}[ht]
  \begin{center}
    \begin{tabular}{c}
      \includegraphics[width=0.7\textwidth]{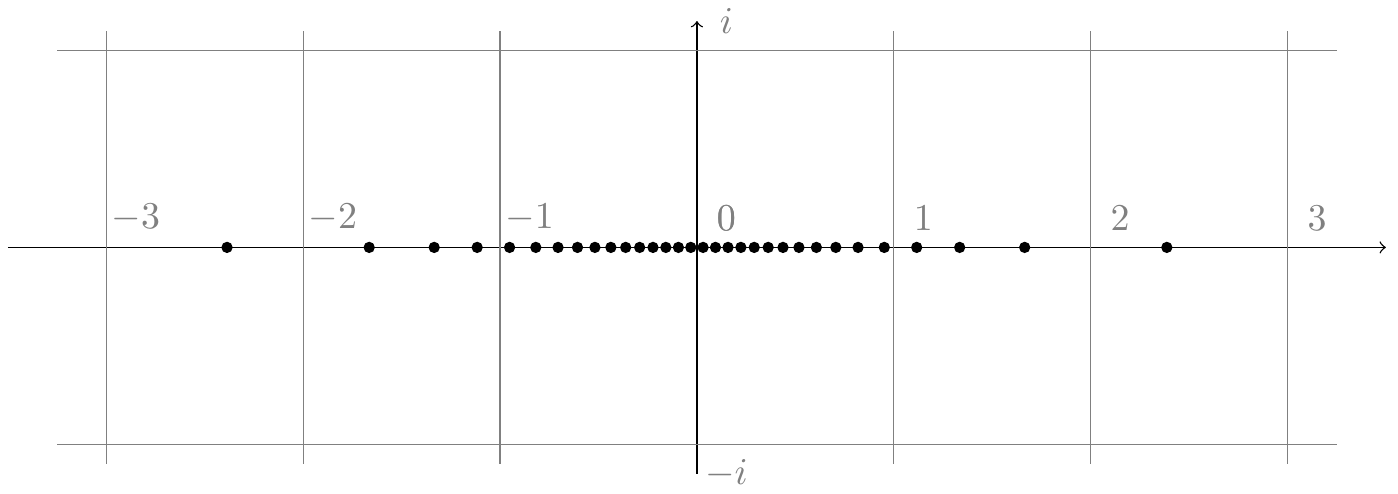} \\
      ground state, spin $0$, momentum $P_0 \, = \, 0$ \\ \\
      \includegraphics[width=0.7\textwidth]{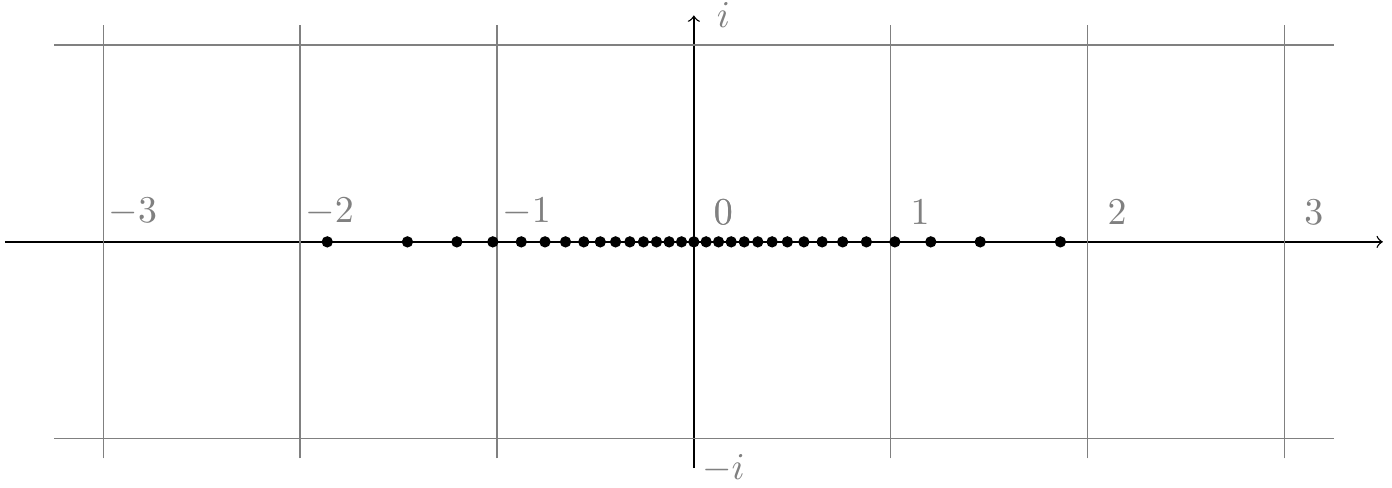} \\
      $\ket{P_1\left[ \phi_{\frac{1}{2}} \otimes \overline{\phi}_{\frac{1}{2}} \right]^+ }$, spin $1$, momentum $P= \frac{\pi}{a_0}$ \\ \\
      \includegraphics[width=0.7\textwidth]{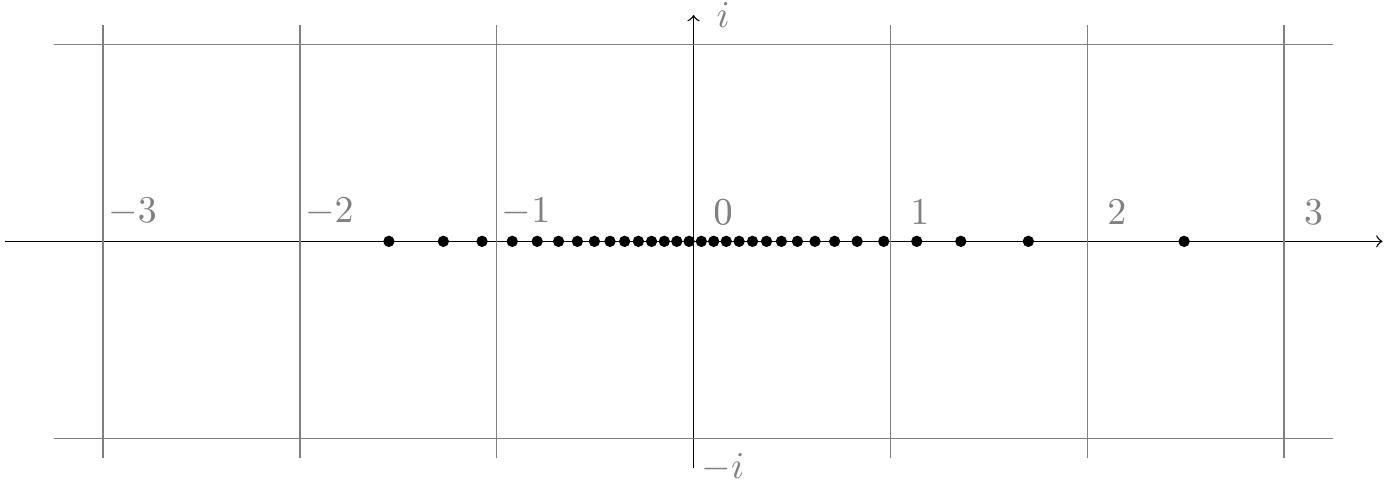} \\
      $ \ket{J^+}$ , spin $1$, momentum $P = \frac{2 \pi }{L}$
    \end{tabular}
  \end{center}
  \caption{Configurations of Bethe roots for $k=1$, and $N=64$, corresponding to the ground state, and to the two highest weight vectors of interest.}
\end{figure}

\begin{figure}[ht]
  \begin{tabular}{cc}
    \includegraphics[width=0.48\textwidth]{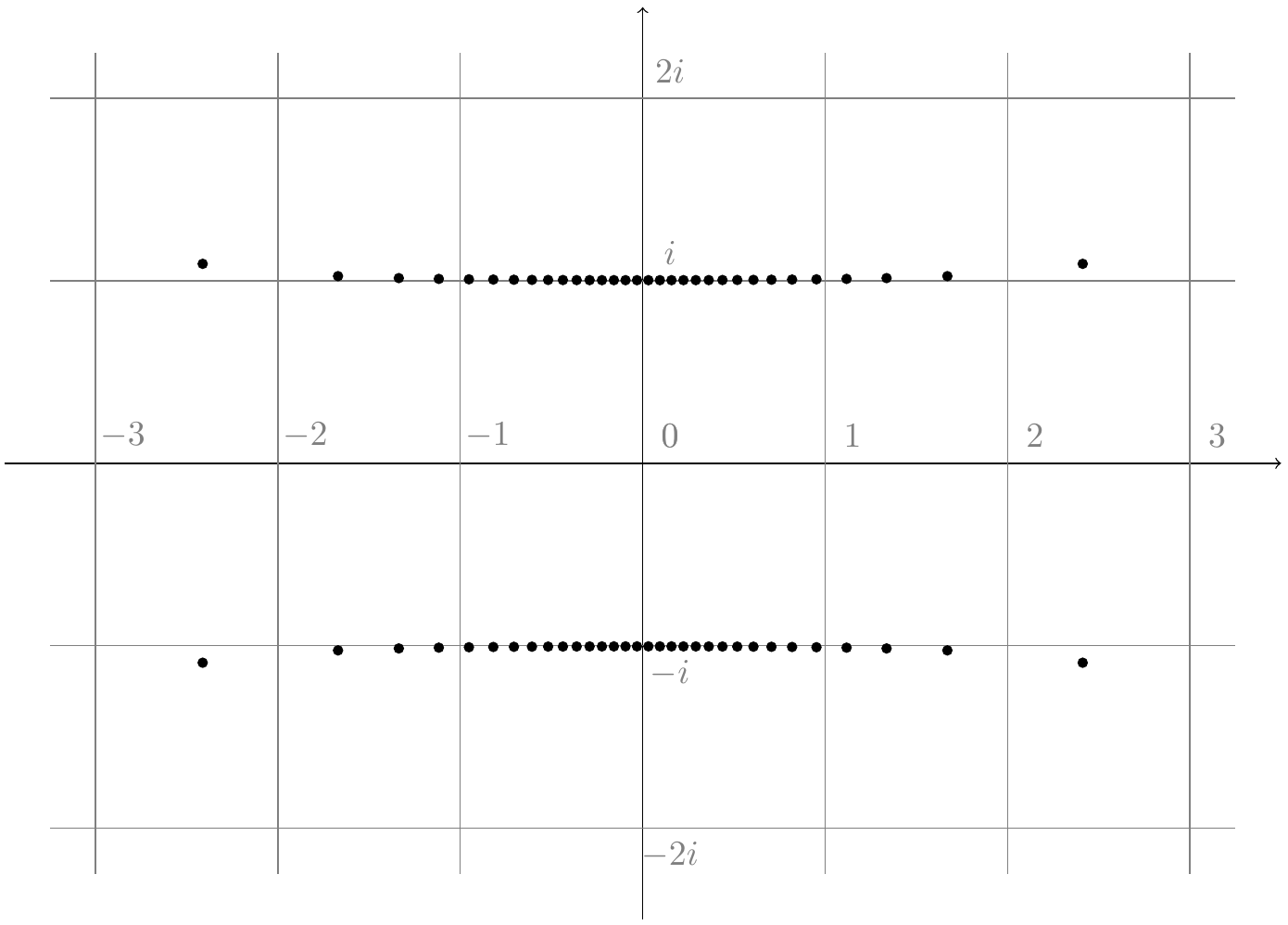} &
    \includegraphics[width=0.48\textwidth]{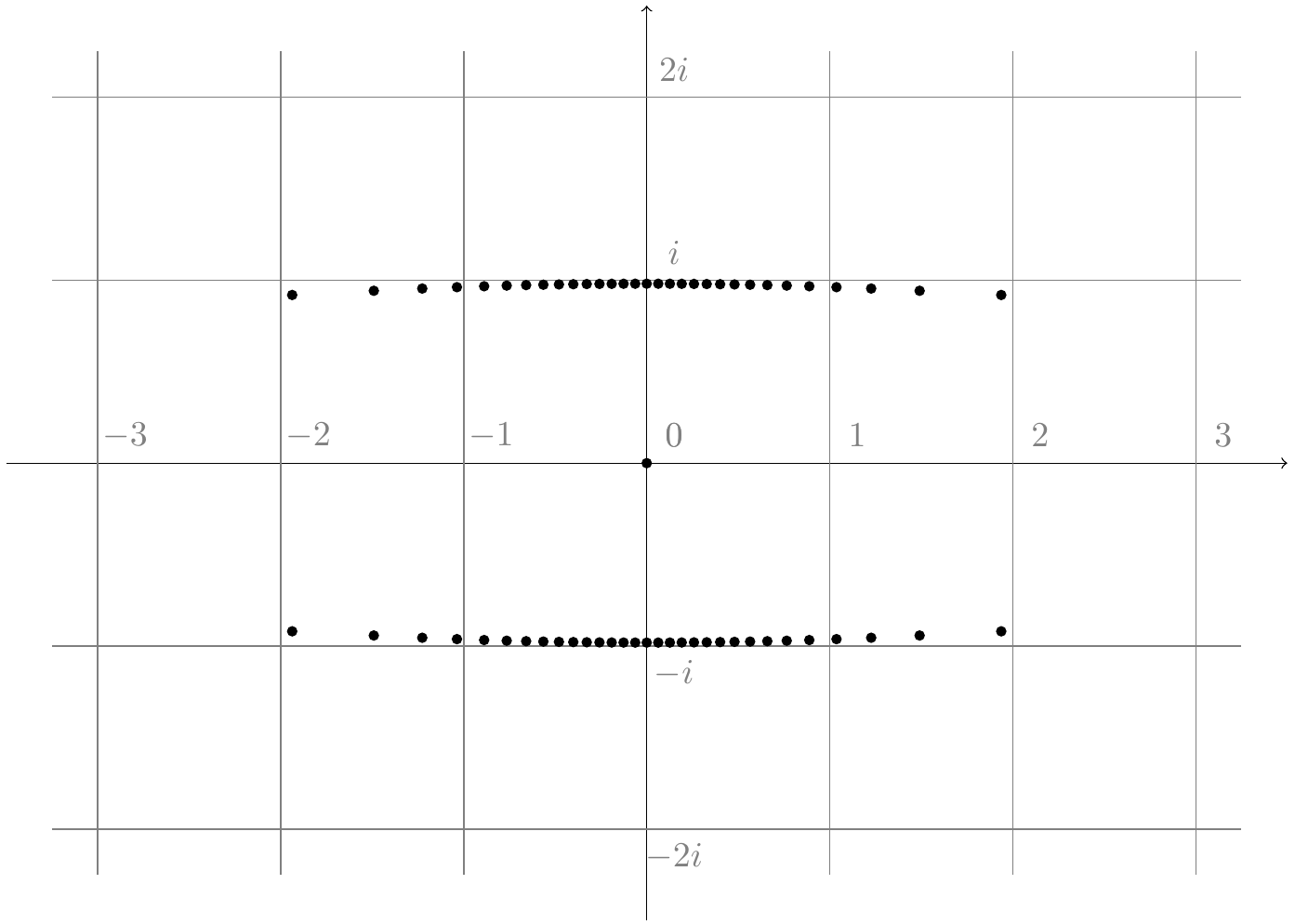} \\
    ground state, spin $0$, momentum $P_0 = 0$  &    $\ket{P_1\left[ \phi_{\frac{1}{2}} \otimes \overline{\phi}_{\frac{1}{2}} \right]^+ }$, spin $1$, momentum $\frac{\pi}{a_0}$ \\ \\
    \includegraphics[width=0.48\textwidth]{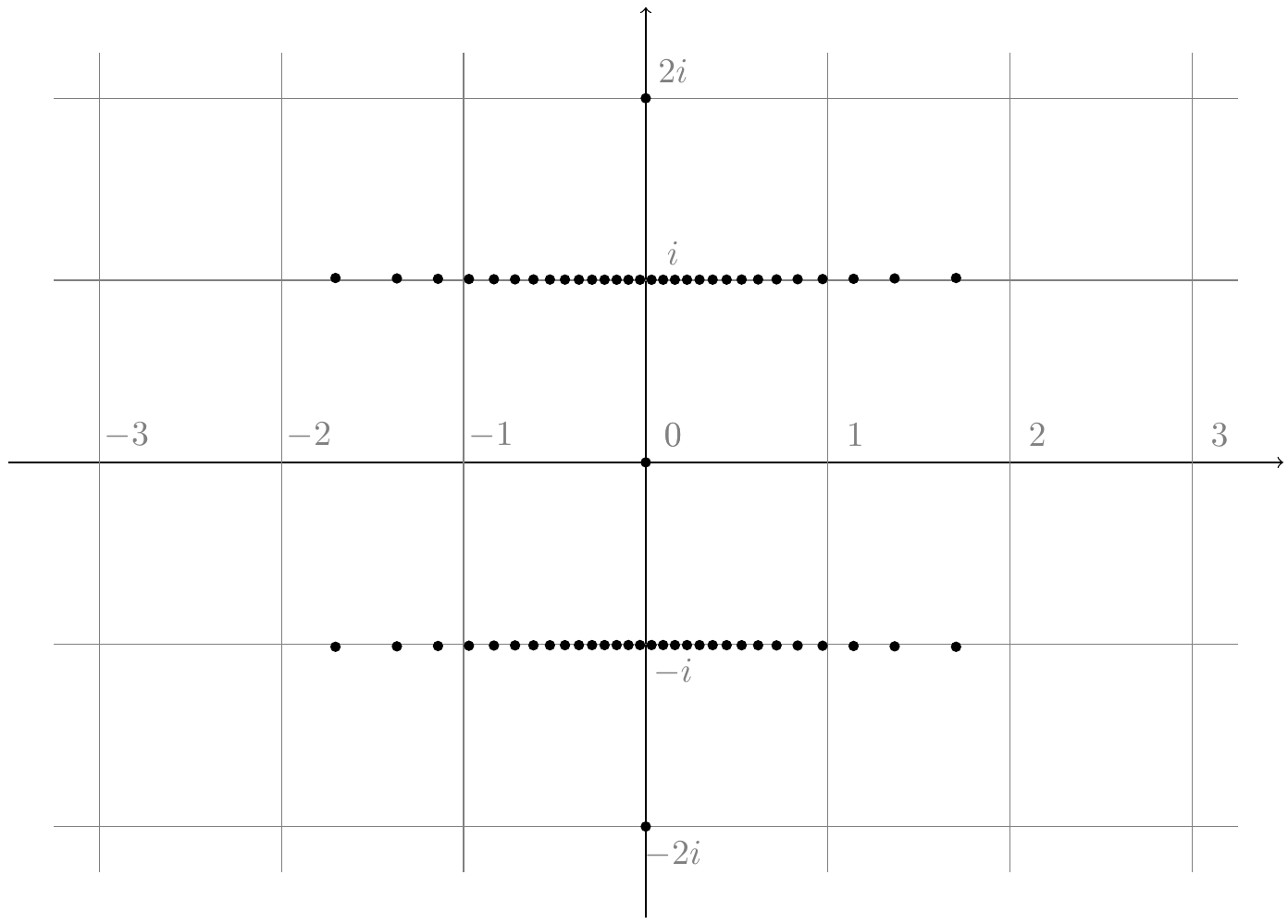} &
    \includegraphics[width=0.48\textwidth]{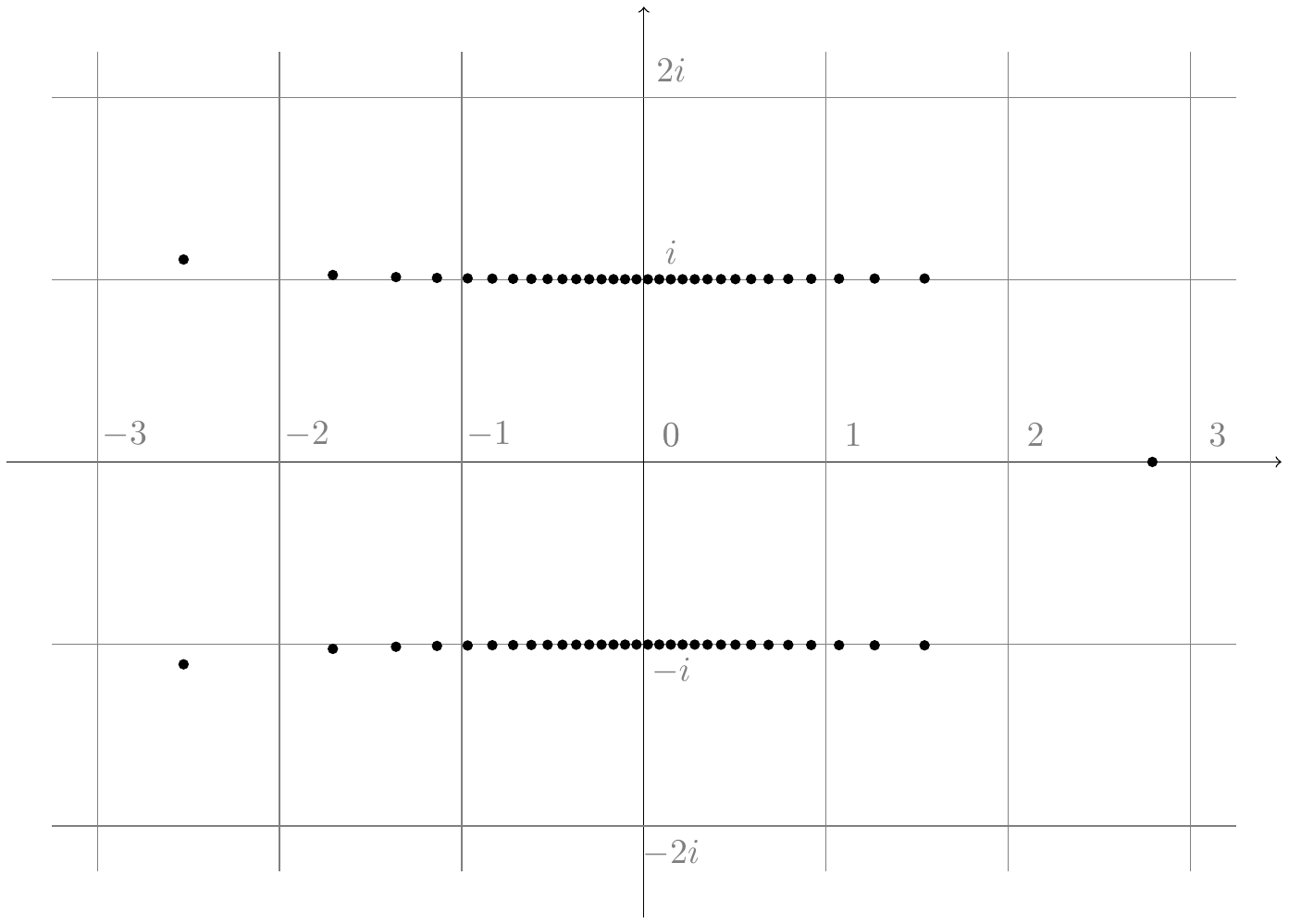} \\
    $\ket{P_1\left[ \phi_{1} \otimes \overline{\phi}_{1} \right]^+ }$, spin $1$, momentum $0$   &  $ \ket{J^+}$, spin $1$, momentum $\frac{2 \pi}{L}$
  \end{tabular}
  \caption{Configurations of Bethe roots for $k=2$, and $N=64$, corresponding to the ground state, and to the three highest weight vectors of interest. Note the presence of two eigenvalues at $\pm 2i$ in the configuration that corresponds to $\ket{P_1\left[ \phi_{1} \otimes \overline{\phi}_{1} \right]^+ }$.}
\end{figure}

\newpage

\subsection{Useful formulas for $k=1$}

\subsubsection*{Eigenvalue of the transfer matrix $T^{(1/2)}(u)$}
In terms of the elements of the monodromy matrix, we have
\begin{equation}
	T^{(1/2)}(u) \, = \, A(u) + D(u).
\end{equation}
This, together with the 'RTT' relations and the Bethe equations,
allows one to show that the eigenvalue of $T^{(1/2)}[u=(1-i\mu)/2]$ corresponding to
a Bethe state $\ket{ \{  \lambda_q \} }$ is
\begin{equation}
    \Lambda^{(1/2)}\left(
      u=\frac{1-i\mu}{2}, \{\lambda_q\}
    \right) = \left[
     \prod_{q=1}^M \frac{\lambda_q-\mu -2i}{\lambda_q-\mu} \, + \, \left(\frac{\mu+i}{\mu-i} \right)^N  \prod_{q=1}^M \frac{\lambda_q-\mu +2i}{\lambda_q-\mu}
  \right] .
\end{equation}

\subsubsection*{Norms of Bethe states and overlaps}

The norm of a Bethe state $\ket{\{ \lambda_q \}}$ (where $\{ \lambda_q\}$ is a solution of the Bethe equations) can be expressed in terms of a determinant,
\begin{equation}
  \langle \{\lambda_q\} | \{\lambda_q\} \rangle = 
  (2 i)^M \, \left( \prod_{a \neq b} \frac{\lambda_a-\lambda_b -2 i}{\lambda_a - \lambda_b} \right)\times \det G \,,
\end{equation}
where the entries of the $M \times M$ Gaudin matrix $G$ are:
\begin{equation}
  G_{ab} = \begin{cases}
    \displaystyle \frac{-2 i N}{\lambda_a^2+1} + \sum_{k \neq a} \frac{4 i }{(\lambda_a-\lambda_k)^2 + 4}
    & \text{for $a=b$,} \\
    \displaystyle \frac{-4 i }{(\lambda_a-\lambda_k)^2 + 4}
    & \text{for $a \neq b$.}
  \end{cases}
\end{equation}
The overlap between a Bethe eigenstate $\ket{\{\lambda_q\}}$ and an off-shell Bethe state $\ket{\{\mu_p\}}$ (i.e. the $\mu_p$ are not assumed to satisfy the Bethe equations) also admits a determinant expression,
\begin{equation}
  \langle \{\lambda_q\} | \{\mu_p\} \rangle =  \frac{\det T}{\det \Lambda} \,,
\end{equation}
where $\Lambda$ and $T$ are $M \times M$ matrices with elements
\begin{align}
  \Lambda_{ab} &= \frac{1}{\mu_b - \lambda_a} \,, \\
  T_{ab} &= \frac{2i}{(\lambda_a - \mu_b)^2} \left[ \prod_{k \neq a} \frac{\lambda_k - \mu_b -2i}{\lambda_k - \mu_b} \, - \, \left( \frac{\mu_b + i}{\mu_b - i} \right)^N \prod_{k \neq a} \frac{\lambda_k - \mu_b +2i}{\lambda_k - \mu_b} \right] \,.
\end{align}

\subsubsection*{Form factor for $k=1$}
The matrix element $\bra{\{ \lambda \}} S^+_{x=0} \ket{ \{ \mu \} }$
between two (not normalized) Bethe states has been computed in \cite{KMT}. Here there are $M$ $\lambda$-roots, and $M+1$ $\mu$-roots. The result, which we use in our numerical evaluation, is:
\begin{equation}
	\bra{\{ \lambda \}} S^+_{x=0} \ket{ \{ \mu \} } \, = \, \frac{ \prod_r (\mu_r + i )}{\prod_k (\lambda_k + i)}  \frac{{\rm det } H}{ \prod_{j>l} (\mu_j - \mu_l)  \prod_{p<q} (\lambda_p - \lambda_q) } \, ,
\end{equation}
where $H$ is an $(M+1) \times (M+1)$ matrix with entries
\begin{equation}
	H_{ab} \, = \, \left\{ \begin{array}{lcl}
		\displaystyle \frac{-2i}{ \mu_a - \lambda_b } \left[  \prod_{k \neq a} (\mu_k - \lambda_b - 2i)   \, - \, \left( \frac{\lambda_b + i}{\lambda_b - i} \right)^N  \prod_{k \neq a} (\mu_k - \lambda_b + 2i)    \right]  && {\rm if} \; b <M+1 \\ \\
		\displaystyle \frac{-2i}{ \mu_a^2 + 1}	\qquad \qquad {\rm if } \; b = M+1 \, .
	\end{array} \right. 
\end{equation}

\subsection{Useful formulas for $k=2$}

\subsubsection*{Eigenvalues of the transfer matrix $T^{(1)}(u)$}
The transfer matrix is expressed in terms of the entries of the monodromy matrix as
\begin{equation}
  T^{(1)}(u) = A_+ A_- + D_+ D_- + \frac{1}{2}
  \left(
    A_+ D_- + D_+ A_- + B_+ C_- + C_+ B_-
  \right) \,,
\end{equation}
where we have used the notations $A_\pm = A(u \pm 1/2)$, and similarly for $B,C,D$.
Setting $u=-i\mu/2$, the eigenvalue associated to the Bethe state $\ket{\{ \lambda_q \}}$ is
\begin{eqnarray}
  \Lambda^{(1)}(u, \{ \lambda_q\}) & = & \Lambda^{(1/2)}(u+1/2, \{ \lambda_q\}) \, \Lambda^{(1/2)}(u-1/2, \{ \lambda_q\})   \, - \, \left( \frac{\mu+3i}{\mu - i}\right)^N.
\end{eqnarray}

\subsubsection*{Norms and overlaps}

The norm of a Bethe state $\ket{\{ \lambda_q \}}$ (where $\{ \lambda_q\}$ is a solution to the Bethe equations for $k=2$) is
\begin{equation}
  \langle \{\lambda_q\} | \{\lambda_q\} \rangle = 
  (-2 i)^M \, \left( \prod_{a \neq b} \frac{\lambda_a-\lambda_b -2 i}{\lambda_a - \lambda_b} \right)\times \det G \,,
\end{equation}
where the entries of the Gaudin matrix $G$ are:
\begin{equation}
  G_{ab} = \begin{cases}
    \displaystyle \frac{-4 i N}{\lambda_a^2+1} + \sum_{k \neq a} \frac{4 i }{(\lambda_a-\lambda_k)^2 + 4}
    & \text{for $a=b$,} \\
    \displaystyle \frac{-4 i }{(\lambda_a-\lambda_k)^2 + 4}
    & \text{for $a \neq b$.}
  \end{cases}
\end{equation}
The overlap between a Bethe eigenstate $\ket{\{\lambda_q\}}$ and an off-shell Bethe state $\ket{\{\mu_p\}}$ is:
\begin{equation}
  \langle \{\lambda_q\} | \{\mu_p\} \rangle =  \frac{\det T}{\det \Lambda} \,,
\end{equation}
where $\Lambda$ and $T$ are $M \times M$ matrices with elements
\begin{align}
  \Lambda_{ab} &= \frac{1}{\mu_b - \lambda_a} \,, \\
  T_{ab} &= \frac{2i}{(\lambda_a - \mu_b)^2} \left[ \prod_{k \neq a} \frac{\lambda_k - \mu_b -2i}{\lambda_k - \mu_b} \, - \, \left( \frac{\mu_b + 2i}{\mu_b - 2i} \right)^N \prod_{k \neq a} \frac{\lambda_k - \mu_b +2i}{\lambda_k - \mu_b} \right] \,.
\end{align}

\subsubsection*{Form factor for $k=2$}
The expression for the matrix element $\bra{\{ \lambda \}} S^+_{x=0} \ket{ \{ \mu \} }$ between two Bethe states for $k>1$ is given in \cite{Olalla_Maillet}. Here we adapt this result to our needs. Again, there are $M$ $\lambda$-roots, and $M+1$ $\mu$-roots. We have:
\begin{equation}
	\bra{\{ \lambda \}} S^+_{x=0} \ket{ \{ \mu \} } \, = \, \frac{ \prod_r (\mu_r +2 i ) }{ \prod_k (\lambda_k +2 i) }  \frac{{\rm det } H}{ \prod_{j>l} (\mu_j - \mu_l)  \prod_{p<q} (\lambda_p - \lambda_q) } \, ,
\end{equation}
where $H$ is an $(M+1) \times (M+1)$ matrix with entries
\begin{equation}
	H_{ab} \, = \, \left\{ \begin{array}{lcl}
		\displaystyle \frac{-2i}{ \mu_a - \lambda_b } \left[  \prod_{k \neq a} (\mu_k - \lambda_b - 2i)   \, - \, \left( \frac{\lambda_b +2 i}{\lambda_b -2 i} \right)^N  \prod_{k \neq a} (\mu_k - \lambda_b + 2i)    \right]  && {\rm if} \; b <M+1 \\ \\
		\displaystyle \frac{-4i}{ \mu_a^2 + 1}	\qquad \qquad {\rm if } \; b = M+1 \, .
	\end{array} \right. 
\end{equation}

\bibliographystyle{amsalpha}
\bibliography{biblio}

\newcommand{\etalchar}[1]{$^{#1}$}
\providecommand{\bysame}{\leavevmode\hbox to3em{\hrulefill}\thinspace}
\providecommand{\MR}{\relax\ifhmode\unskip\space\fi MR }
\providecommand{\MRhref}[2]{%
  \href{http://www.ams.org/mathscinet-getitem?mr=#1}{#2}
}
\providecommand{\href}[2]{#2}
\begin{thebibliography}{IWWZJ13}

\bibitem[Aff85]{Affleck1985}
I~Affleck, \emph{Critical behavior of two-dimensional systems with continuous
  symmetries}, Phys. Rev. Lett. \textbf{55} (1985), 1355--1358.

\bibitem[Aff86]{affleck1986exact}
\bysame, \emph{Exact critical exponents for quantum spin chains, non-linear
  $\sigma$-models at $\theta=\pi$ and the quantum {H}all effect}, Nuclear
  Physics B \textbf{265} (1986), no.~3, 409--447.

\bibitem[Aff88]{affleck1988critical}
\bysame, \emph{Critical behaviour of {SU($n$)} quantum chains and topological
  non-linear $\sigma$-models}, Nuclear Physics B \textbf{305} (1988), no.~4,
  582--596.

\bibitem[Aff98]{Affleck_JPA98}
\bysame, \emph{{Exact correlation amplitude for the 1/2 Heisenberg
  antiferromagnetic chain}}, Journal of Physics A Mathematical General
  \textbf{31} (1998), 4573--4581.

\bibitem[AGSZ89]{Affleck1989}
I~Affleck, D~Gepner, H~J Schulz, and T~Ziman, \emph{Critical behaviour of
  spin-s {H}eisenberg antiferromagnetic chains: analytic and numerical
  results}, Journal of Physics A: Mathematical and General \textbf{22} (1989),
  no.~5, 511.

\bibitem[AGT10]{AGT}
L~F Alday, D~Gaiotto, and Y~Tachikawa, \emph{{L}iouville correlation functions
  from four-dimensional gauge theories}, Letters in Mathematical Physics
  \textbf{91} (2010), no.~2, 167--197.

\bibitem[AH87]{affleck1987critical}
I~Affleck and F~D~M Haldane, \emph{Critical theory of quantum spin chains},
  Physical Review B \textbf{36} (1987), no.~10, 5291.

\bibitem[Ber88a]{bernard1988wess_g}
D~Bernard, \emph{On the {W}ess-{Z}umino-{W}itten models on {R}iemann surfaces},
  Nuclear Physics B \textbf{309} (1988), no.~1, 145--174.

\bibitem[Ber88b]{bernard1988wess}
\bysame, \emph{On the {W}ess-{Z}umino-{W}itten models on the torus}, Nuclear
  Physics B \textbf{303} (1988), no.~1, 77--93.

\bibitem[BF91]{Bernard1991}
D~Bernard and G~Felder, \emph{Quantum group symmetries in two-dimensional
  lattice quantum field theory}, Nuclear Physics B \textbf{365} (1991), no.~1,
  98 -- 120.

\bibitem[BPZ84]{BPZ}
A~A Belavin, A~M Polyakov, and A~B Zamolodchikov, \emph{{Infinite conformal
  symmetry in two-dimensional quantum field theory}}, Nucl. Phys. B
  \textbf{241} (1984), 333.

\bibitem[CAM07]{Olalla_Maillet}
O~A Castro-Alvaredo and J-M Maillet, \emph{{Form factors of integrable
  Heisenberg (higher) spin chains}}, Journal of Physics A: Mathematical and
  Theoretical \textbf{40} (2007), no.~27, 7451.

\bibitem[DFMS97]{DMS97}
Ph~Di~Francesco, P~Mathieu, and D~{S\'en\'echal}, \emph{{C}onformal {F}ield
  {T}heory}, Springer-Verlag New York, 1997.

\bibitem[DFSZ88]{di1988generalized}
Ph~Di~Francesco, H~Saleur, and J-B Zuber, \emph{Generalized {C}oulomb-gas
  formalism for two-dimensional critical models based on {SU(2)} coset
  construction}, Nuclear Physics B \textbf{300} (1988), 393--432.

\bibitem[DRR12]{dubail2012edge}
J~Dubail, N~Read, and E~H Rezayi, \emph{Edge-state inner products and
  real-space entanglement spectrum of trial quantum {H}all states}, Physical
  Review B \textbf{86} (2012), no.~24, 245310.

\bibitem[FK80]{fradkin1980disorder}
E~Fradkin and L~P Kadanoff, \emph{Disorder variables and para-fermions in
  two-dimensional statistical mechanics}, Nuclear Physics B \textbf{170}
  (1980), no.~1, 1--15.

\bibitem[Gin90]{ginsparg1990lecture}
P~Ginsparg, \emph{{Applied Conformal Field Theory, Lecture in Les Houches,
  Session XLIX, 1988}}, arXiv preprint hep-th/9108028.

\bibitem[GRAS05]{Sierra_book}
C~G{\'o}mez, M~Ruiz-Altaba, and G~Sierra, \emph{{Quantum groups in
  two-dimensional physics}}, Cambridge University Press, 2005.

\bibitem[HC07]{Hagemans_Caux}
R~Hagemans and J-S Caux, \emph{{Deformed strings in the Heisenberg model}},
  Journal of Physics A: Mathematical and Theoretical \textbf{40} (2007),
  no.~49, 14605.

\bibitem[Hen99]{henkel1999conformal}
M~Henkel, \emph{Conformal invariance and critical phenomena}, Springer, 1999.

\bibitem[IWWZJ13]{Ikhlef2013}
Y~Ikhlef, R~Weston, M~Wheeler, and P~Zinn-Justin, \emph{{Discrete
  holomorphicity and quantized affine algebras}}, J. Phys. A: Math. Theor.
  \textbf{46} (2013), 265205.

\bibitem[KC71]{kadanoff1971determination}
L~P Kadanoff and H~Ceva, \emph{Determination of an operator algebra for the
  two-dimensional {I}sing model}, Physical Review B \textbf{3} (1971), no.~11,
  3918.

\bibitem[KMT99]{KMT}
N~Kitanine, J-M Maillet, and V~Terras, \emph{{Form factors of the XXZ
  Heisenberg spin-1/2 finite chain}}, Nuclear Physics B \textbf{554} (1999),
  647--678.

\bibitem[KRS81]{KRS81}
P~P Kulish, N~Y Reshetikhin, and E~K Sklyanin, \emph{{Y}ang-{B}axter equation
  and representation theory: I}, Letters in Mathematical Physics \textbf{5}
  (1981), no.~5, 393--403.

\bibitem[KS94]{SaleurKoo}
W.M. Koo and H.~Saleur, \emph{Representations of the virasoro algebra from
  lattice models}, Nuclear Physics B \textbf{426} (1994), no.~3, 459 -- 504.

\bibitem[KZ84]{knizhnik1984current}
V~G Knizhnik and A~B Zamolodchikov, \emph{Current algebra and {W}ess-{Z}umino
  model in two dimensions}, Nuclear Physics B \textbf{247} (1984), no.~1,
  83--103.

\bibitem[LT03]{LukyanovTerras}
S~Lukyanov and V~Terras, \emph{{Long-distance asymptotics of spin-spin
  correlation functions for the XXZ spin chain}}, Nuclear Physics B
  \textbf{654} (2003), 323--356.

\bibitem[Luk98]{Lukyanov97}
S~Lukyanov, \emph{{Low energy effective Hamiltonian for the XXZ spin chain}},
  Nuclear Physics B \textbf{522} (1998), 533--549.

\bibitem[LW03]{LudwigWiese}
A~W~W Ludwig and K~J Wiese, \emph{{The 4-loop $\beta$-function in the 2D
  non-Abelian Thirring model, and comparison with its conjectured exact form}},
  Nuclear Physics B \textbf{661} (2003), no.~3, 577--607.

\bibitem[MCA{\etalchar{+}}14]{mong2014parafermionic}
R~S~K Mong, D~J Clarke, J~Alicea, N~H Lindner, and P~Fendley,
  \emph{Parafermionic conformal field theory on the lattice}, arXiv preprint
  arXiv:1406.0846 (2014).

\bibitem[MR91]{MooreRead}
G~Moore and N~Read, \emph{{Nonabelions in the fractional quantum Hall effect}},
  Nucl. Phys. B \textbf{360} (1991), 362.

\bibitem[Mus10]{mussardo2010statistical}
G~Mussardo, \emph{Statistical {F}ield {T}heory}, Oxford Univ. Press, 2010.

\bibitem[NCS11]{NCS11}
A~E~B Nielsen, J~I Cirac, and G~Sierra, \emph{{Quantum spin Hamiltonians for
  the $SU(2)_k$ WZW model}}, Journal of Statistical Mechanics: Theory and
  Experiment \textbf{2011} (2011), no.~11, P11014.

\bibitem[RR99]{ReadRezayi}
N~Read and E~H Rezayi, \emph{{Beyond paired quantum Hall states: Parafermions
  and incompressible states in the first excited Landau level}}, Phys. Rev. B
  \textbf{59} (1999), 8084.

\bibitem[RS01]{Read2001}
N.~Read and H.~Saleur, \emph{Exact spectra of conformal supersymmetric
  nonlinear sigma models in two dimensions}, Nuclear Physics B \textbf{613}
  (2001), no.~3, 409 -- 444.

\bibitem[Sla89]{Slavnov1989}
N.~A. Slavnov, \emph{{Calculation of scalar products of wave functions and form
  factors in the framework of the algebraic Bethe Ansatz}}, Theor. Math. Phys.
  \textbf{79} (1989), 502.

\bibitem[TRSG12]{Thomale2012}
R~Thomale, S~Rachel, P~Schmitteckert, and M~Greiter, \emph{{Family of spin-$S$
  chain representations of SU(2)${}_{k}$ Wess-Zumino-Witten models}}, Phys.
  Rev. B \textbf{85} (2012), 195149.

\bibitem[TS72]{Takahashi1972}
M~Takahashi and M~Suzuki, \emph{One-dimensional anisotropic heisenberg model at
  finite temperatures}, Progress of Theoretical Physics \textbf{48} (1972),
  no.~6, 2187--2209.

\bibitem[VC14]{Caux_spin_1}
R~Vlijm and J-S Caux, \emph{{Computation of dynamical correlation functions of
  the spin-1 Babujan--Takhtajan chain}}, Journal of Statistical Mechanics:
  Theory and Experiment \textbf{2014} (2014), no.~5, P05009.

\bibitem[Wit84]{witten1984non}
E~Witten, \emph{Non-{A}belian bosonization in two dimensions}, Communications
  in Mathematical Physics \textbf{92} (1984), no.~4, 455--472.

\end{thebibliography}

\end{document}